\documentclass[preprint,superscriptaddress,nofootinbib]{revtex4-1}
\usepackage[english]{babel}
\usepackage{graphicx}
\usepackage[letterpaper,top=2cm,bottom=2cm,left=2cm,right=2cm,marginparwidth=1.25cm]{geometry}
\usepackage{braket}
\usepackage{tabularx}
\usepackage{float}
\usepackage{amsmath}
\usepackage{physics}
\usepackage{graphicx}
\usepackage{xcolor}
\usepackage{comment}
\usepackage{caption}
\usepackage{subcaption}
\usepackage[colorlinks=true, allcolors=blue]{hyperref}
\usepackage{lipsum}

\begin{document}
\title{Phase space distributions in information theory}

\author{Vikash Kumar Ojha}
\email{vko@svnit.ac.in}
\affiliation{Department of Physics, Sardar Vallabhbhai National
Institute of Technology, Surat, 395 007, India.}
\author{Ramkumar Radhakrishnan}
\email{rradhak2@ncsu.edu}
\affiliation{Department of Physics, North Carolina State University, Raleigh, NC 27695, USA.}
\author{Siddharth Kumar Tiwari}
\email{sidkt2015@gmail.com}
\affiliation{Department of Physics, Sardar Vallabhbhai National
Institute of Technology, Surat, 395 007, India.}
\author{Mariyah Ughradar}
\email{ds22ph003@phy.svnit.ac.in}
\affiliation{Department of Physics, Sardar Vallabhbhai National
Institute of Technology, Surat, 395 007, India.}
\begin{abstract}
We use phase space distributions—specifically, the Wigner distribution (WD) and Husimi distribution (HD)—to investigate certain information-theoretic measures as descriptors for a given system. We extensively investigate and analyze Shannon, Wehrl and Rényi entropies, its divergences, mutual information and other correlation measures within the context of these phase space distributions. The analysis is illustrated with an anharmonic oscillator and is studied with respect to perturbation parameter ($\lambda$) and states ($n$). The entropies associated with the Wigner distribution are observed to be lower than those of the Husimi distribution, which aligns with the findings regarding the marginals. Moreover, the real components of the entropies associated with the Wigner distribution tend to approach the entropic uncertainty bound more closely compared to those of the corresponding Husimi distribution. Moreover, we quantify the precise amount of information lost when opting for the Husimi distribution over the Wigner distribution for characterizing the specified system. Since it is not always positive definite, the entropies cannot always be defined. 
\end{abstract}
\maketitle
\section{Introduction}\label{sec:introduction}
Originally used to study how communication systems work, information theory has become important in understanding things like quantum science. It combines ideas from Classical Information Theory, Quantum Mechanics and Computer Science \cite{djordjevic2021quantum}. A lot of research has been done using the applications of information-theoretic measures to analyse various quantum systems \cite{atre2004quantum,chatzisavvas2005information,lin2015shannon,najafizade2016nonrelativistic,ghosal2016information,edet2021shannon,bera2020relaxation,gadre1985some,sen2005characteristic}. Among them, Rényi entropies have been utilized in the study of quantum systems as well\cite{romera2008renyi,martinez2021shannon,linden2013structure,ou2019shannon}. This is important because the study of Rényi entropy plays a significant role in understanding quantum entanglement \cite{brydges2019probing,hastings2010measuring,adesso2012measuring,wang2016entanglement}. Nevertheless, the majority of these studies have focused on representations in either coordinate or momentum space by analyzing their concerned densities. This emphasis is attributed to their importance about the entropic uncertainty bound \cite{bialynicki1975uncertainty,hertz2019continuous}. Additionally, some studies explore the use of information-based methods for these distributions \cite{wehrl1979relation,grabowski1984wehrl,salazar2023phase,floerchinger2021wehrl}.\\ \\
Phase space or quasi-probability distributions have been studied in quantum mechanics. The Wigner distribution (WD), also referred to as the Wigner function (WF), is a tool to represent states in its configuration space. It extends the concepts of statistical mechanics to quantum mechanics, when searching quantum corrections to thermodynamics \cite{wigner1932quantum}. Despite providing such a connection, it has a few drawbacks. One such drawback is its lack of positive definiteness, which raises fundamental questions regarding the uncertainty principles and their relationship with the phase space representation. In \cite{moyal1949quantum}, it was demonstrated that the quantum analog of the Liouville equation is the von Neumann equation, which governs the time evolution of the density operator of a quantum system. In recent times, it has also become an excellent tool for studying strongly interacting systems, particularly various aspects of Quantum Chromodynamics \cite{mukherjee2014quark,mukherjee2015wigner,Ojha_2023,Jana_2025}, as it represents the probability density of partons under certain limits \cite{lorce2011quark}. It's also a valuable tool for examining systems in both momentum and coordinate space, particularly in optics and information theories \cite{bastiaans1979wigner,radhakrishnan2022wigner,forbes2000wigner}.\\ \\ 
In the past, there have been several efforts have been made to handle the negativity present in the Wigner distribution. One approach is to apply a Gaussian convolution to the Wigner distribution, resulting in the Husimi distribution \cite{husimi1940some,leonhardt1997measuring}, which is a positive definite function. In contrast, the marginals of the Wigner distribution accurately depict the space densities, and it is experimentally accessible \cite{spasibko2017experimental,kurtsiefer1997measurement,landon2018quantitative}.\\ \\  
This study aims to investigate and analyze the Shannon, Fisher, Rényi, and residual entropies when employed with Wigner distribution, Husimi distribution as well as their marginals. Additionally, other measures such as Fisher information measures are discussed. Related relative entropies, cumulative entropies along with statistical correlation measures, are also explored. The anharmonic oscillator is studied as it allows for investigating perturbations, thus enabling generalization for arbitrary perturbations. One of our goals is to examine how all the entropies and measures vary with perturbation parameter. All these are carried out for ground and first excited states for the anharmonic oscillator concerning the parameter $\lambda$. Here, we'll evaluate different formulations of Shannon and mutual information entropies for both the Wigner distribution, Husimi distribution, aiming to look their compatibility with various interpretations. \\ \\ 
The article follows this structure: In Section \ref{sec:phasespace}, we initiate by explaining the phase space distributions and the entropies linked with them. Specifically, we discuss the Wigner distribution, Husimi distribution and their respective entropies. In Section \ref{sec:infotheory}, we explore various aspects of Information Theory such as the Mutual Information measure, Relative Entropy, etc. In Section \ref{sec:aho}, we discuss these aspects using the anharmonic oscillator and interpret the results. Definitions of all topics will be provided in the following section before we proceed to discuss and examine our results. We set $\hbar = m = c = 1$ throughout this article and note that all subscripts denote the particular state of the system, while superscripts indicate the order of perturbation. The limits on all integrals are $[-\infty,\infty]$ unless otherwise stated.
\section{Phase space distributions and its entropies}\label{sec:phasespace}
In classical mechanics, the description of a particle system involves phase space distributions, depicting particle density at particular points in the configuration space at a particular time. However, for quantum states, the non-commutative nature of coordinate and momentum operators prevents their simultaneous determination \cite{heisenberg1927heisenberg}. Thus, the concept of phase space distributions isn't directly applicable in quantum mechanics. However, Wigner's phase-space formulation \cite{wigner1932quantum}, known as the Wigner distribution of quantum states, provide a comprehensive structure that mirrors classical statistics. In $h\rightarrow 0$ limit, textcolor{blue}{Wigner distribution} converges to classical phase space distributions. Within this framework, quantum states and operators are described by continuous functions characterized by canonical variables $x$ and $p$, commonly known as coordinate and momentum, respectively. The Wigner-Weyl transform \cite{case2008wigner} establishes a direct mapping between quantum operators ($\hat{W}$) and quantum phase-space distributions ($W(x,p)$). This transformation enables the unique association of each quantum operator with a corresponding distribution in phase space, expressed as:
\begin{equation}
    W(x,p) = \frac{1}{\pi \hbar}\int dq \hspace{0.35em}e^{2ipq/\hbar} \langle x-q|\hat{W}|x+q\rangle,
\end{equation}
We set the value of $\hbar$ to be equal to $1$ in the remainder of this paper. In terms of wave function $(\psi(x))$ it is given by
\begin{equation}
    W(x,p) = \frac{1}{\pi} \int_{-\infty}^{\infty} dy\hspace{0.5em} \psi^{*}(x-y)\psi(x+y)e^{-2ipy}.
\end{equation}
satisfying 
\begin{equation}
    \int_{-\infty}^{\infty} \int_{-\infty}^{\infty} \hspace{0.15em}dxdp\hspace{0.15em} W(x,p) = 1.
\end{equation}
This Wigner distribution serves as an approximation to a probability distribution in phase space, in accordance with quantum mechanics. It can also be characterized as a quasi-probability distribution, given that it shares many properties with a probability distribution \cite{radhakrishnan2022wigner}. The marginal distributions are obtained as follows:
\begin{equation}
    \hat{W}_{x}(x) = \bra{x}\hat{W}\ket{x} \hspace{2em},\hspace{2em} \hat{W}_{p}(p) = \bra{p}\hat{W}\ket{p}
\end{equation}
such that,
\begin{equation}
     \rho^{W}_{x}(x) = \int_{-\infty}^{\infty} dp\hspace{0.15em} W(x,p) \hspace{2em},\hspace{2em} \rho^{W}_{p}(p) = \int_{-\infty}^{\infty} dx\hspace{0.15em} W(x,p).
\end{equation}
The computation of the observable ($\hat{\mathcal{O}}$) in the state $\hat{W}$ is easily determined using the overlap formula with its Wigner distribution\cite{leonhardt2010essential}.
\begin{equation}
    <\hat{\mathcal{O}}> = \text{Tr}[\hat{\mathcal{O}}\hat{W}] = 2\pi \int_{-\infty}^{\infty} \int_{\infty}^{\infty} \hspace{0.15em} dx dp\hspace{0.15em} \hat{\mathcal{O}}(x,p)\hspace{0.15em} W(x,p).
\end{equation}
Its notable feature is its lack of positive definiteness \cite{kenfack2004negativity}, leading to fundamental inquiries regarding the uncertainty principle and its relationship with the phase space depiction. Numerous efforts have been made in the past to address the negativity of the Wigner distribution. For instance, the Hudson theorem \cite{hudson1974wigner} asserts that every pure non-Gaussian state will exhibit a Wigner distribution that permits negative regions. The Hudson theorem provides insights into the characteristics of Wigner distribution for specific quantum states. The negativity in the Wigner distribution is what we pay due to the principle of uncertainty, which prevents us from defining non-commuting coordinates \cite{kenfack2004negativity}. As a result, commonly used probability distribution measures, such as Shannon differential entropy, generally lose their clear definition when applied using WD. Shannon entropies of the marginals (wave function) are defined as 
\begin{equation}
    S_{x}^{W} = S_{x} = -\int_{-\infty}^{\infty} \hspace{0.15em}dx \hspace{0.15em} \rho^{W}_{x}(x)\ln[\rho^{W}_{x}(x)],
\end{equation}
\begin{equation}
     S_{p}^{W} = S_{p} = -\int_{-\infty}^{\infty}  \hspace{0.15em}dp \hspace{0.15em} \rho^{W}_{p}(p)\ln[\rho^{W}_{p}(p)].
\end{equation}
An effective method to make Wigner distribution functional is by applying a Gaussian smoothing to them. The result of this process is a well-defined probability distribution in configuration space known as  Husimi distribution \cite{leonhardt2010essential}. It is always positive definite and can be utilized to derive measures that rely on probability distributions. Despite its positive definiteness, the Husimi distribution \cite{husimi1940some} has its shortcomings. Performing a Gaussian convolution on the Wigner distribution results in a loss of significant information related to the system. This can be obtained by the Weistrass transform applied to the Wigner distribution\cite{husimi1940some,harriman1988some}.
\begin{equation}
   H(x,p) = \frac{1}{\pi}\int_{-\infty}^{\infty}\int_{-\infty}^{\infty}dx^{\prime}dp^{\prime}\hspace{0.5em} W(x^{\prime},p^{\prime})e^{-\frac{(x-x^{\prime})^{2}}{2}}e^{-2(p-p^{\prime})^{2s^{2}}}, 
\end{equation}
where $W(x^{\prime},p^{\prime})$ is the Wigner distribution, and $s$ is an arbitrary parameter that we will set to unity in this context. Consequently, measures obtained from the Husimi distribution cannot be considered an accurate representation of the system's true nature \cite{wehrl1979relation}. Another limitation is its absence of desirable properties found in the Wigner distribution, like the overlap formula \cite{leonhardt2010essential}. In contrast, the Husimi distribution is distinguished by its positivity and circumvents the negativity present in the Wigner distribution. Nevertheless, the Husimi distribution lacks the identical local structure as the Wigner distribution due to its transformation via a Gaussian filter \cite{appleby1999generalized}. In this context of information theory, the loss of information from the Wigner to the Husimi distribution primarily manifests as a loss of local structure. The Gaussian smearing involved in the transformation tends to blur out sharp peaks and valleys, leading to a smoother representation of the quantum state. While this smoothing process improves the interpretability of the distribution and facilitates the calculation of certain quantities, it also sacrifices some level of precision in capturing the exact localization and momentum properties of the quantum system \cite{pennini2004escort}. Overall, while the Husimi distribution offers advantages in terms of positivity and ease of interpretation, it comes at the cost of sacrificing some of the detailed information present in the original Wigner distribution. The Shannon entropy of the Husimi distribution, known as Wehrl entropy \cite{wehrl1979relation}. Due to its positive definiteness, the Shannon entropy associated with it is a real-valued quantity. Nevertheless, it is positive definite, and it comes with an associated entropy called the Wehrl entropy ($S_{H}$), defined as follows:
\begin{equation}
    S_{H} = -\int_{-\infty}^{\infty} \int_{-\infty}^{\infty} \hspace{0.15em} dx dp\hspace{0.15em} H(x,p)\ln H(x,p),
\end{equation}
where $H(x,p)$ is the Husimi distribution, satisfying the condition
\begin{equation}
    \int_{-\infty}^{\infty} \int_{-\infty}^{\infty} \hspace{0.15em}dxdp\hspace{0.15em} H(x,p) = 1.
\end{equation}
Its marginals are 
\begin{equation}
    \rho^{H}_{x}(x) = \int_{-\infty}^{\infty} \hspace{0.15em}dp\hspace{0.15em} H(x,p) \hspace{2em},\hspace{2em} \rho^{H}_{p}(p)    = \int_{-\infty}^{\infty} \hspace{0.15em}dx\hspace{0.15em} H(x,p).
\end{equation}
Entropies of the wave function marginals are given as 
\begin{equation}
    S_{x}^{H} = -\int_{-\infty}^{\infty} \hspace{0.15em}dx \hspace{0.15em} \rho^{H}_{x}(x)\ln[\rho^{H}_{x}(x)],
\end{equation}
\begin{equation}
     S_{p}^{H} =  -\int_{-\infty}^{\infty}  \hspace{0.15em}dp \hspace{0.15em} \rho^{H}_{p}(p)\ln[\rho^{H}_{p}(p)].
\end{equation}
Another approach to eliminating the negativity of the Wigner distribution involves utilizing cumulative or survival distributions. A cumulative distribution function (CDF) represents the probability that a random variable will have a value less than or equal to a specific point. We employ this technique to remove the Wigner negativity. 
\begin{equation} \label{eq:survival}
    s_{W}(a,b) = \int_{a}^{\infty} \int_{b}^{\infty}dxdp\hspace{0.5em}W(x,p),
\end{equation}
whose survivals are
\begin{equation}
    s_{x}^{W}(a) = \int_{a}^{\infty}dx\hspace{0.5em}\rho(x),
\end{equation}
\begin{equation}
    s_{p}^{W}(b) = \int_{b}^{\infty}dp\hspace{0.5em}\rho(p).
\end{equation}
Regions that get integrated out are the regions of the distribution that corresponds to the negative entropy. Using this we can obtain cumulative residual entropy ($C$), which is a measure used in information theory to quantify the uncertainty or randomness of a sequence of events over time \cite{rao2004cumulative}. This is given as
\begin{equation}
    C^{x}_{W} = -\int_{-\infty}^{\infty} da \hspace{0.5em} s_{x}^{W}(a)\ln s_{x}^{W}(a),
\end{equation}
\begin{equation}
    C^{p}_{W} = -\int_{-\infty}^{\infty} da \hspace{0.5em} s_{p}^{W}(b)\ln s_{p}^{W}(b).
\end{equation}
An entropic uncertainty relationship can be formed between them.  We argue that the Wigner entropy, despite applying only to Wigner-positive states, serves as a fitting parameter for quantifying  uncertainty in configuration space. It offers insights into the uncertainty of the marginal distributions of the \(x\) and \(p\) variables, as well as their correlations in the configuration space. Unlike Wehrl entropy \cite{lieb1978proof}, it does not reflect the classical entropy resulting from a specific measurement outcome. The Wigner entropy possesses intriguing characteristics, also, it maintains its invariance under symplectic transformations in configuration space \cite{wehrl1979relation}. It is important to emphasize that an effective measure of phase space uncertainty should retain this invariance, as symplectic transformations are also transformations that preserve area in phase space \cite{cover1991network}. However, the Wehrl entropy is higher for squeezed states compared to coherent states \cite{lieb1978proof}. 
\section{Aspects of Information theory}\label{sec:infotheory}
After computing the phase space distributions, our next step involves extending our analysis to include the calculation of entropies within the regime of information theory. We begin by computing Kullback Leibler (KL) divergence, also known as relative entropy \cite{van2014renyi}, which serves as a fundamental concept utilized to measure the disparity between two probability distributions. Specifically, it measures how one probability distribution diverges from another, expected distribution. We can compute the relative entropy between the probability distributions (Wigner distribution, 
 Husimi distribution) obtained from the system's wave function using both the Wigner distribution and Husimi distribution.  Kullback-Leibler (KL) divergence is defined as 
\begin{equation}
        S_{\text{Relative}} = S_{KL} = \int_{-\infty}^{\infty} dx\hspace{0.5em} P(x) \ln\bigg[\frac{P(x)}{Q(x)}\bigg]. 
    \end{equation}
For density matrices $P$ and $Q$ in a Hilbert space  $\mathcal{H}$, the relative entropy from  $Q$  to $P$ is defined as
\begin{equation}
    S_{\text{Relative}} = S_{KL} = Tr\bigg[P(\log P-\log Q)\bigg].
\end{equation}
Similarly, cumulative residual entropies give rise to Jeffreys divergence ($R$) \cite{seal2020fuzzy}. This acts as the counterpart to the relative entropy described in equation \ref{eq:survival}. It is utilized to gauge the divergence between the survival densities of the marginals in the Wigner distribution, and Husimi distribution.
\begin{equation}
    R = \int_{-\infty}^{\infty} da\Bigg[ s_{x}^{W}(a)\ln \bigg(\frac{s^{x}_{W}(a)}{s^{x}_{H}(a)}\bigg) + s^{x}_{H}(a)\ln \bigg(\frac{s^{x}_{H}(a)}{s^{x}_{W}(a)}\bigg)  \Bigg].
\end{equation}
In Information theory, the smallest relative entropy value across all separable states $Q$ is the degree of entanglement in the state $P$. We explained the concept of relative entropy and its role in quantifying differences between probability distributions, we can seamlessly transition to discussing mutual information. Mutual information \cite{cover1999elements}, in the context of information theory, serves as a measure of the amount of information shared between two random variables or sets of variables \cite{kraskov2004estimating}. It captures the extent to which the knowledge of one variable reduces uncertainty about the other. Formally, mutual information is defined in terms of the joint and probability distributions of the variables in question. By quantifying the statistical dependence or correlation between variables, mutual information provides valuable insights into the relationships within a system. Interestingly, there exists a close connection between mutual information and relative entropy \cite{mackay2003information}. It can be formulated in terms of relative entropy, underscoring their complementary functions within information theory. While relative entropy quantifies the disparity between probability distributions, mutual information centers on the shared information among variables. The correlation between coordinate and momentum (mutual information) in the WF can be defined as \cite{cover1999elements}
\begin{equation} \label{eq:mutualwigner}
    I^{W}_{\text{Mutual}} = \int_{-\infty}^{\infty} \int_{-\infty}^{\infty} dx dp\hspace{0.5em} W(x,p) \ln\bigg[\frac{W(x,p)}{
    \rho^{W}_{x}(x)\rho^{W}_{p}(p)}\bigg], 
\end{equation}
where $\rho^{W}_{x}(x), \rho^{W}_{p}(p)$ are the marginals obtained using the Wigner distribution. In the previous section, we examined Shannon entropies derived from the Wigner distribution. We utilize these entropies to calculate mutual information, which serves as a measure of the shared information between two variables. Mutual information using Wigner distribution is given by \cite{salazar2023phase}
\begin{equation}
    I^{W}_{\text{Mutual}} = S^{W}_{x}+S^{W}_{p}-S_{W},
\end{equation}
$S^{W}_{x}+S^{W}_{p}$ obeys the entropic uncertainty relationship \cite{bialynicki1975uncertainty,beckner1975inequalities}
\begin{equation}
    S^{W}_{x}+S^{W}_{p} \geq 1+\ln\pi.  
\end{equation}
We extend the same for the Husimi distribution to obtain mutual information in this context \cite{grabowski1984wehrl}. In the context of the Husimi distribution, mutual information is defined as \cite{floerchinger2021wehrl} 
\begin{equation}
        I^{H}_{\text{Mutual}} = \int_{-\infty}^{\infty} \int_{-\infty}^{\infty} dx dp\hspace{0.5em} H(x,p) \ln\bigg[\frac{H(x,p)}{
    \rho^{H}_{x}(x)\rho^{H}_{p}(p)}\bigg], 
    \end{equation}
where $\rho^{H}_{x}(x), \rho^{H}_{p}(p)$ are the marginals obtained using the Husimi distribution. Mutual information can be also obtained using Husimi distribution
\begin{equation}
    I^{H}_{\text{Mutual}} = S^{H}_{x}+S^{H}_{p}-S_{H}\geq 0,
\end{equation}
The equivalent of mutual information is the cross-cumulative residual entropy($\mathcal{C}$) (\cite{rao2004cumulative}), which assesses correlation in a survival distribution. It is defined for the Wigner distribution on as
\begin{equation} \label{eq:crosscumulative}
    \mathcal{C}^{W} = C^{W}_{x} -\epsilon,
\end{equation}
where $$\epsilon = -\int_{-\infty}^{\infty} \int_{-\infty}^{\infty} dbdx \int_{b}^{\infty} dp W(x,p)\ln\bigg[\frac{\int_{b}^{\infty} dp W(x,p)}{\rho(x)}\bigg]$$
which is the correction term. This can be obtained using Husimi distribution and also in momentum space. So far, we've discussed methods of deriving densities associated with the Wigner distribution, which are characterized by their positivity. These densities are preferred because they avoid complications regarding their application in entropic definitions. Densities that lack positivity can lead to entropies with complex values. Another method involves employing non-positive definite densities in entropic expressions, avoiding the generation of complex-valued quantities. One of our motivations for undertaking this work is to explore entanglement entropy \cite{wang2016entanglement}, and one approach to this is to obtain Rényi entropy. Rényi entropy is a measure that quantifies the uncertainty or randomness associated with a probability distribution \cite{van2014renyi}. It generalizes Shannon entropy, offering a family of entropy measures indexed by a parameter $\alpha$. Let us illustrate it with an example, if we have a system initially at thermal equilibrium and then rapidly decrease its temperature by a factor of $\alpha$, the maximum work the system can perform as it returns to equilibrium at the new temperature, divided by the change in temperature, is equivalent to the Rényi entropy of the system in its original state (\cite{kraskov2004estimating}). This holds true for both classical and quantum systems. Mathematically, this can be articulated as the Rényi entropy being the $\alpha$ deformation of the conventional concept of entropy, specifically as the $\alpha^{-1}$ derivative of the system's free energy with respect to temperature. It is defined in as
\begin{equation}
    R_{x}^{\alpha} = \frac{1}{1-\alpha} \ln \bigg[\int_{-\infty}^{\infty} dx\hspace{0.5em} \bigg\{|\rho^{W}_{x}(x)|^{2}\bigg\}^{\alpha} \bigg], 
\end{equation}
\begin{equation}
    R_{p}^{\beta} = \frac{1}{1-\beta} \ln \bigg[\int_{-\infty}^{\infty} dp\hspace{0.5em} \bigg\{|\rho^{W}_{p}(p)|^{2}\bigg\}^{\beta} \bigg],
\end{equation}
where $\alpha$ and $\beta$ are parameters, $\rho^{W}_{x}(x), \rho^{W}_{p}(p)$ are the marginals obtained using Wigner distribution. When these parameters ($\alpha$ and $\beta$) approach a numerical value of $1$, the expressions of Rényi entropies reduce to the Shannon entropies in their respective spaces. 
\begin{equation}
    \lim_{\alpha\rightarrow 1} R_{x}^{\alpha} = S^{W}_{x} \hspace{2em}, \hspace{2em} \lim_{\beta\rightarrow 1} R_{p}^{\beta} = S^{W}_{p},
\end{equation}
where $S^{W}_{x}, S^{W}_{p} $ are the Shannon entropies in coordinate and momentum space obtained using Wigner distribution. Now, we will study the Rényi entropy of the Wigner distribution. In the following discussion, we employ the same notation ($\alpha, \beta$) for this parameter as used in the entropy computation for coordinate and momentum space. The choice $\alpha = 2 = \beta$ is referred to as collision entropy, which bears importance in the context of the Wigner distribution (\cite{muller2013quantum}). This is illustrated as follows:
\begin{equation}
    R^{W}_{2} = -\ln\bigg[\int_{-\infty}^{\infty} \int_{-\infty}^{\infty} dx dp \hspace{0.5em} W(x,p)^{2}\bigg].
\end{equation}
The entropy is defined despite the Wigner distribution being negative when the specific value of $\alpha = 2$ is chosen. The Rényi uncertainty bound is given as \cite{zozor2007classes}
\begin{equation}
     R_{p}^{\beta} +  R_{x}^{\alpha} \geq -\frac{1}{2(1-\beta)}\ln\bigg(\frac{\beta}{\pi}\bigg)-\frac{1}{2(1-\alpha)}\ln\bigg(\frac{\alpha}{\pi}\bigg),
\end{equation}
with the restriction that 
\begin{equation}
    \frac{1}{\alpha} + \frac{1}{\beta} = 2.
\end{equation}
Using such a relationship with varying parameter values in each space complicates the comparison between entropies obtained through probability distributions and a properly defined phase space entropy. It raises the question of which parameter value to prioritize for the phase space entropy. To facilitate a fair comparison, a consistent parameter value should be used across all entropic definitions. Moreover, it's essential for this parameter to be an even value to guarantee a phase space Rényi entropy that is real-valued. However, there is an additional importance when the value of $\alpha = \beta = 2$. The Rényi uncertainty relation is expressed as:
\begin{equation} \label{eq:renyiuncertain}
    R^{W}_{2} = S_{x}^{2, W} + S_{p}^{2, W} \geq \ln2\pi.
\end{equation}
Here, the terms on the right-hand side denote the Rényi entropies of the probability distribution. Rényi divergence (RD), also referred to as Rényi relative entropy, serves as a measure of the discrepancy between two probability distributions. It is an extension of Kullback-Leibler (KL) divergence and is defined as a function of a parameter $\alpha$, akin to Rényi entropy. For two probability distributions $P(x)$ and $Q(x)$, the Rényi divergence of order $\alpha$ from $Q$ to $P$ is given by:
\begin{equation}
    S_{RD} = \frac{1}{\alpha-1}\ln\bigg[\int_{-\infty}^{\infty} dx\hspace{0.5em}\frac{P(x)^{\alpha}}{Q(x)^{\alpha-1}}\bigg].
\end{equation}
For this particular choice of $\alpha = 2$, Rényi mutual information ($I^{W}_{2}$) is defined for the Wigner distribution as
\begin{equation}
    I^{W}_{2} = \ln \bigg[\int_{-\infty}^{\infty} \int_{-\infty}^{\infty} dxdp\hspace{0.5em} \frac{W[x,p]^{2}}{\rho^{W}_{x}(x)\rho^{W}_{p}(p)}\bigg],
\end{equation}
where $\rho^{W}_{x}(x), \rho^{W}_{p}(p)$ are the wave functions obtained using the Wigner distribution in coordinate and momentum space respectively. The corresponding measures for the Husimi distribution (Rényi entropy, mutual information) are obtained by replacing the Wigner distribution with the HF and modifying the probability densities accordingly in the previously discussed equations. The uncertainty relations formulated in relation to the Husimi distribution \cite{gonccalves2011renyi} are
\begin{equation}
    \text{For}\hspace{0.3em} \alpha = 2, \hspace{1.5em} R^{H}_{2} = S_{x}^{2,H} + S_{p}^{2,H} \geq \ln2\pi.
\end{equation}
Another divergence called Cauchy - Schwarz divergence ($D_{CS}$) \cite{hoang2015cauchy} has a form similar to mutual information. It's a measure of how two variables or vectors deviate from being perfectly aligned. It is particularly defined for one variable. For given distributions $f_{1}(x)$ and $f_{2}(x)$, Cauchy - Schwarz divergence ($D_{CS}$) is defined as 
\begin{equation}
    D_{CS} = -\ln\Bigg[\frac{\int dx [f_{1}(x)f_{2}(x)]^{2}}{\sqrt{[\int dx f_{1}(x)^{2}][\int dx f_{2}(x)^{2}]}}\Bigg].
\end{equation}
The mutual information concerned with the CS divergence is called Cauchy - Schwarz mutual information. 
\section{Anharmonic oscillator} \label{sec:aho}
\begin{figure}[H]
\centering
\begin{subfigure}{0.3\textwidth}
  \centering
  \includegraphics[width=\linewidth]{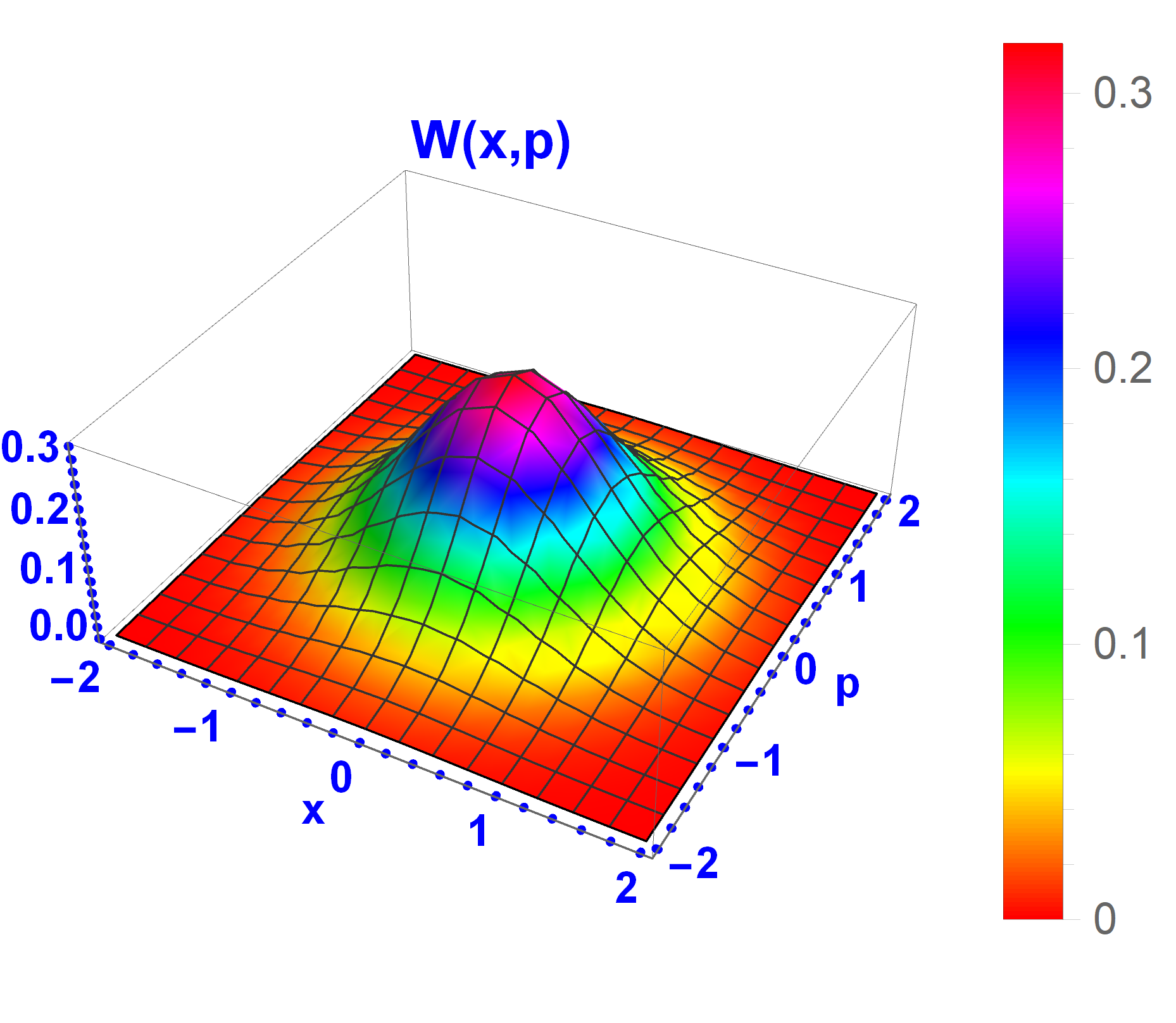}
  \caption{}
  \label{fig:image1}
\end{subfigure}
\begin{subfigure}{0.3\textwidth}
  \centering
  \includegraphics[width=\linewidth]{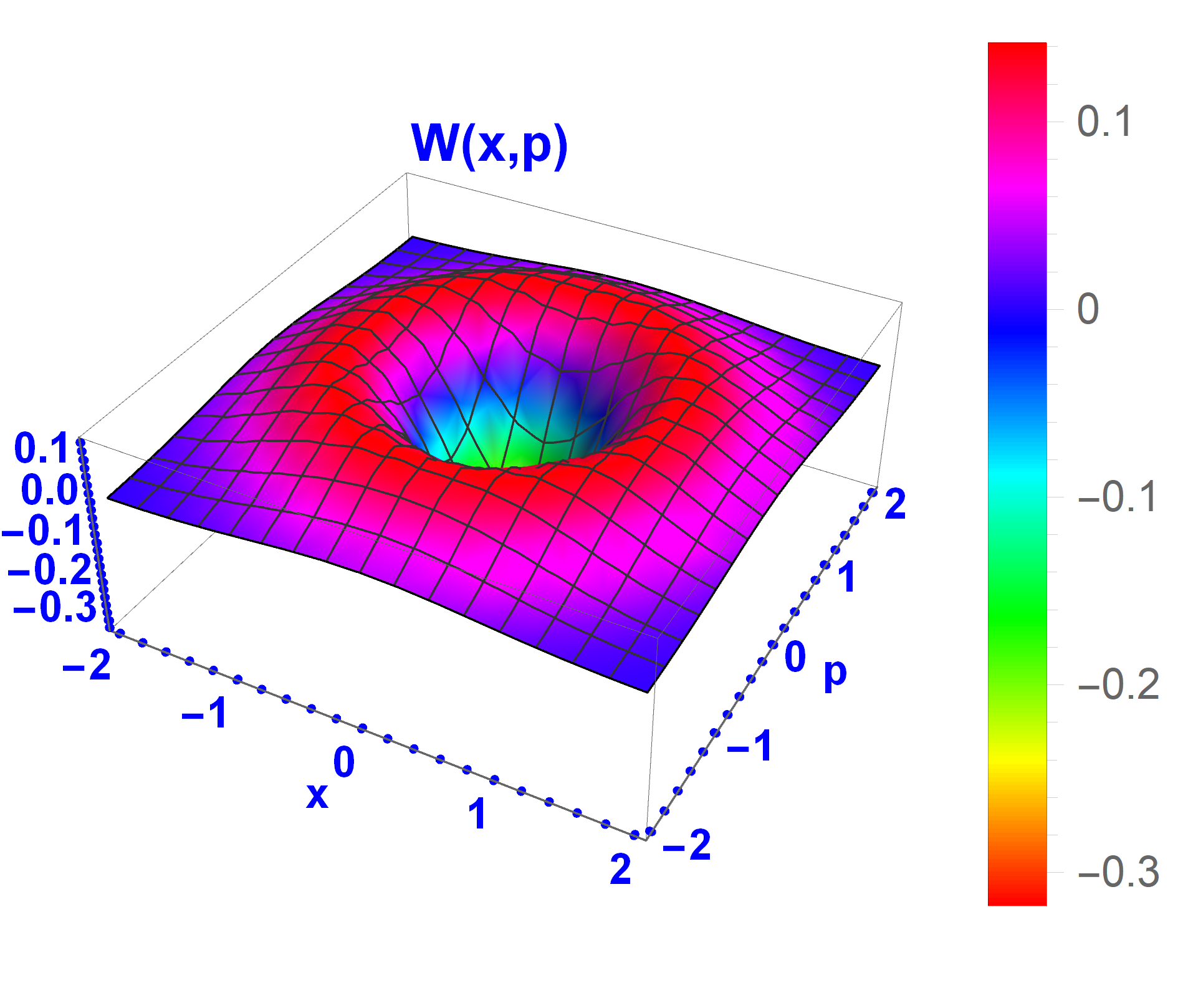}
  \caption{}
  \label{fig:image2}
\end{subfigure}
\begin{subfigure}{0.3\textwidth}
  \centering
  \includegraphics[width=\linewidth]{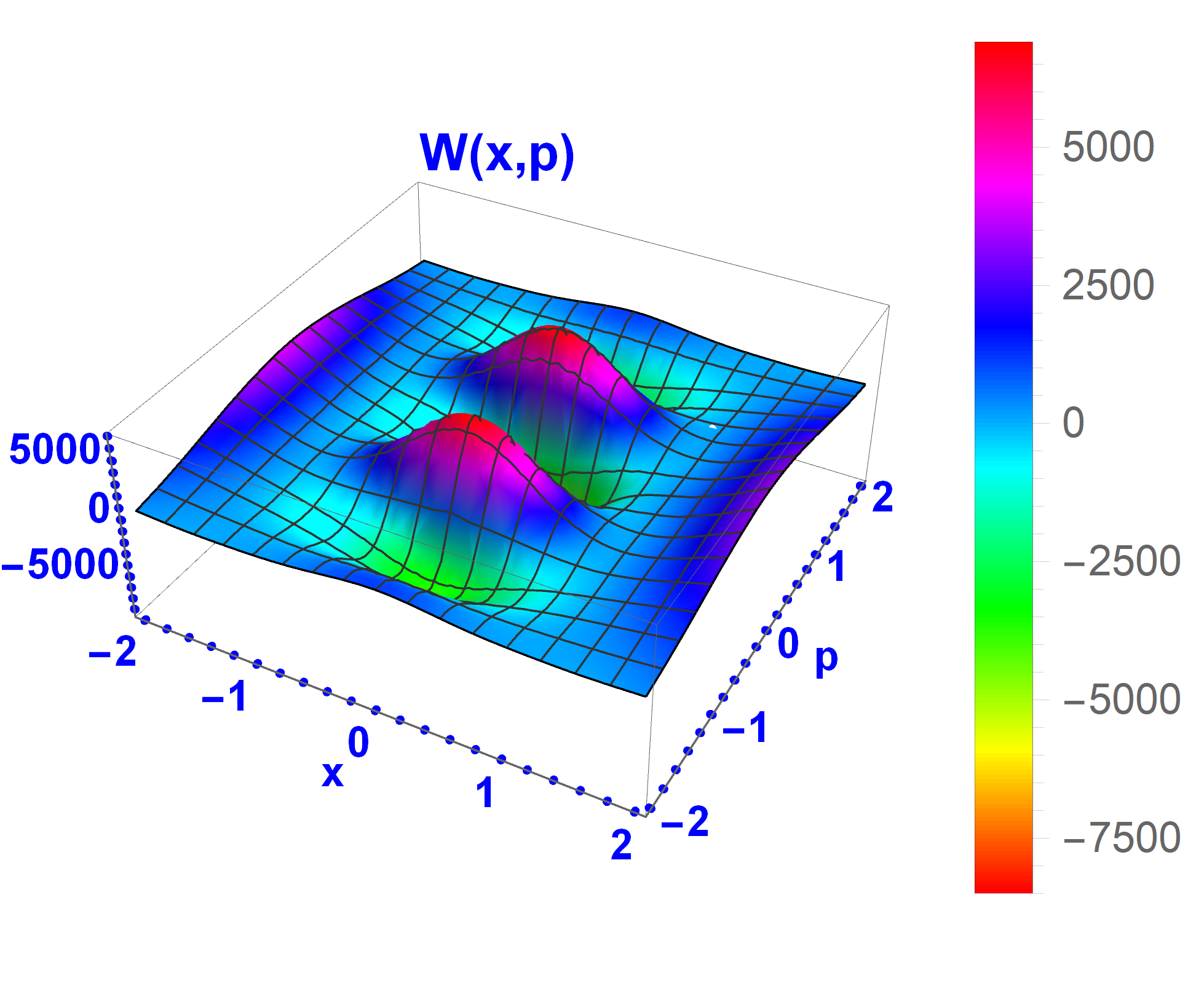}
  \caption{}
  \label{fig:image3}
\end{subfigure}
\begin{subfigure}{0.3\textwidth}
  \centering
  \includegraphics[width=\linewidth]{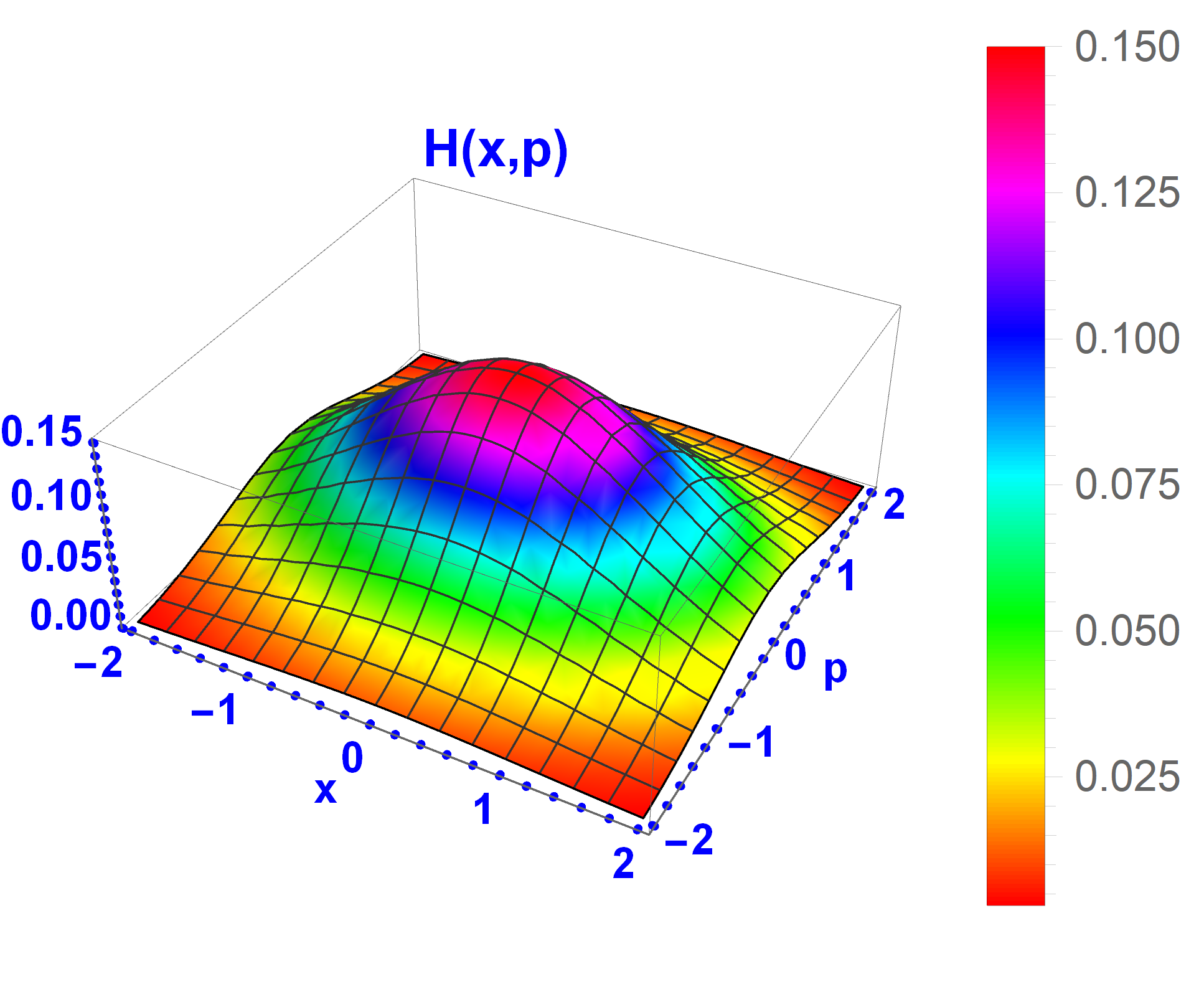}
  \caption{}
  \label{fig:image4}
\end{subfigure}
\begin{subfigure}{0.3\textwidth}
  \centering
  \includegraphics[width=\linewidth]{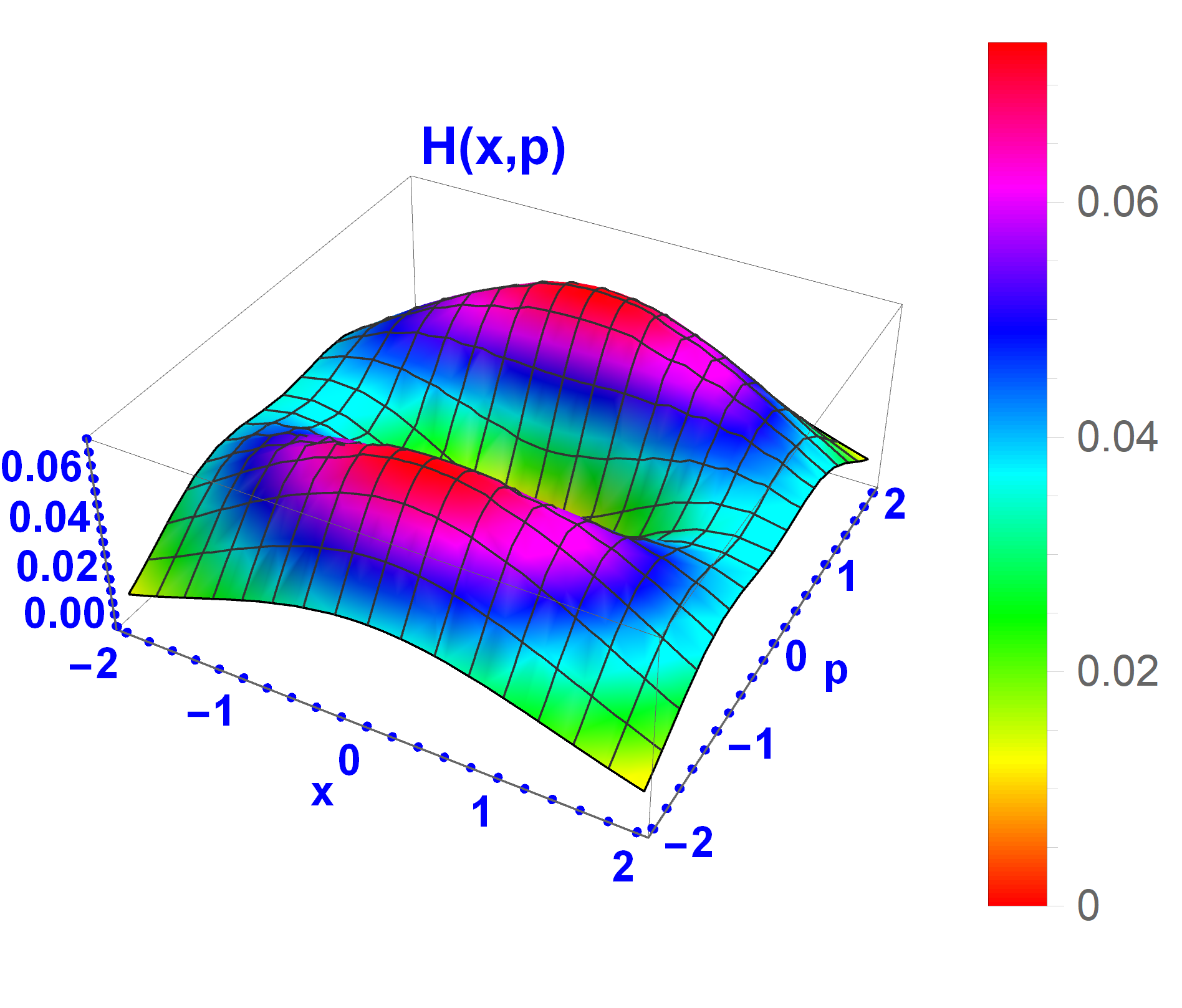}
  \caption{}
  \label{fig:image5}
\end{subfigure}
\begin{subfigure}{0.3\textwidth}
  \centering
  \includegraphics[width=\linewidth]{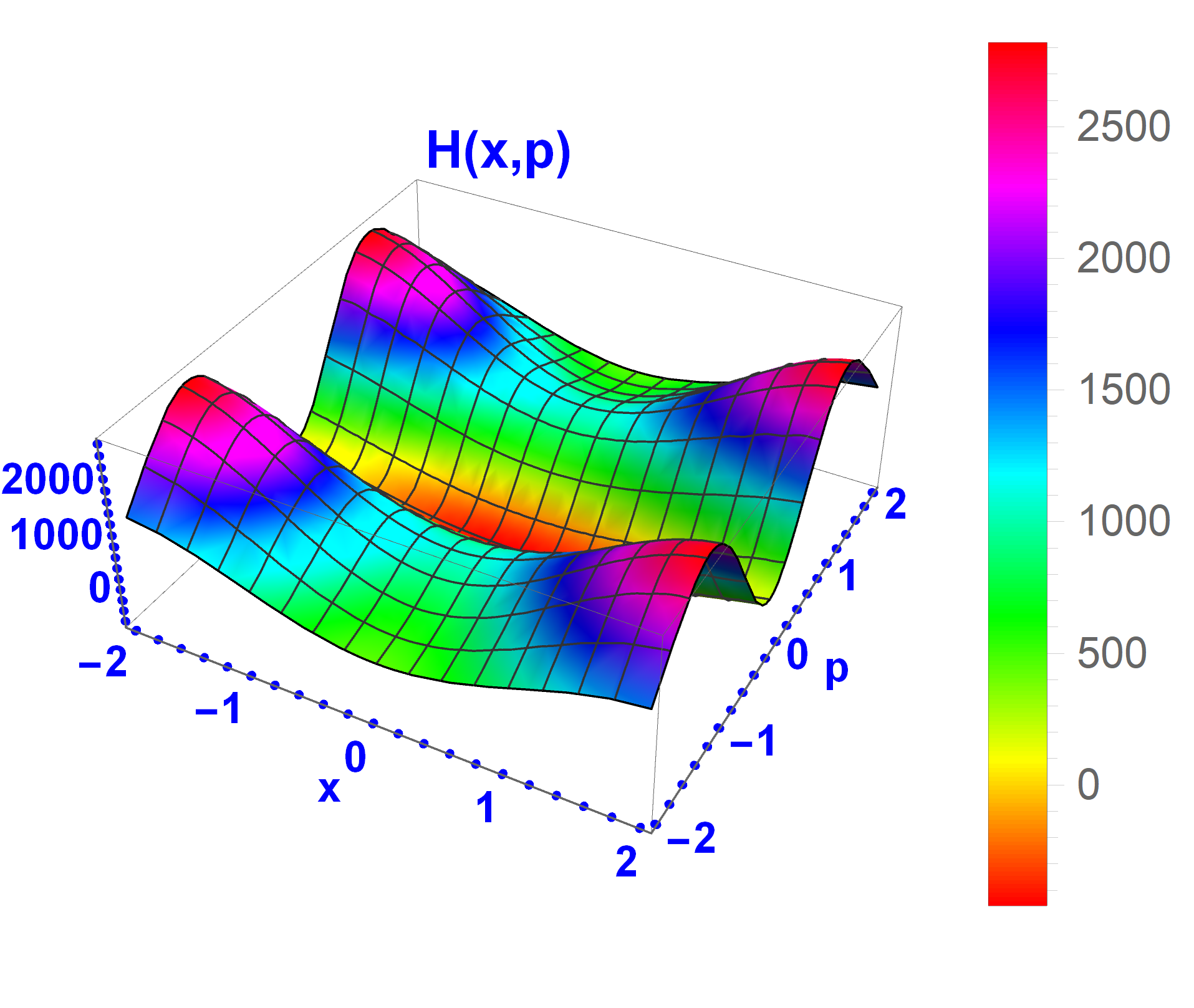}
  \caption{}
  \label{fig:image6}
\end{subfigure}
\caption{Plots (a)–(c) represent the Wigner distribution $W(x, p)$, while plots (d)–(f) depict the Husimi distribution $H(x, p)$, for states (a) and (d) with $n = 0$, (b) and (e) with $n = 1$, and (c) and (f) with $n=5$. The densities are plotted for $\lambda = 0.0001$.}
\label{fig:wignerhusimi}
\end{figure}

In this study, our primary focus shifts towards examining the anharmonic oscillator with a quartic potential. Our objective is to analyze and differentiate between the entropic and correlation measures relative to different states for different values of the parameters $\lambda$. The general Hamiltonian for an anharmonic oscillator is given by
\begin{equation}
    \mathcal{H} = \frac{p^{2}}{2m} + \frac{x^2}{2} + \frac{\lambda}{4} x^{4}.
\end{equation}
We will keep $m=1$ for the sake of simplicity. The wave function of an anharmonic oscillator is obtained using Perturbation theory and is given in ground state ($n=0$) as
    \begin{equation}
   \psi_{0}(x) = A_{0}\Bigg[1-\frac{\lambda}{4}\bigg(\frac{1}{4}(4x^4-12x^2+3)+3(2x^2-1)\bigg)\Bigg]e^{\frac{-x^2}{2}},
\end{equation}
where $A_{0} = \frac{1}{\sqrt{1-1.25\times 10^{-16}\lambda+1.21\lambda^{2}}}\bigg(\frac{1}{\pi}\bigg)^{1/4}$. In momentum space 
\begin{equation}
\phi_{0}(p) = B_{0}\Bigg[16-\lambda \bigg(4p^4-36p^2+15\bigg)\Bigg]e^{\frac{-p^2}{2}},
\end{equation}
where $B_{0} = \frac{0.0469}{\sqrt{0.998+1.216\lambda^{2}}}$. For $n=1$,
\begin{equation}
   \psi_{1}(x) = A_{1}\Bigg[1-\frac{\lambda}{32}\bigg\{10(2x^{2}-3)+\frac{1}{2}(4x^{4}-20x^{2}+15)\bigg\}\Bigg]xe^{-\frac{x^2}{2}},
\end{equation}
where $A_{1} = \frac{\sqrt{2}}{\pi^{1/4}\sqrt{1+2.505\times 10^{-16}\lambda+0.6152\lambda^{2}}}$. In momentum space
\begin{equation}
\phi_{1}(p) = B_{1}\Bigg[4p^{4}\lambda-60p^{2}\lambda+ 75\lambda-64\Bigg]p e^{-\frac{p^{2}}{2}},
\end{equation}
where $B_{1} = \frac{-0.0165 i}{\sqrt{1+2.505\times 10^{-16}\lambda+0.6152\lambda^{2}}}\sqrt{\frac{1.625 +4.071\times 10^{-16}\lambda+\lambda^{2}}{1.606+8.049\times 10^{-16}\lambda + 0.9883\lambda^{2}}}$. The corresponding Wigner distribution and Husimi distribution in different states are given in figure \ref{fig:wignerhusimi}.
We're particularly interested in examining the behavior of the Shannon entropies of the Wigner distribution and Husimi distribution. While the Wehrl entropy is real and Wigner entropy is complex. Additionally, we're exploring the entropies of their individual components. One might expect that the Shannon entropies of the Husimi distribution would be higher than those of the Wigner distribution. However, it's important to measure them when calculating statistical correlation (mutual information). This measures how much $x$ and $p$ variables are related. We'll also compare the sums of entropies from the Wigner distribution and Husimi distribution parts. These sums are used to create uncertainty relations based on entropy. We'll compare them to the entropies of the original distributions. Moreover, other types of entropies, like the Rényi entropy, are studied with the aim of obtaining Wigner entropies and correlation measures that are real-valued. We'll compare how these different measures behave to make sure our results are consistent. This is important for understanding and interpreting our findings accurately.
\section{Results and Discussion}
We begin by examining the local characteristics of Wigner distribution and Husimi distribution, along with their associated entropies and marginal densities for different states. Information measures are obtained using these results. In the subsequent section, we will analyze other entropic and correlation measures for different states. 
\subsection{Phase space distributions and marginals}
\begin{figure}[H]
\begin{subfigure}{0.3\textwidth}
  \centering
  \includegraphics[width=\linewidth]{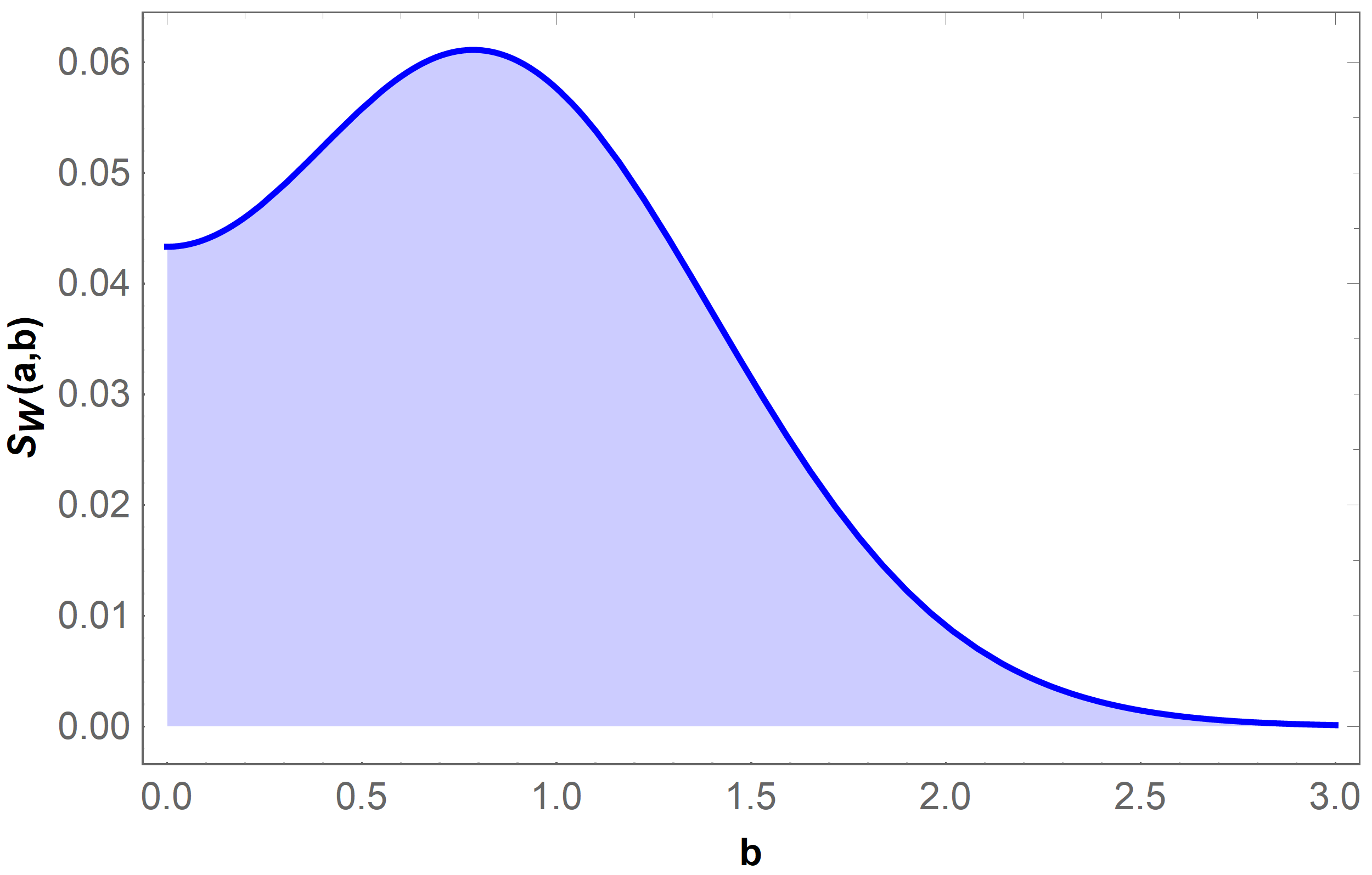}
  \caption{}
  \label{fig:image7}
\end{subfigure}
\begin{subfigure}{0.3\textwidth}
  \centering
  \includegraphics[width=\linewidth]{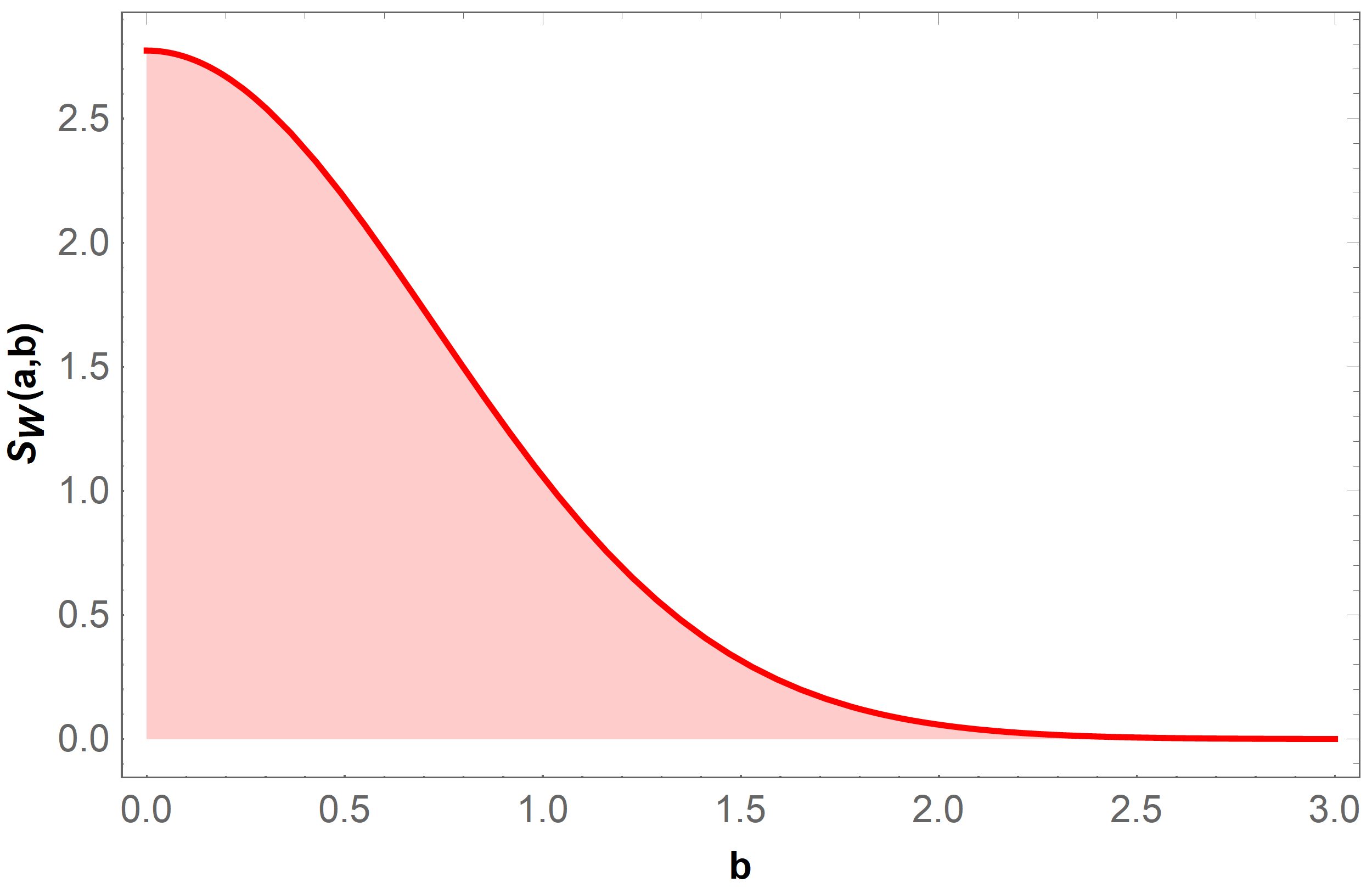}
  \caption{}
   \label{fig:image8}
\end{subfigure}
\begin{subfigure}{0.3\textwidth}
  \centering
  \includegraphics[width=\linewidth]{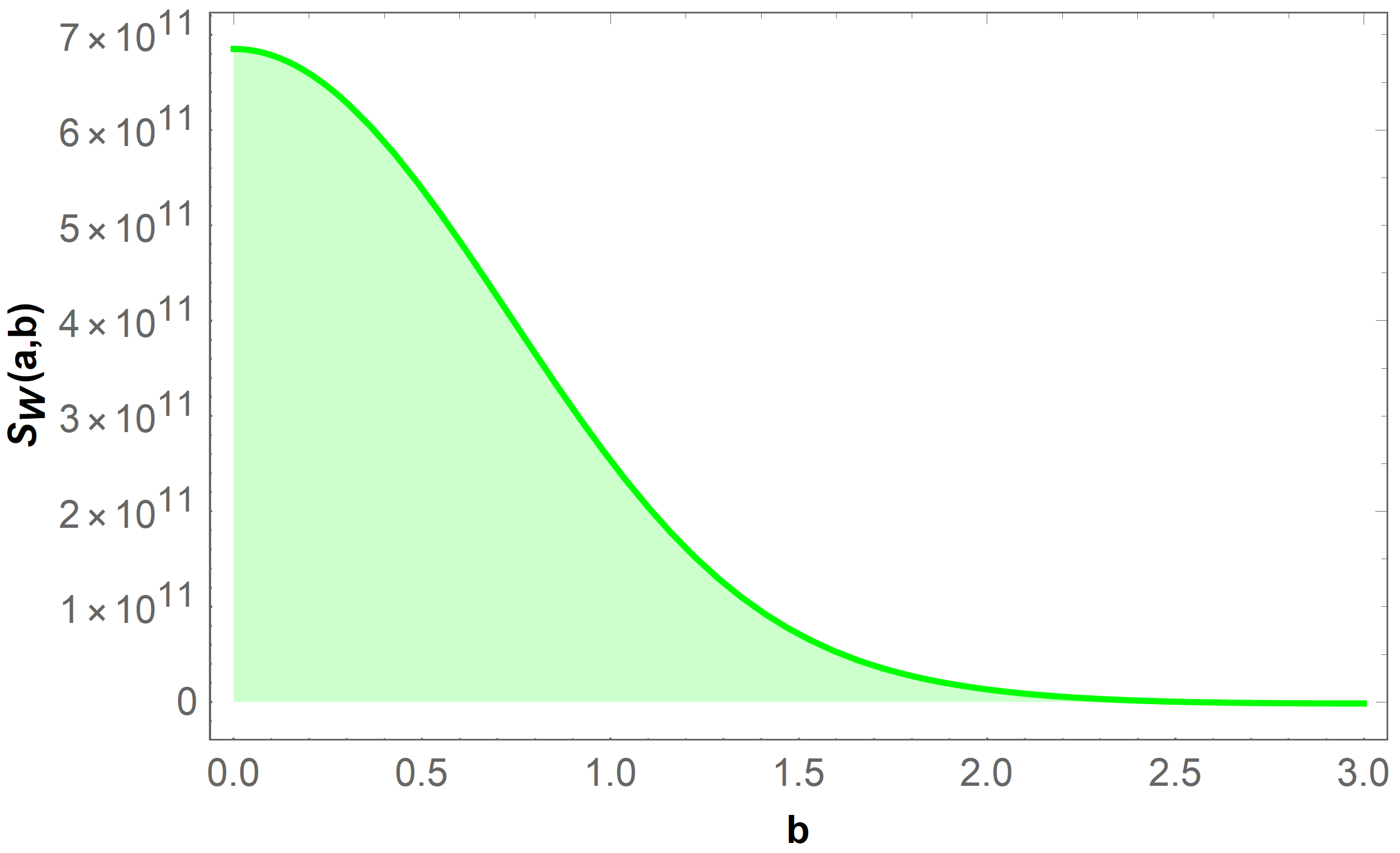}
  \caption{}
   \label{fig:image9}
\end{subfigure}
\begin{subfigure}{0.3\textwidth}
  \centering
  \includegraphics[width=\linewidth]{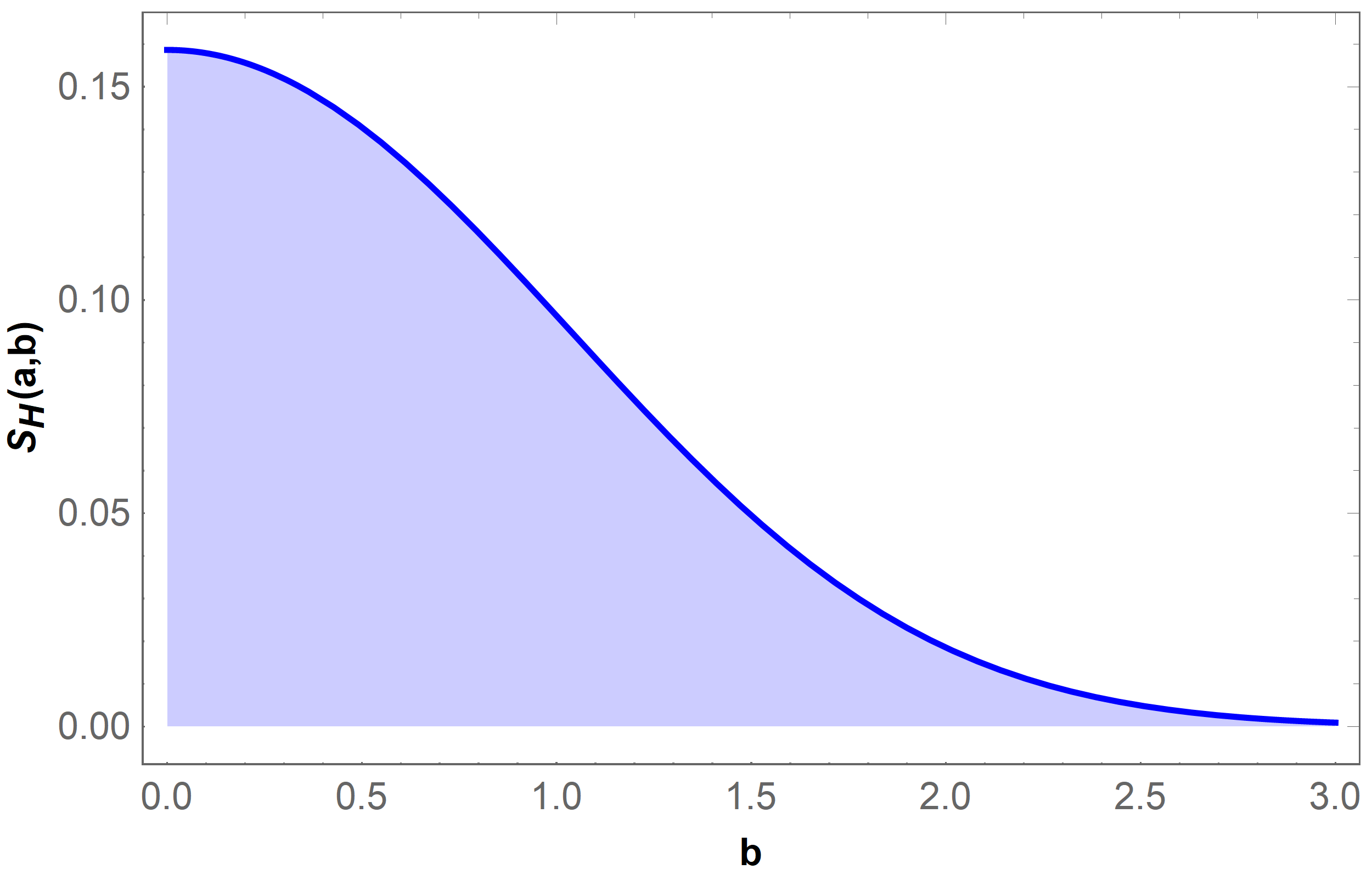}
  \caption{}
   \label{fig:image10}
\end{subfigure}
\begin{subfigure}{0.3\textwidth}
  \centering
  \includegraphics[width=\linewidth]{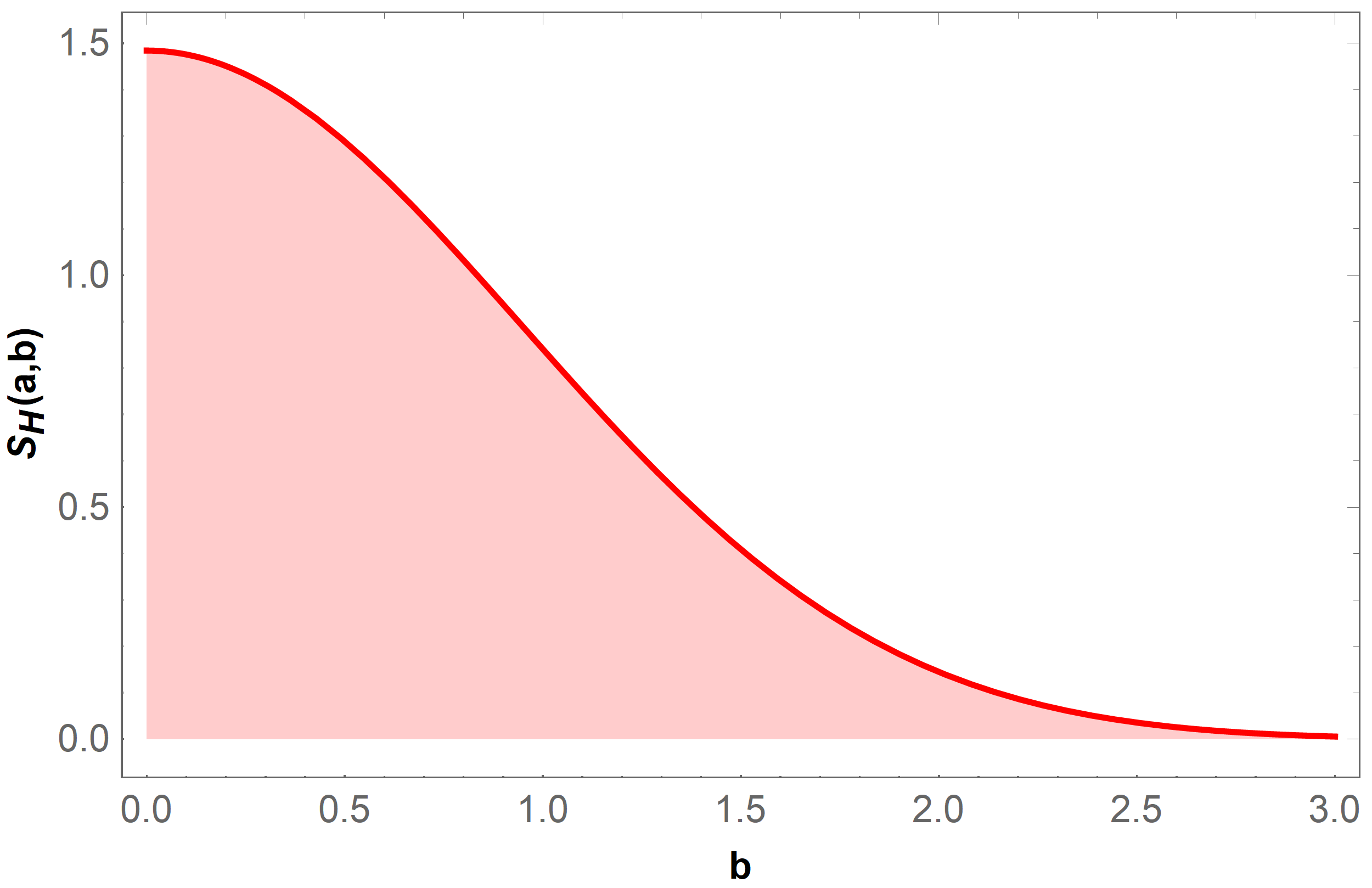}
  \caption{}
  \label{fig:image11}
\end{subfigure}
\begin{subfigure}{0.3\textwidth}
  \centering
  \includegraphics[width=\linewidth]{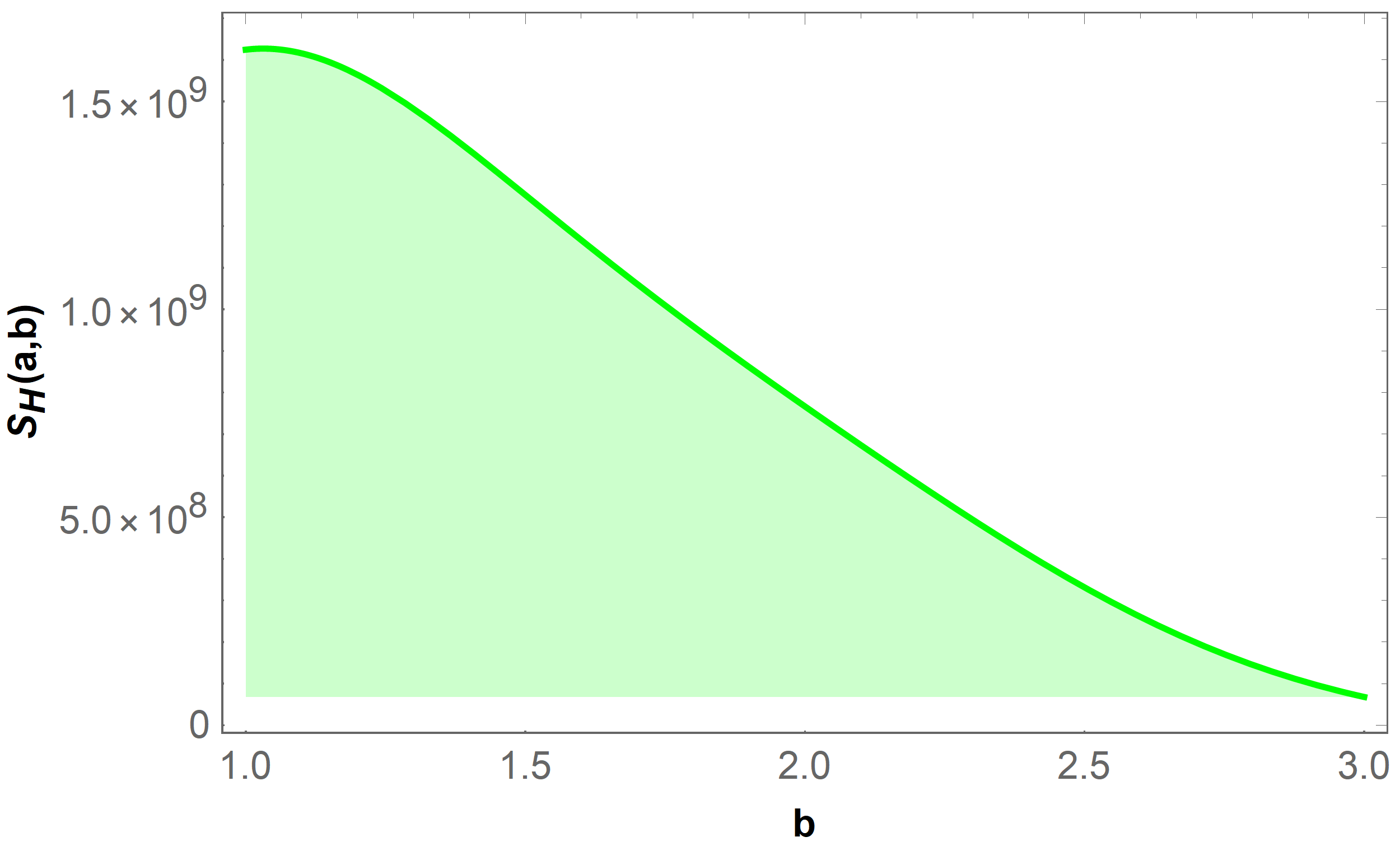}
  \caption{}
    \label{fig:image12}
\end{subfigure}
 \caption{Plots (a)-(c) are the survival functions of the Wigner distribution $s_{W}(a,b)$ and (d)-(f) are the survival functions of the Husimi distribution $s_{H}(a,b)$ for states with (a) and (d) for $n=0$, (b) and (e) for $n=1$ and (c) and (f) for $n=5$. The plots are shown as a function of $b$ by fixing the values of $a$ and parameter $\lambda$.}
 \label{fig:image2main}
\end{figure}

\begin{figure}[H]
\begin{subfigure}{0.3\textwidth}
  \centering
  \includegraphics[width=\linewidth]{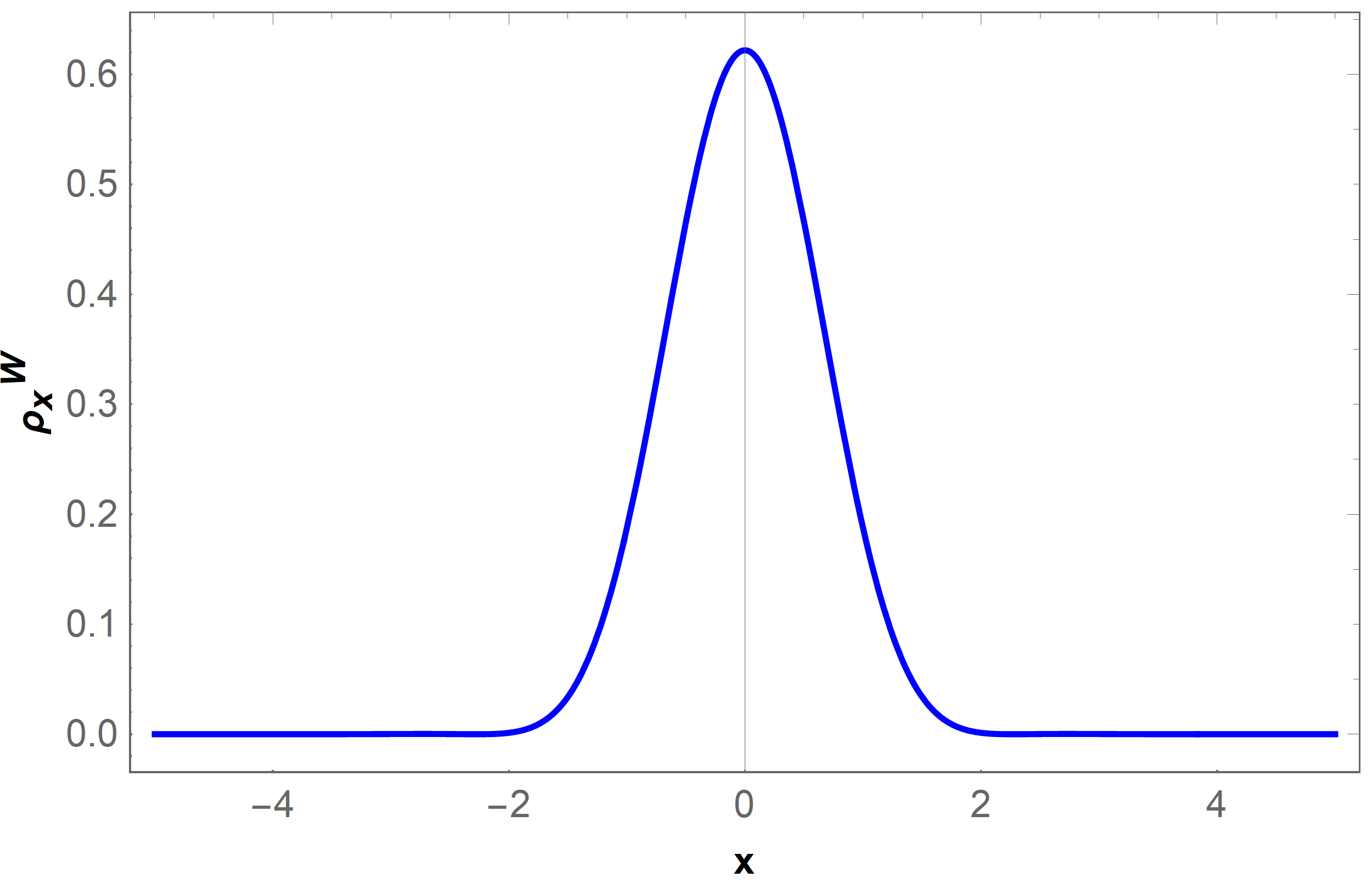}
  \label{fig:image13}
  \caption{}
\end{subfigure}
\begin{subfigure}{0.3\textwidth}
  \centering
  \includegraphics[width=\linewidth]{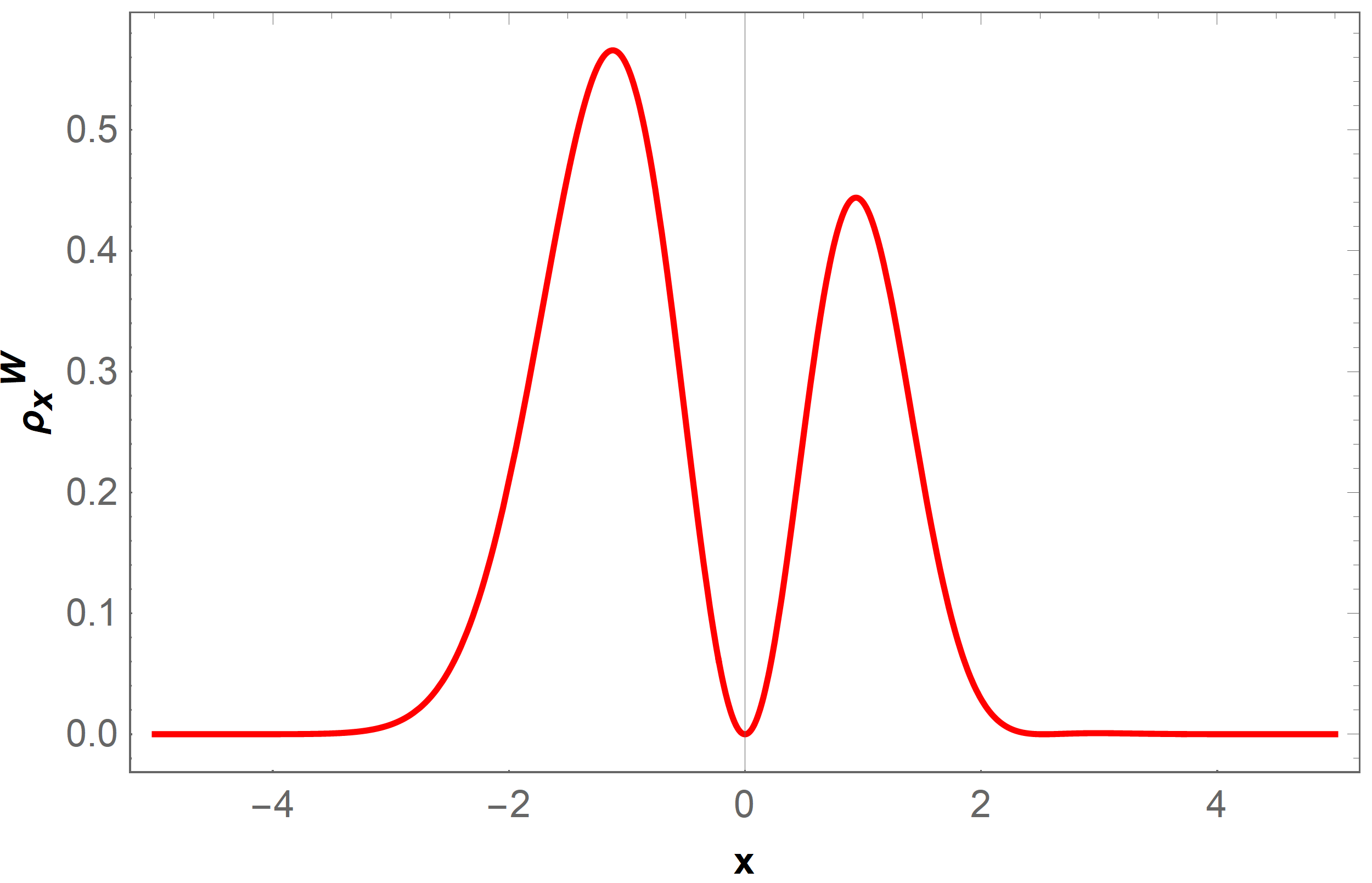}
    \caption{}
  \label{fig:image14}
\end{subfigure}
\begin{subfigure}{0.3\textwidth}
  \centering
  \includegraphics[width=\linewidth]{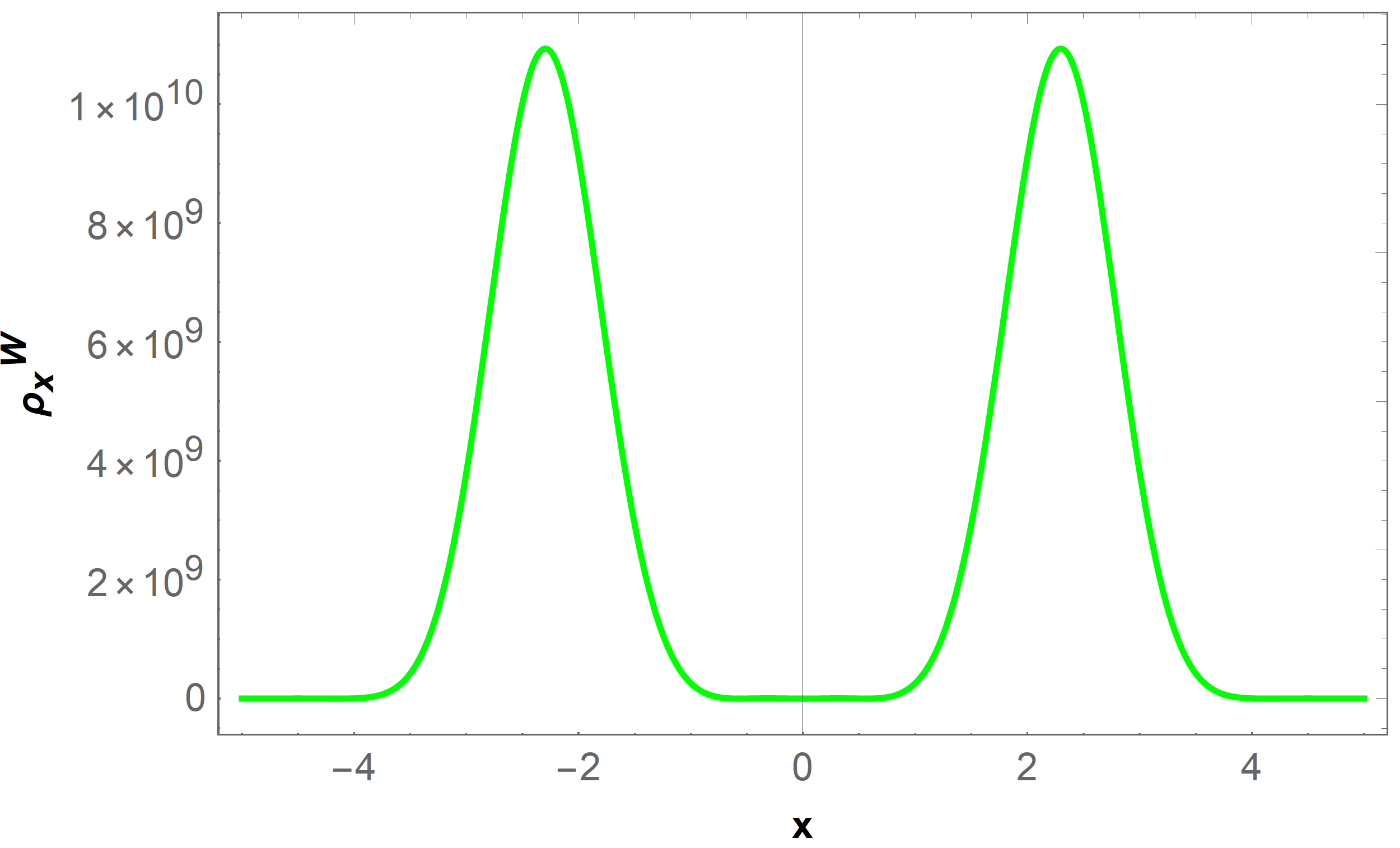}
    \caption{}
  \label{fig:image15}
\end{subfigure}
\begin{subfigure}{0.3\textwidth}
  \centering
  \includegraphics[width=\linewidth]{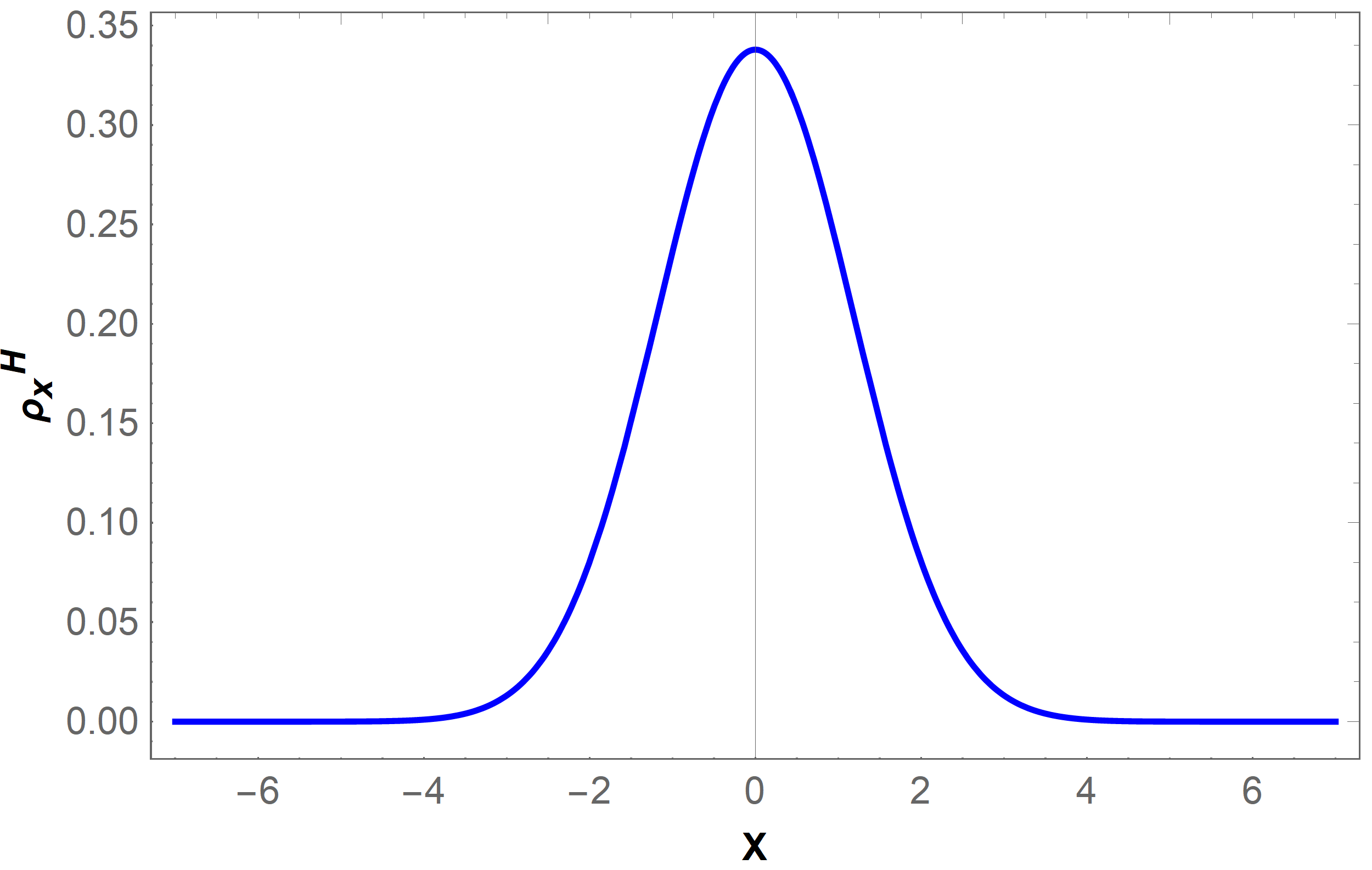}
    \caption{}
  \label{fig:image16}
\end{subfigure}
\begin{subfigure}{0.3\textwidth}
  \centering
  \includegraphics[width=\linewidth]{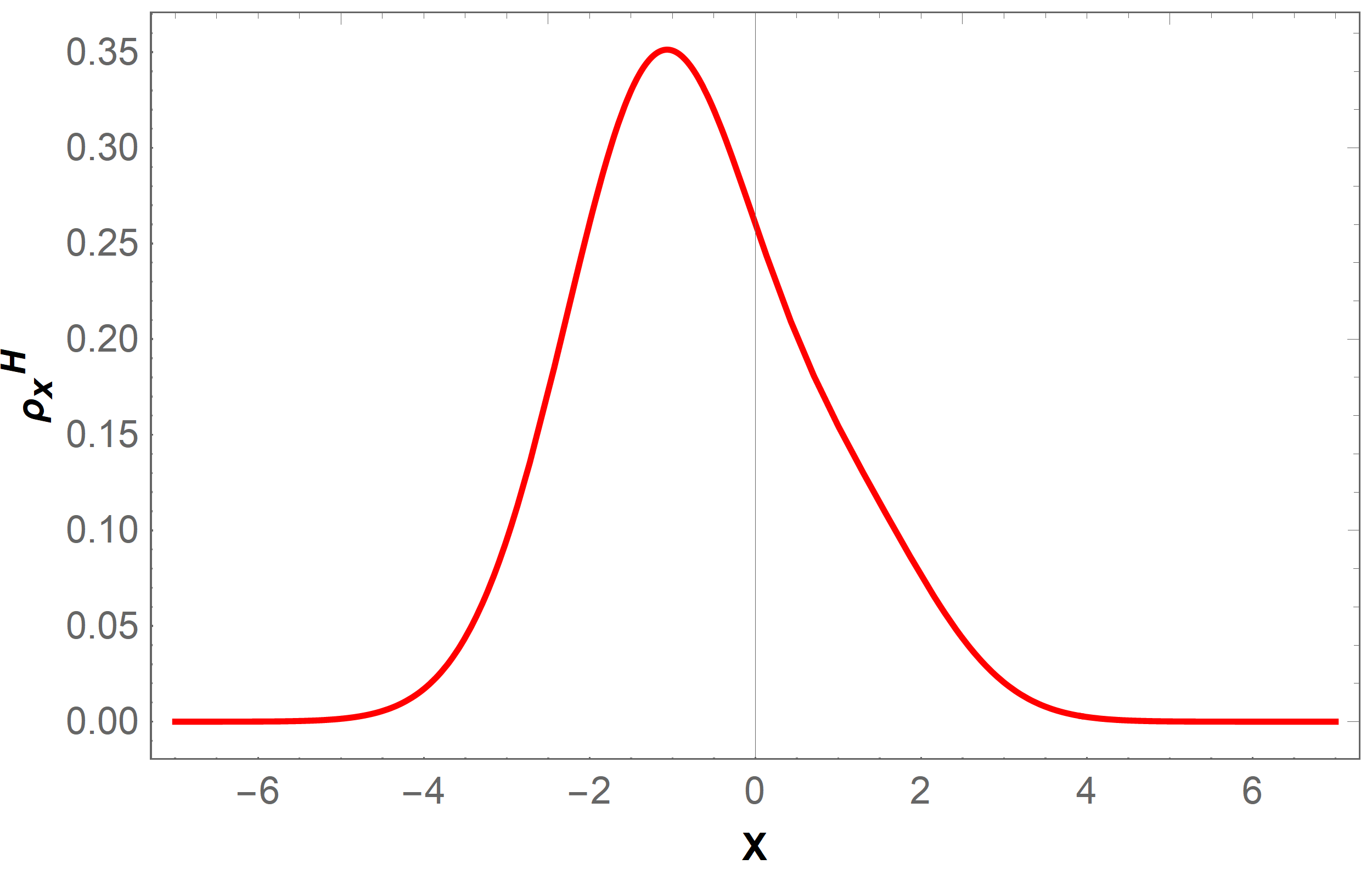}
    \caption{}
  \label{fig:image17}
\end{subfigure}
\begin{subfigure}{0.3\textwidth}
  \centering
  \includegraphics[width=\linewidth]{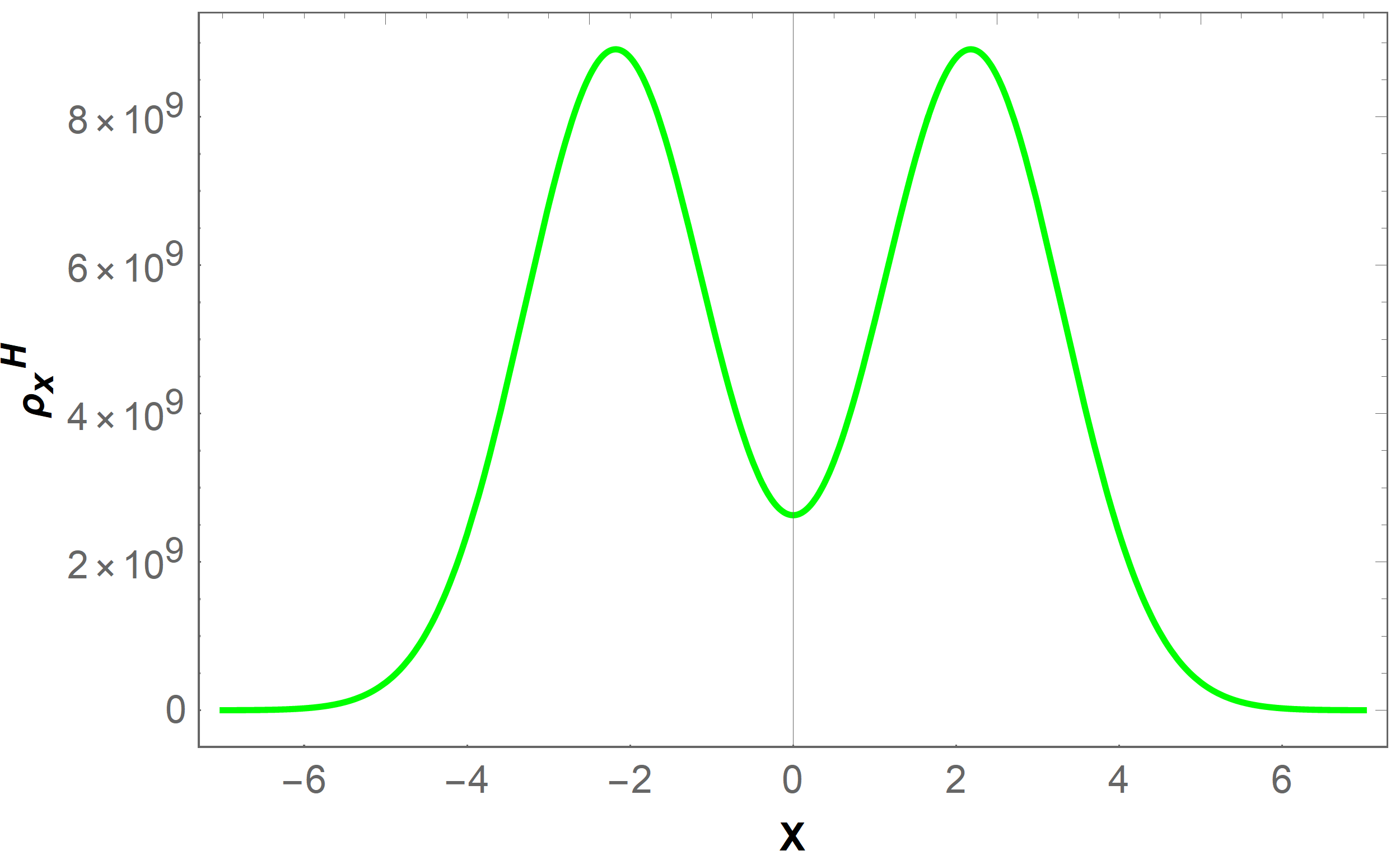}
    \caption{}
  \label{fig:image18}
\end{subfigure}
\caption{Plots (a) through (c) depict the coordinate space marginals ($\rho^{W}_{x}$) obtained using Wigner distribution, while plots (d) through (f) illustrate the coordinate space marginals obtained using Husimi distribution ($\rho^{H}_{x}$) for states with (a) and (d) for $n=0$, (b) and (e) for $n=1$, and (c) and (f) for $n=5$.
}
 \label{fig:image3main}
\end{figure}

\begin{figure}[H]
\begin{subfigure}{0.3\textwidth}
  \centering
  \includegraphics[width=\linewidth]{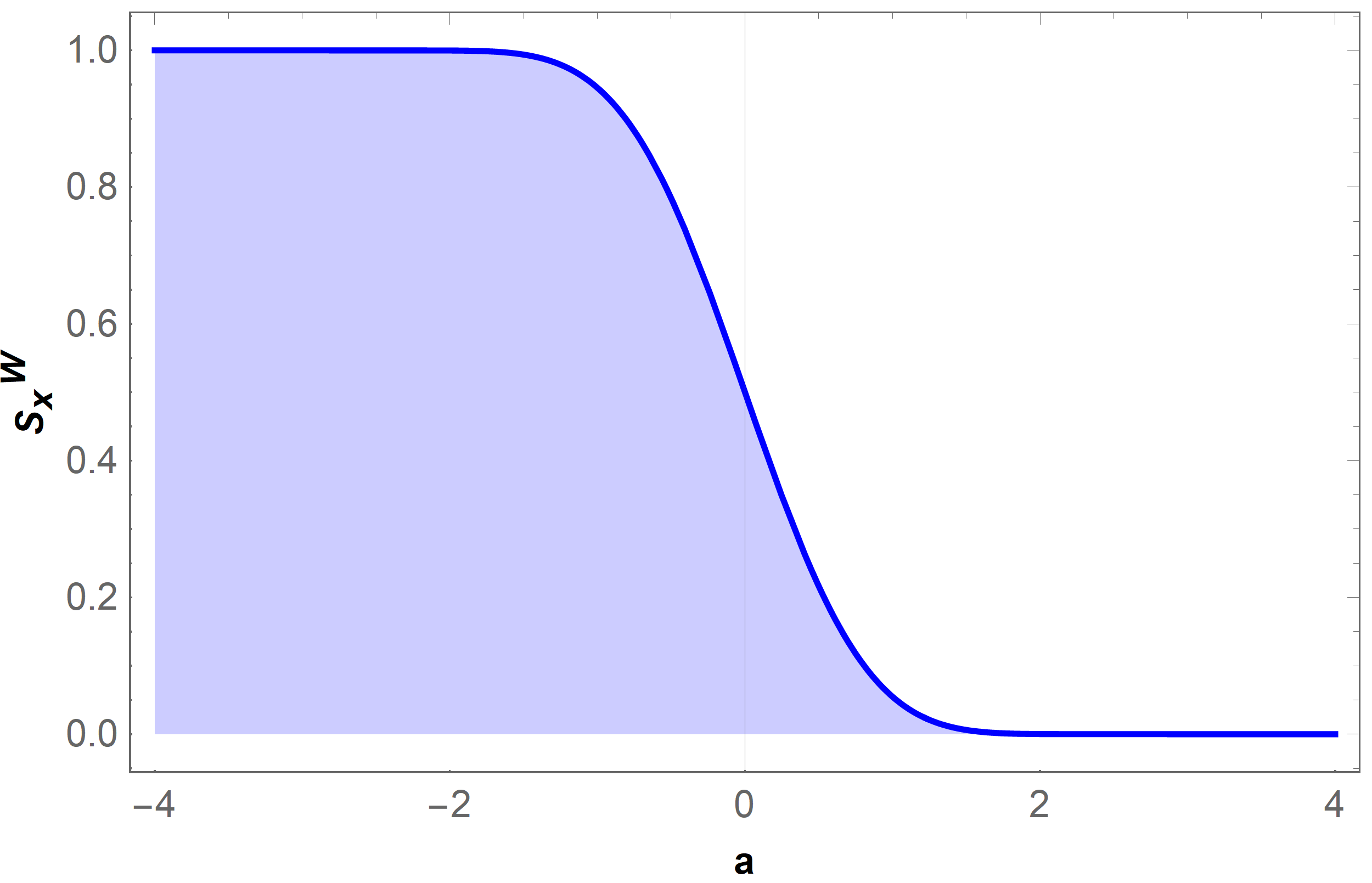}
  \caption{}
  \label{fig:image19}
\end{subfigure}
\begin{subfigure}{0.3\textwidth}
  \centering
  \includegraphics[width=\linewidth]{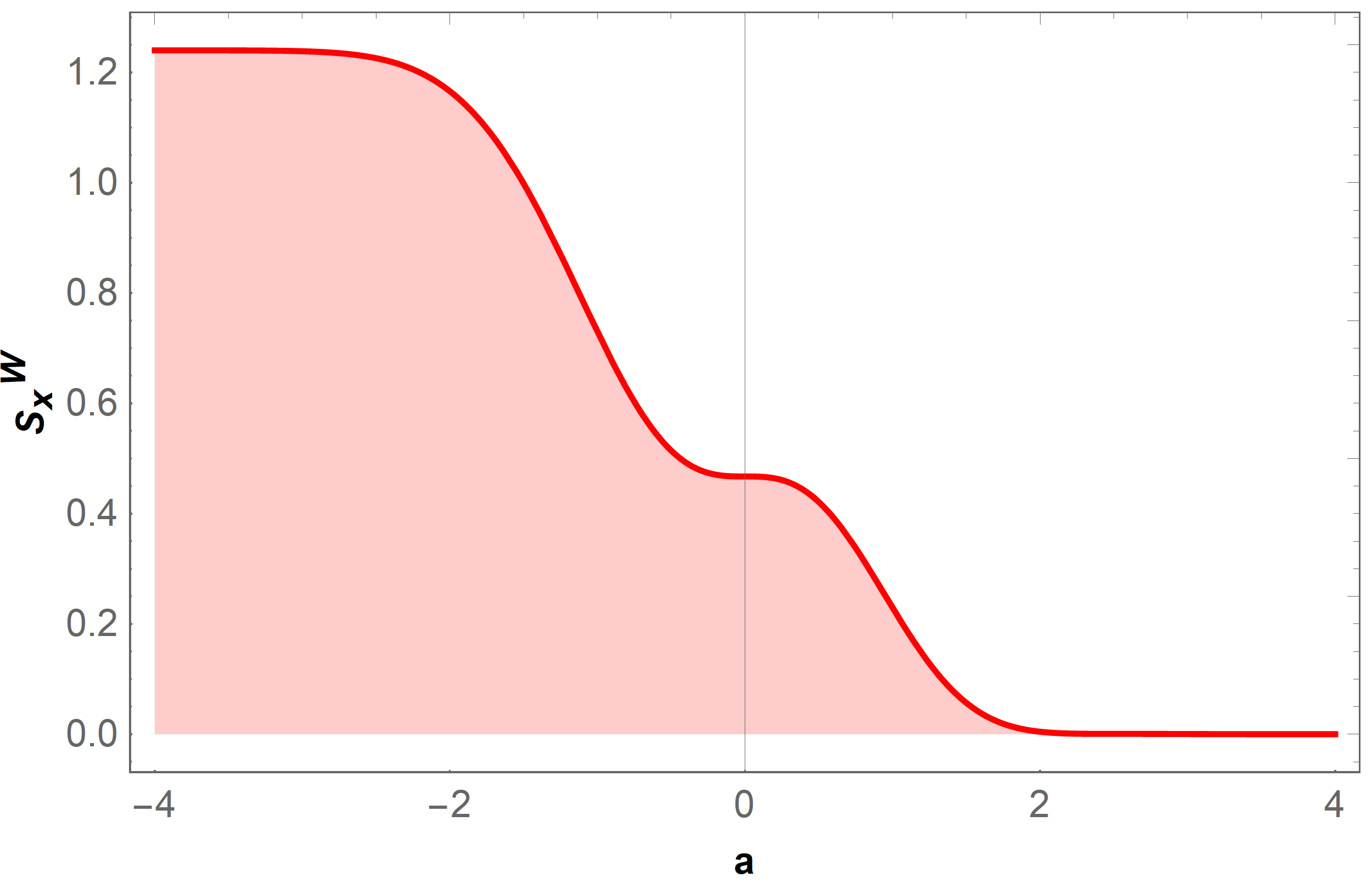}
  \caption{}
   \label{fig:image20}
\end{subfigure}
\begin{subfigure}{0.3\textwidth}
  \centering
  \includegraphics[width=\linewidth]{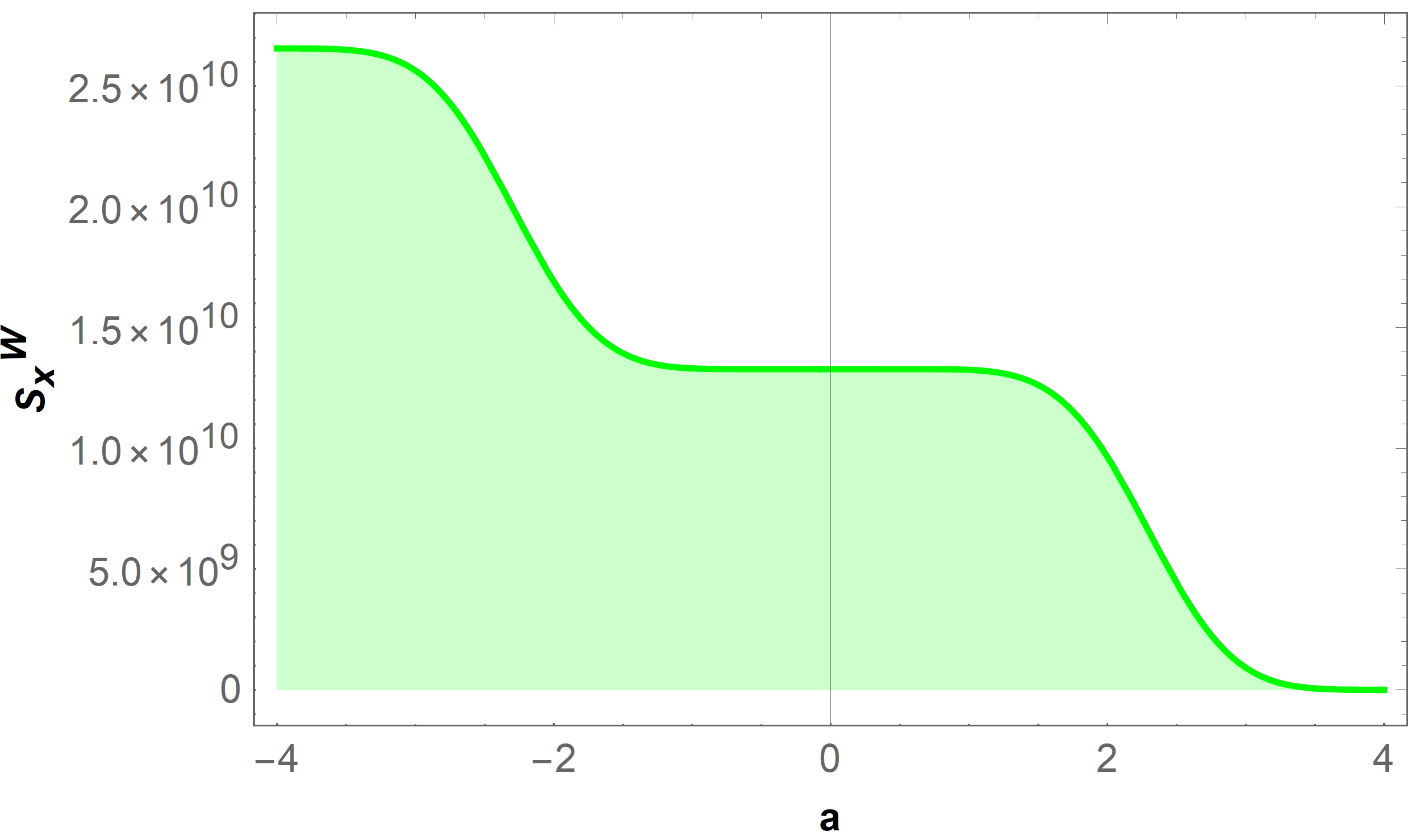}
  \caption{}
   \label{fig:image21}
\end{subfigure}
\begin{subfigure}{0.3\textwidth}
  \centering
  \includegraphics[width=\linewidth]{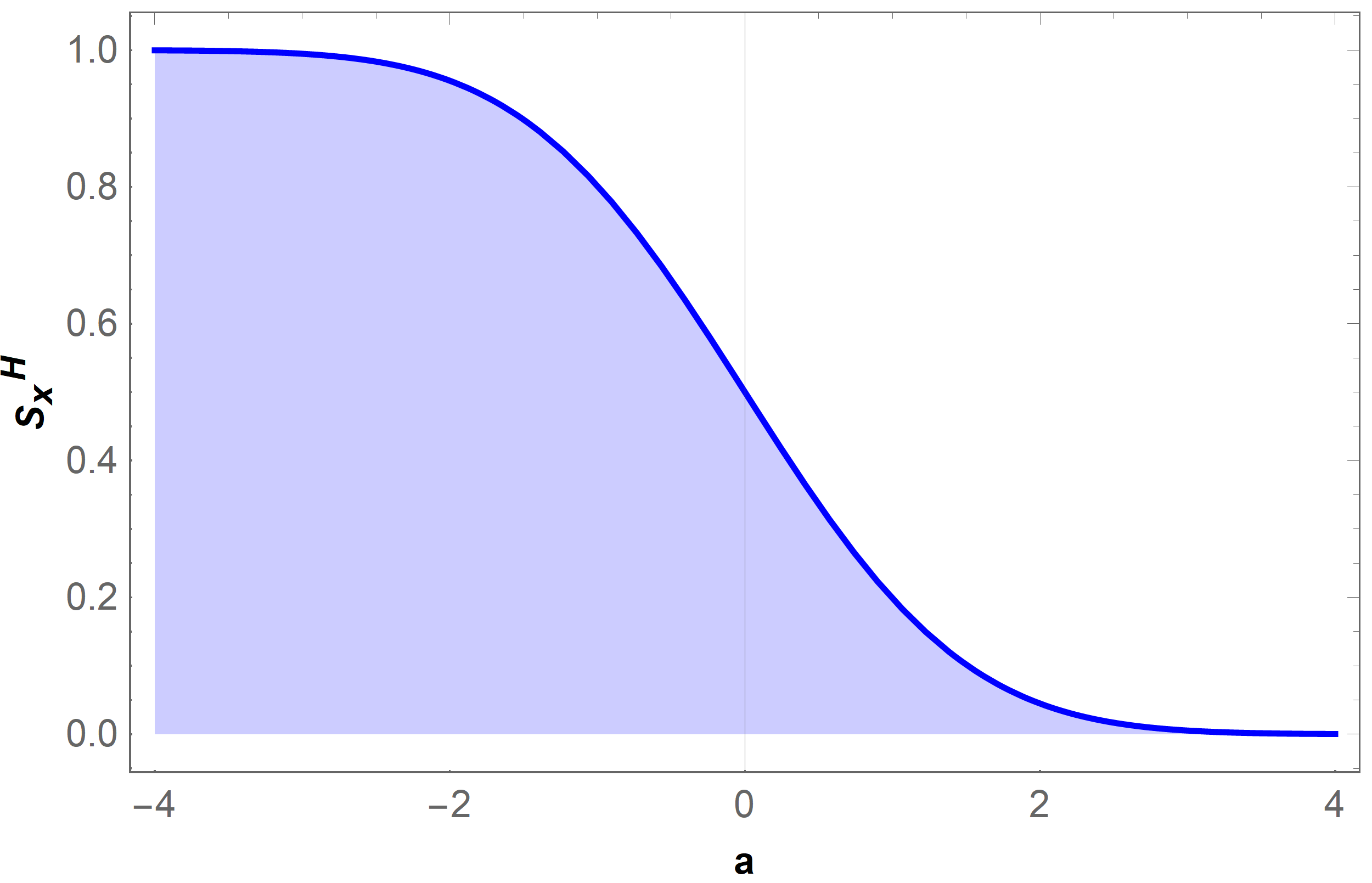}
  \caption{}
   \label{fig:image22}
\end{subfigure}
\begin{subfigure}{0.3\textwidth}
  \centering
  \includegraphics[width=\linewidth]{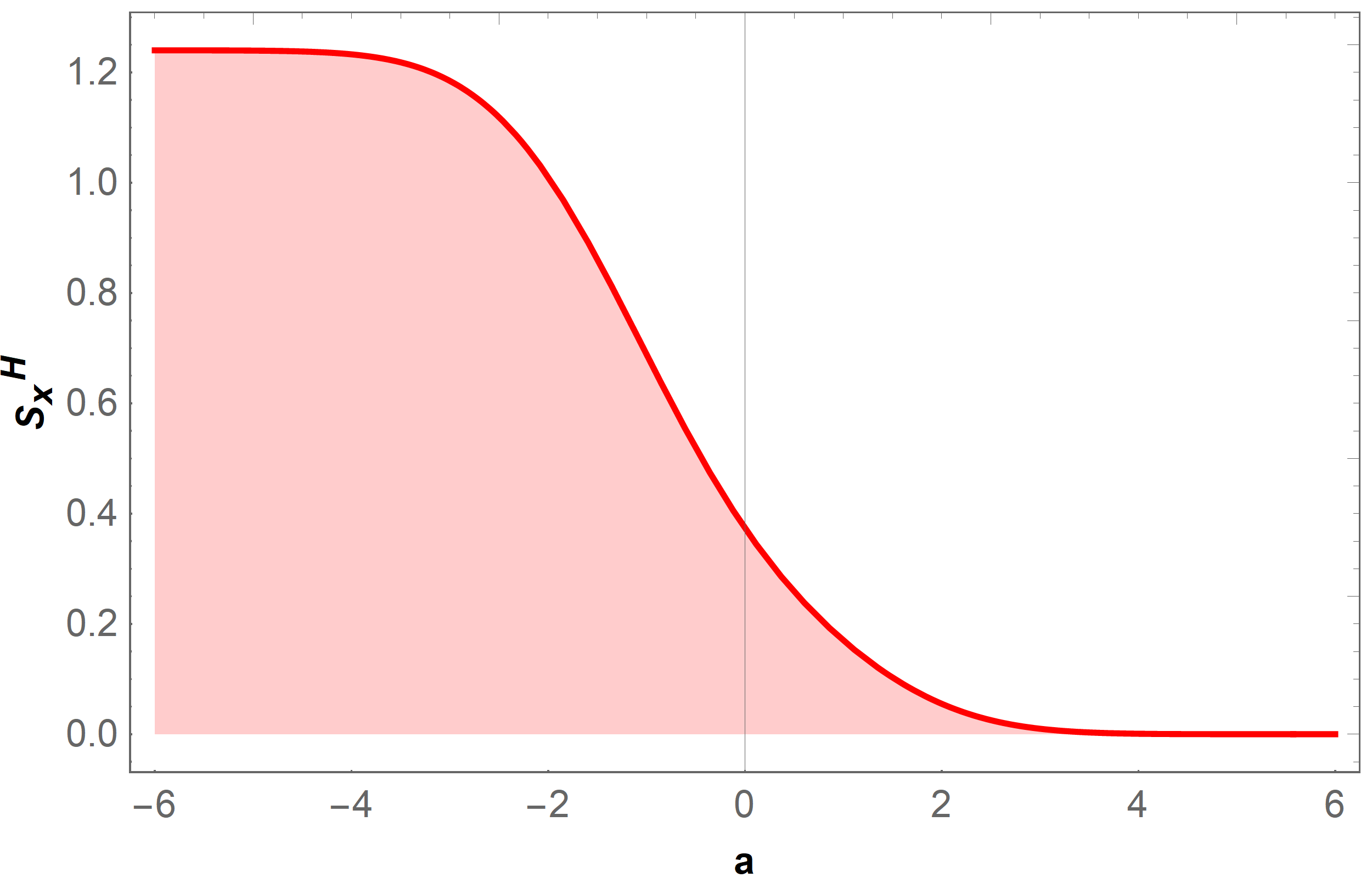}
  \caption{}
  \label{fig:image23}
\end{subfigure}
\begin{subfigure}{0.3\textwidth}
  \centering
  \includegraphics[width=\linewidth]{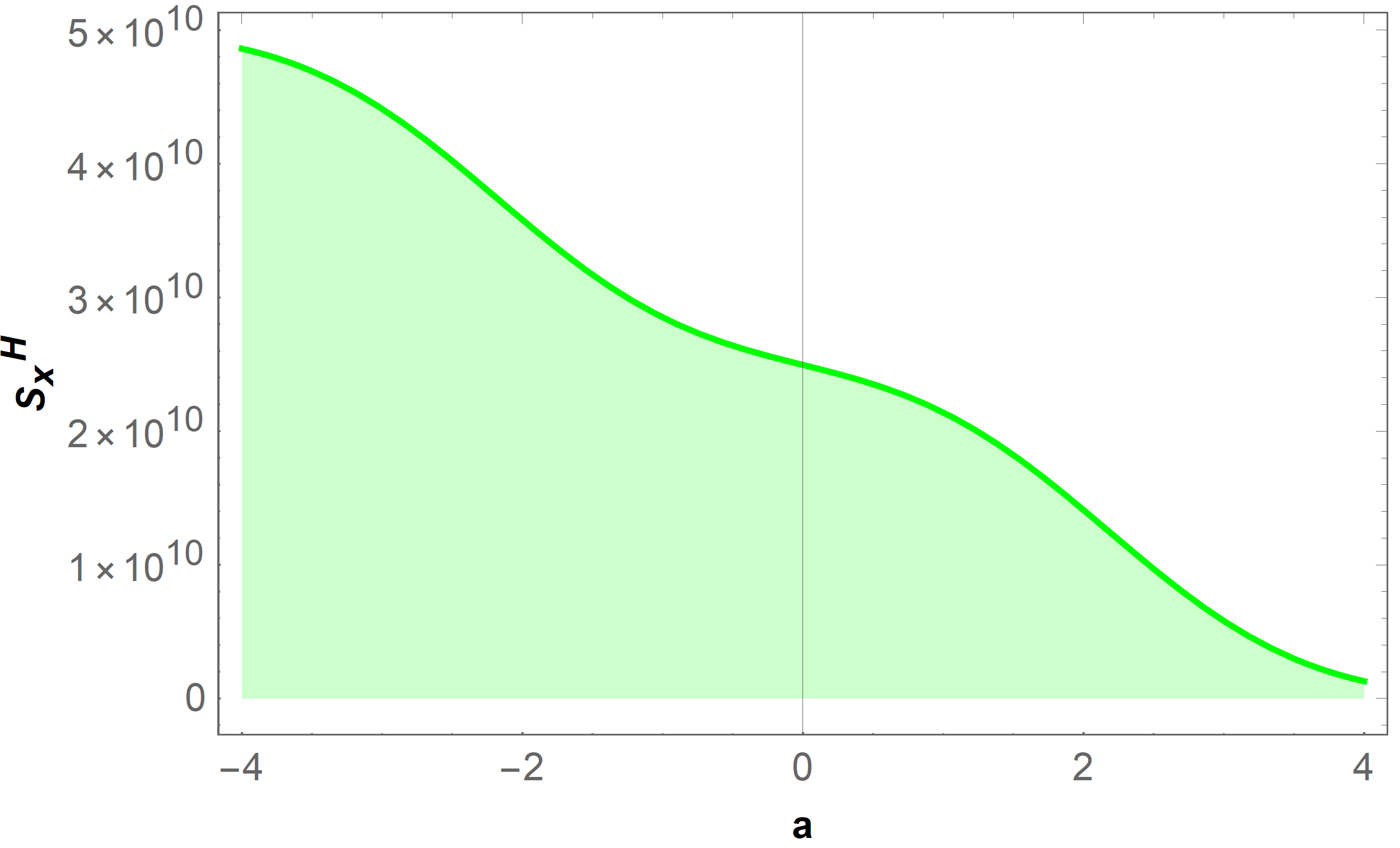}
  \caption{}
    \label{fig:image24}
\end{subfigure}
 \caption{Plots (a)-(c) represent the survival functions of the Wigner distribution marginals $s^{W}_{x}(a,b)$, while (d)-(f) depict the survival functions of the Husimi distribution marginals $s^{H}_{x}(a,b)$ for states with (a) and (d) for $n=0$, (b) and (e) for $n=1$, and (c) and (f) for $n=5$. The plots are shown as a function of $a$ while fixing the values of $b$ and the parameter $\lambda$.
}
 \label{fig:image4main}
\end{figure}

We start by obtaining the Wigner distribution and Husimi distribution for different states concerning the constant parameter $\lambda$, as described in Fig. \ref{fig:wignerhusimi}. From this illustration, we can note that the Wigner distribution and Husimi distribution exhibit nearly identical patterns for both the ground state $(n=0)$ and the first excited state $(n=1)$. However, for higher excited states (greater values of $n$), the difference becomes more prominent. Once we obtain these distributions, we plot survival functions (fig. \ref{fig:image2main}) from the Wigner distribution ($S_{W}(a,b)$) and Husimi distribution ($S_{H}(a,b)$). We choose to plot these functions concerning the parameter $b$, achieved by fixing the values of $a$ and the parameter $\lambda$. The plots plotted using the distinctly have some negative regions. These negative regions are absent in the plots concerning the Husimi distribution. Due to the positive definiteness of the Husimi distribution, all curves of survival functions are monotonically decreasing functions. These changes are distinctly visible. 
The marginal densities are shown in figure \ref{fig:image3main}. We show the coordinate ($\rho_{x}$) space marginal densities. It's clear from these illustrations that the nodal structure intensifies with higher values of $n$. These marginal distributions are derived from both the Wigner distribution and Husimi distribution. In the scenario of the harmonic oscillator \cite{salazar2023phase}, the nodal structure is lacking in the Husimi distribution marginals, whereas it is evident in the Wigner distribution marginals. For the harmonic oscillator, the number of nodal structures increases for the values of $n$ concerning the Wigner marginal densities. However, this is not the case with the marginals obtained using the Husimi distribution. The nodal structures are fewer in number when compared with the nodal structures of Wigner marginal densities.
Once we obtain the marginals, we need to derive the survival function for the marginals (fig. \ref{fig:image4main}). The present nodal structures present in the survival function from the distributions is not the same as that present in the case of its marginals. The lumps present in the survival function of the Wigner distribution are not those present in the survival function of its marginals. However, this is not the case with the Husimi distribution, because the Husimi marginal densities themselves lack the nodal structure to the same degree as the Wigner distribution.
\subsection{Entropies associated with the Wigner and Husimi distributions, as well as their respective marginals}
\begin{figure}[H]
\begin{subfigure}{0.3\textwidth}
  \centering
  \includegraphics[width=\linewidth]{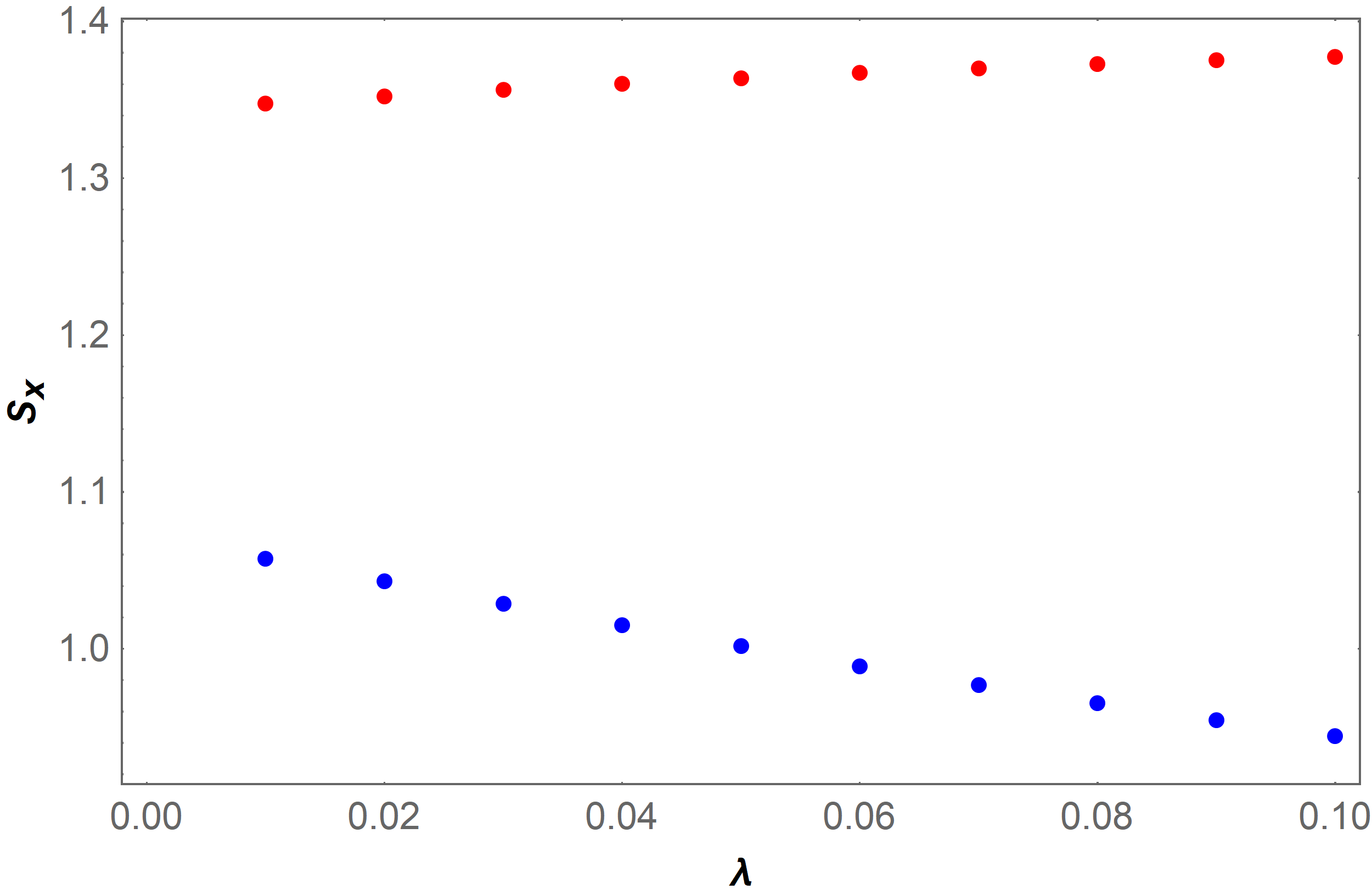}
  \caption{}
  \label{fig:image25}
\end{subfigure}
\begin{subfigure}{0.3\textwidth}
  \centering
  \includegraphics[width=\linewidth]{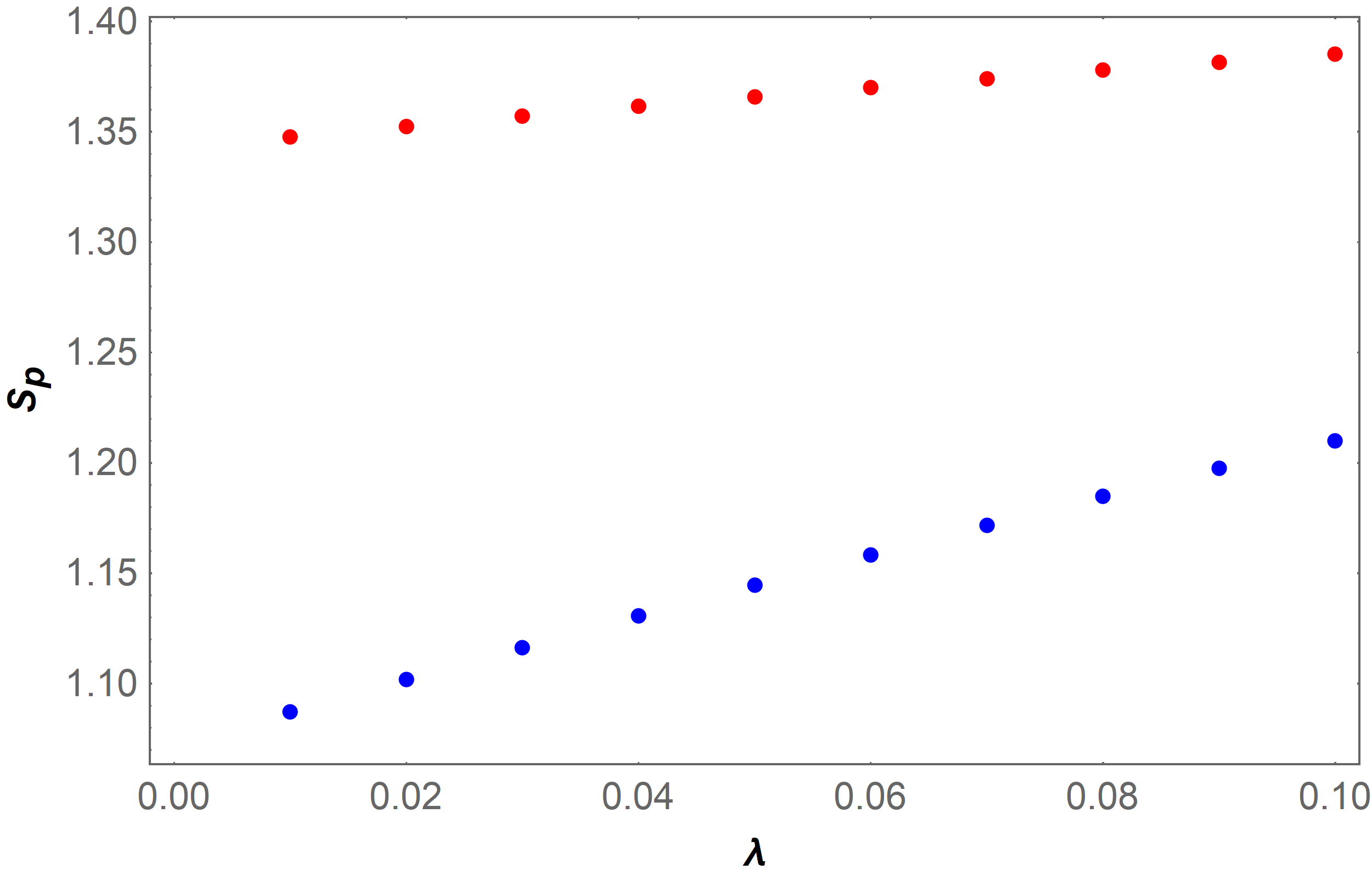}
  \caption{}
  \label{fig:image26}
\end{subfigure}
\begin{subfigure}{0.3\textwidth}
  \centering
  \includegraphics[width=\linewidth]{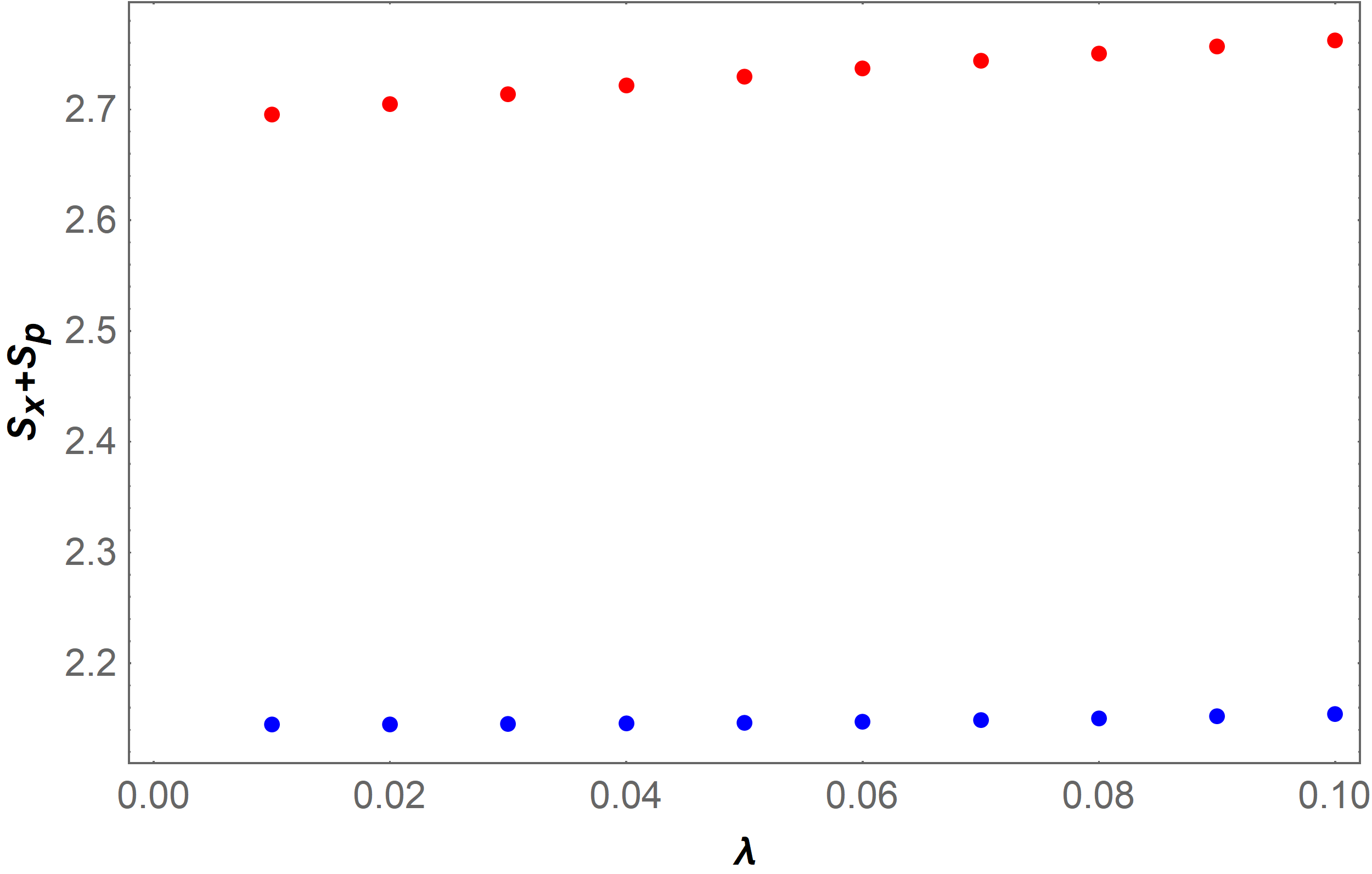}
  \caption{}
  \label{fig:image27}
\end{subfigure}
\begin{subfigure}{0.3\textwidth}
  \centering
  \includegraphics[width=\linewidth]{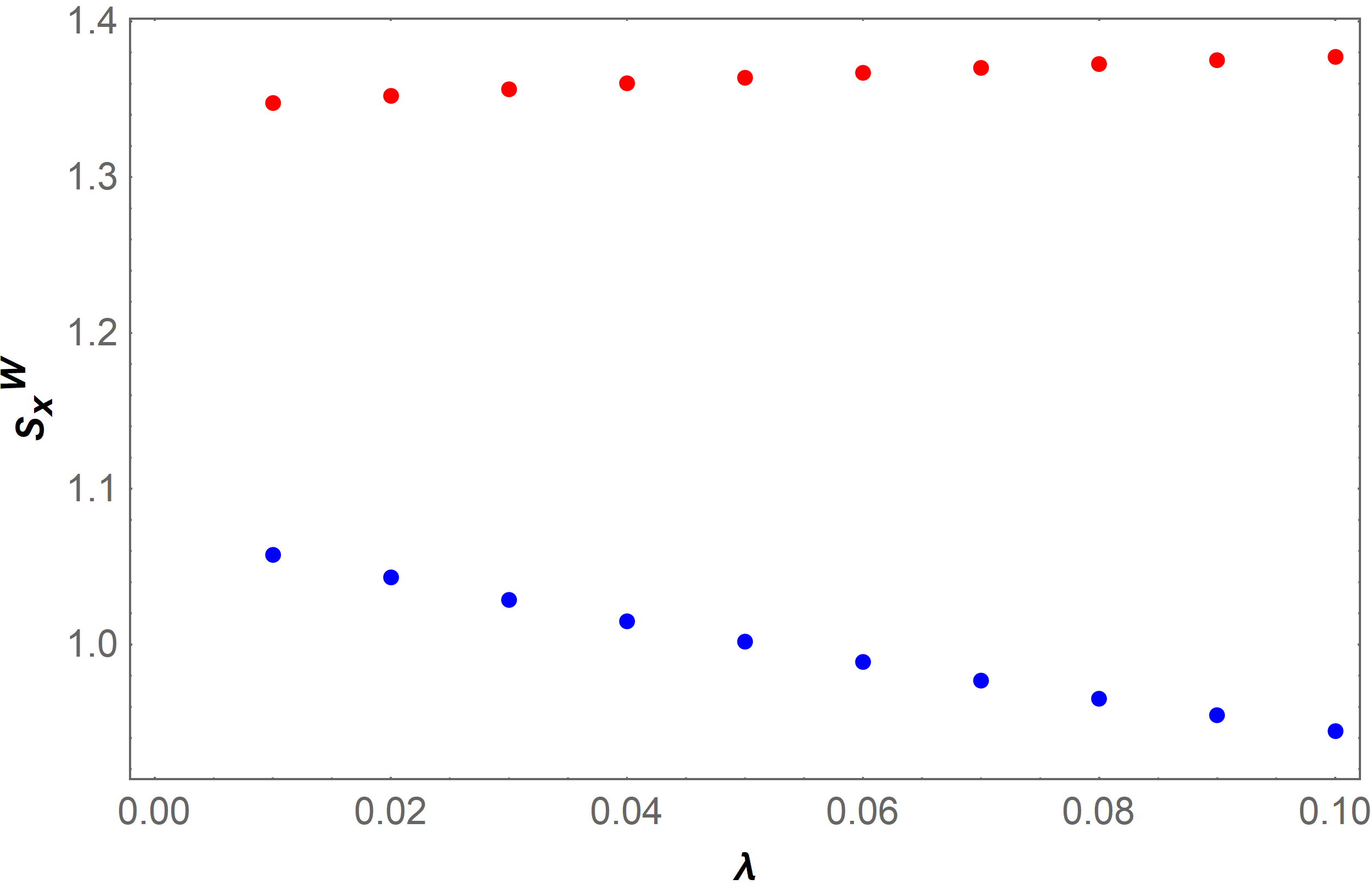}
  \caption{}
  \label{fig:image28}
\end{subfigure}
\begin{subfigure}{0.3\textwidth}
  \centering
  \includegraphics[width=\linewidth]{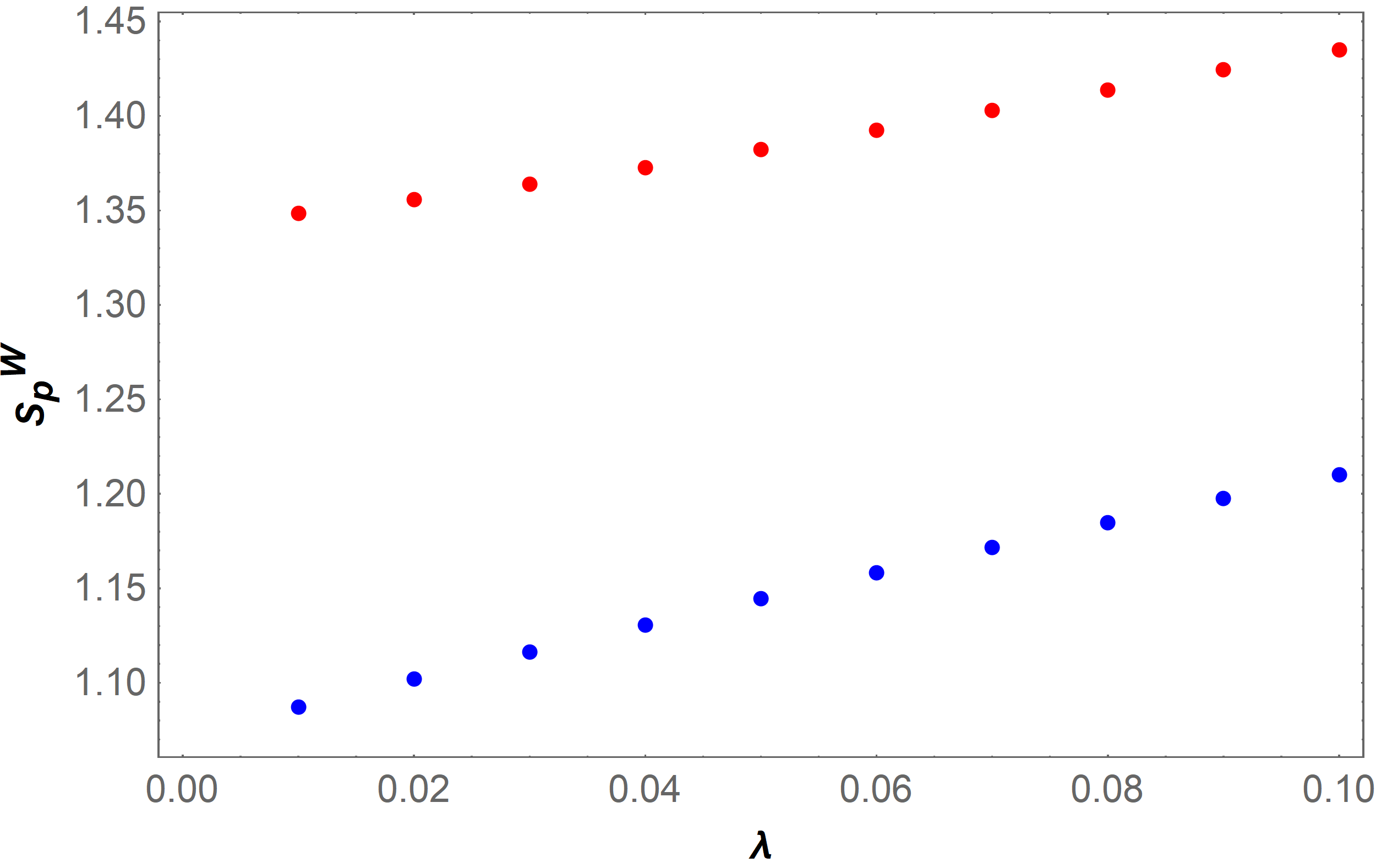}
  \caption{}
  \label{fig:image29}
\end{subfigure}
\begin{subfigure}{0.3\textwidth}
  \centering
  \includegraphics[width=\linewidth]{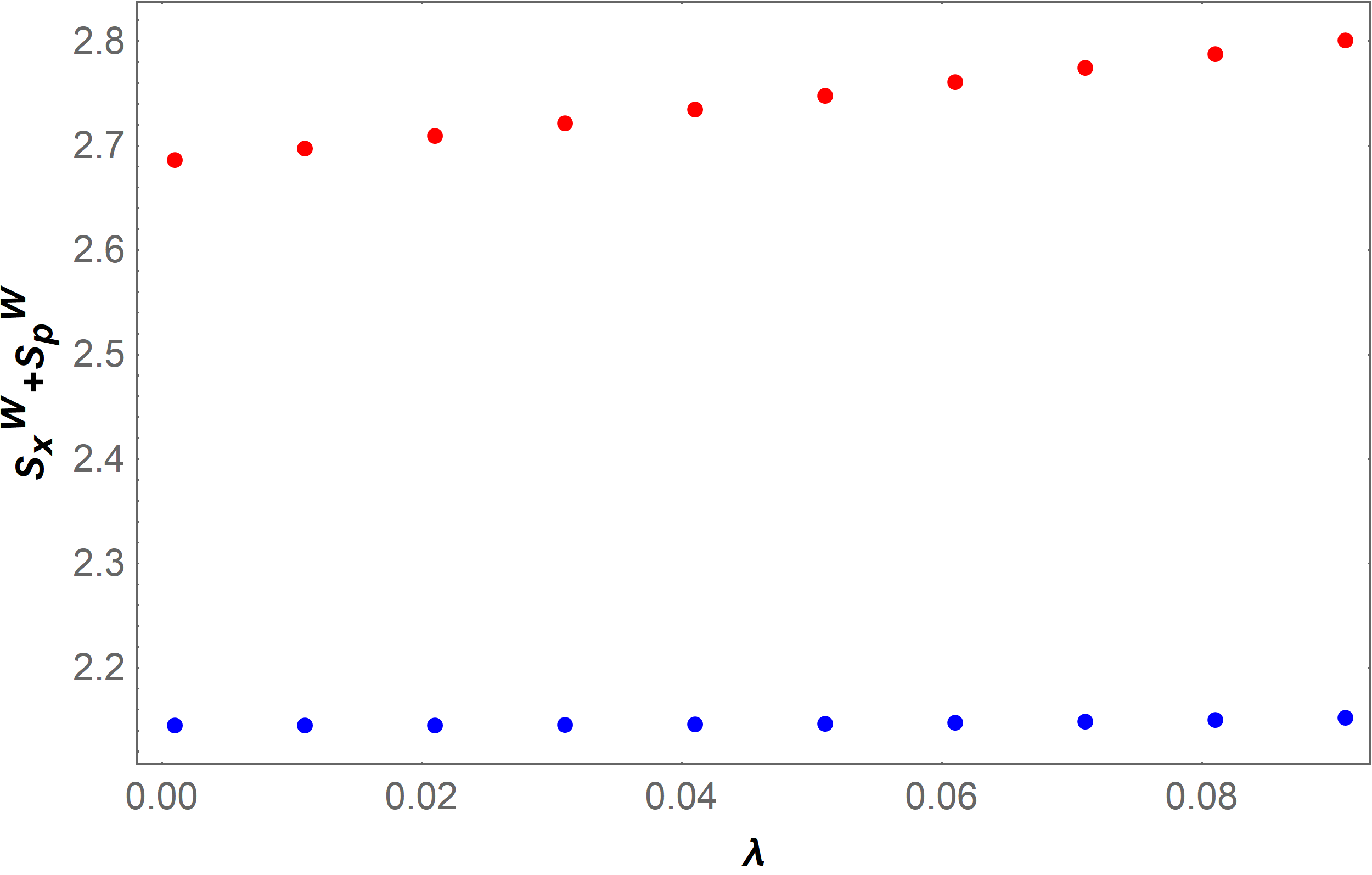}
  \caption{}
  \label{fig:image30}
\end{subfigure}
\begin{subfigure}{0.3\textwidth}
  \centering
  \includegraphics[width=\linewidth]{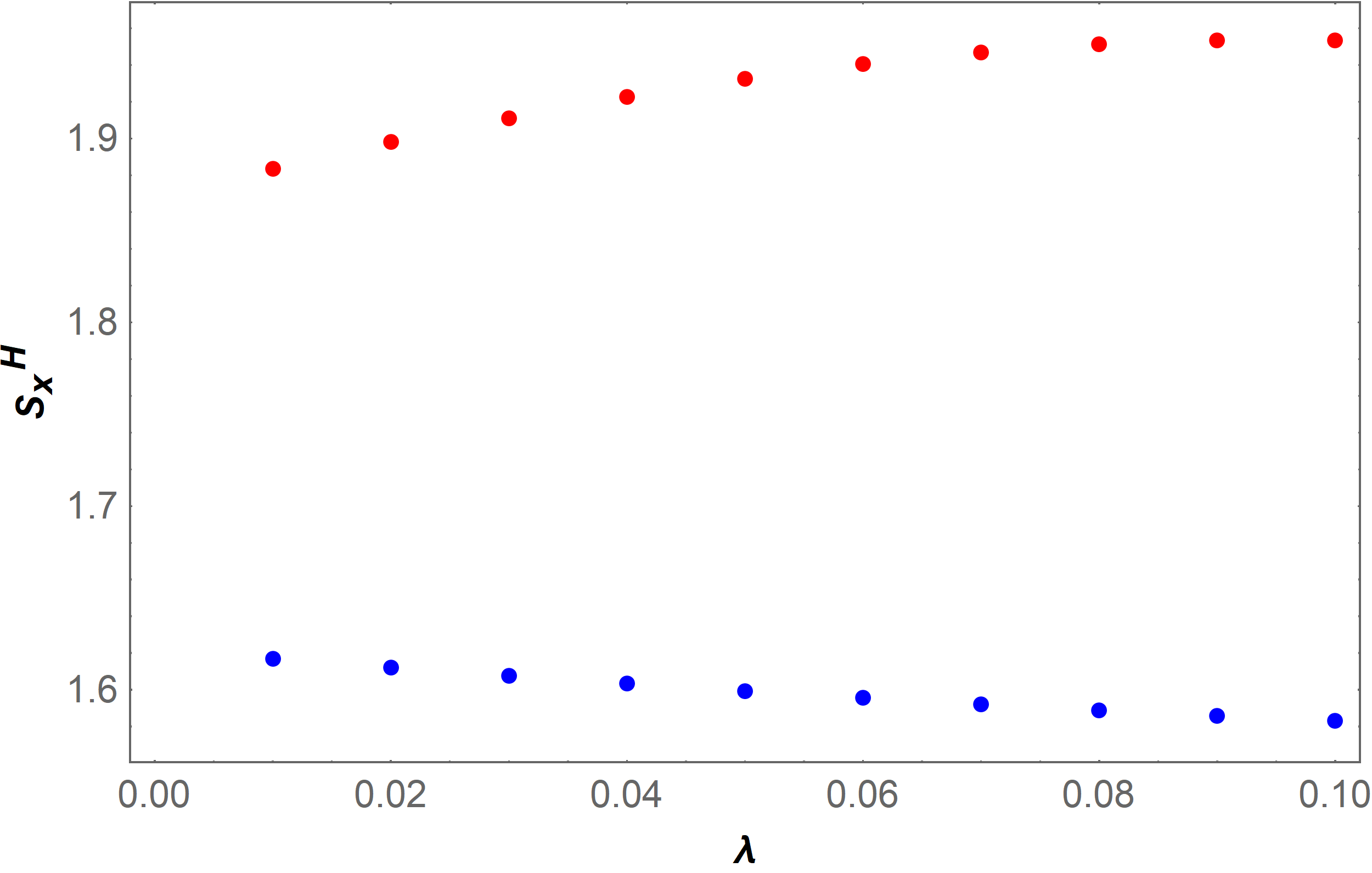}
  \caption{}
  \label{fig:image31}
\end{subfigure}
\begin{subfigure}{0.3\textwidth}
  \centering
  \includegraphics[width=\linewidth]{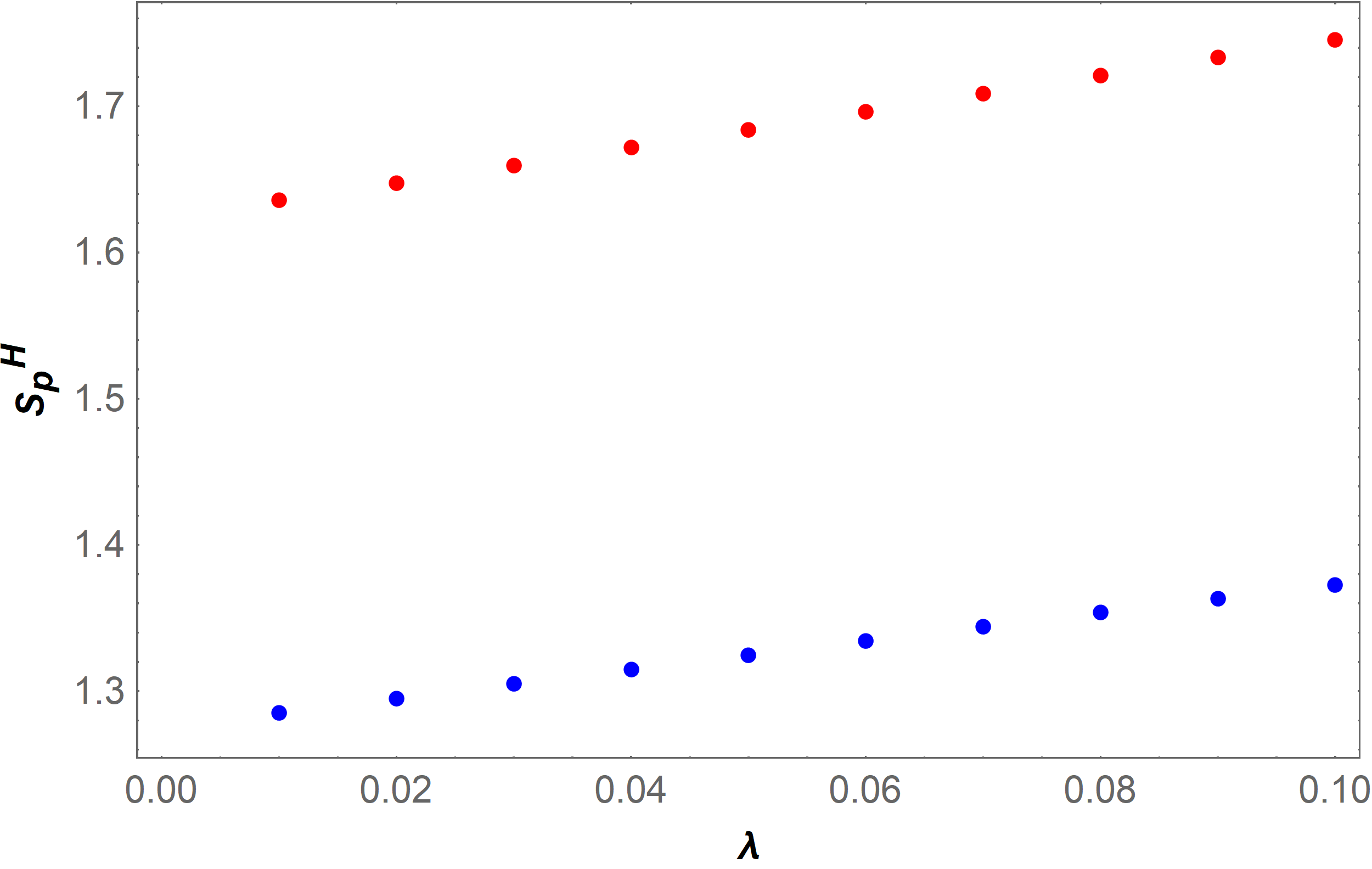}
  \caption{}
  \label{fig:image32}
\end{subfigure}
\begin{subfigure}{0.3\textwidth}
  \centering
  \includegraphics[width=\linewidth]{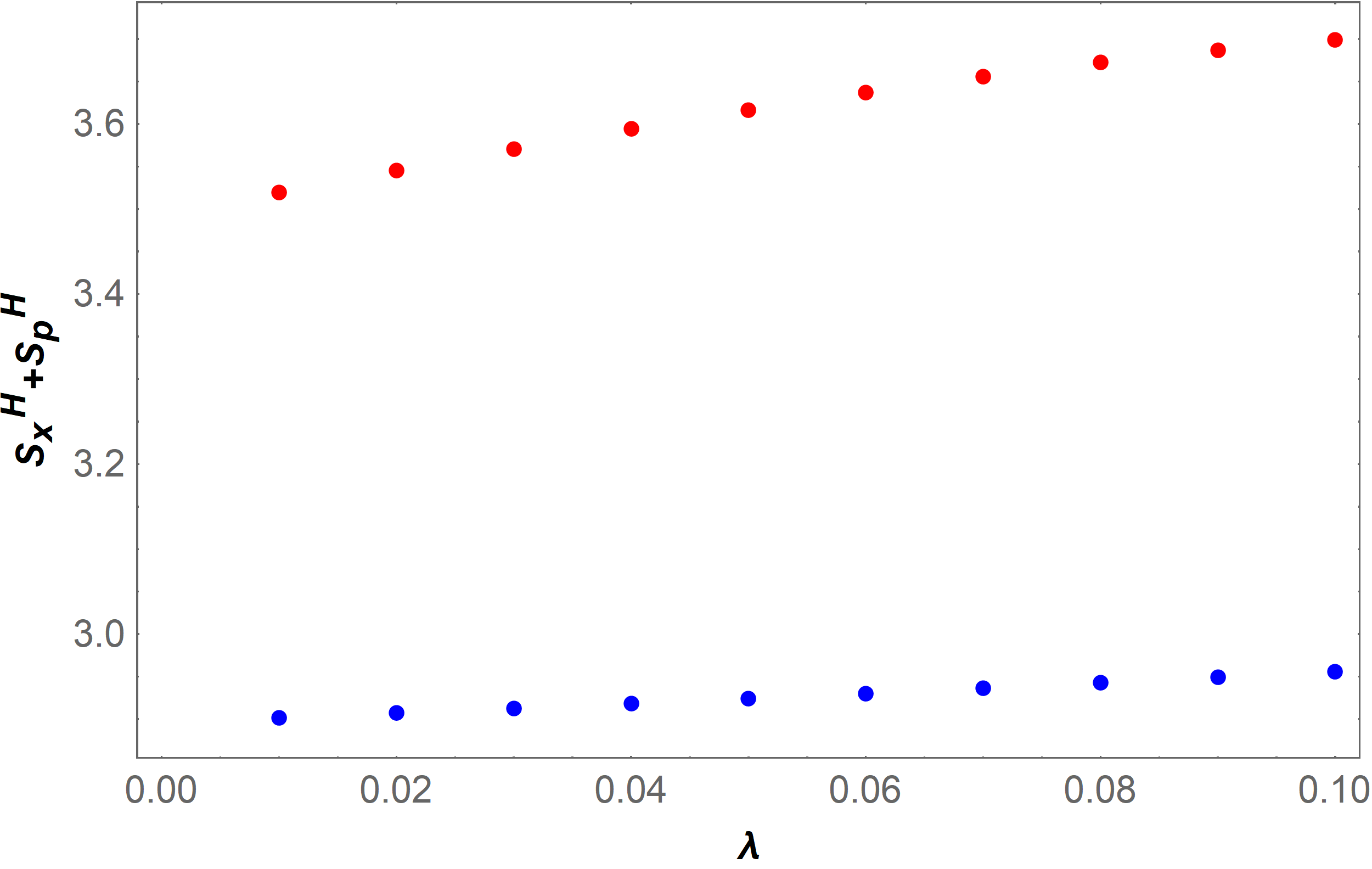}
  \caption{}
  \label{fig:image33}
\end{subfigure}
\caption{(a), (d), and (g) display Shannon entropy plots in coordinate space, (b), (e), and (h) show Shannon entropy plots in momentum space, while (c), (f), and (i) illustrate the entropic uncertainty relation for Shannon-like entropies. Plots (a)-(c) represent Shannon entropy plots using wave function, (d)-(f) depict Shannon entropy plots of Wigner distribution marginals, and (g)-(i) demonstrate Shannon entropy plots of Husimi distribution marginals. Blue dots represent the \(n=0\) state, while red dots represent the \(n=1\) state.}
\label{fig:Shannonentropyplots} 
\end{figure}

\begin{figure}[H]
\centering
\begin{subfigure}{0.3\textwidth}
  \centering
  \includegraphics[width=\linewidth]{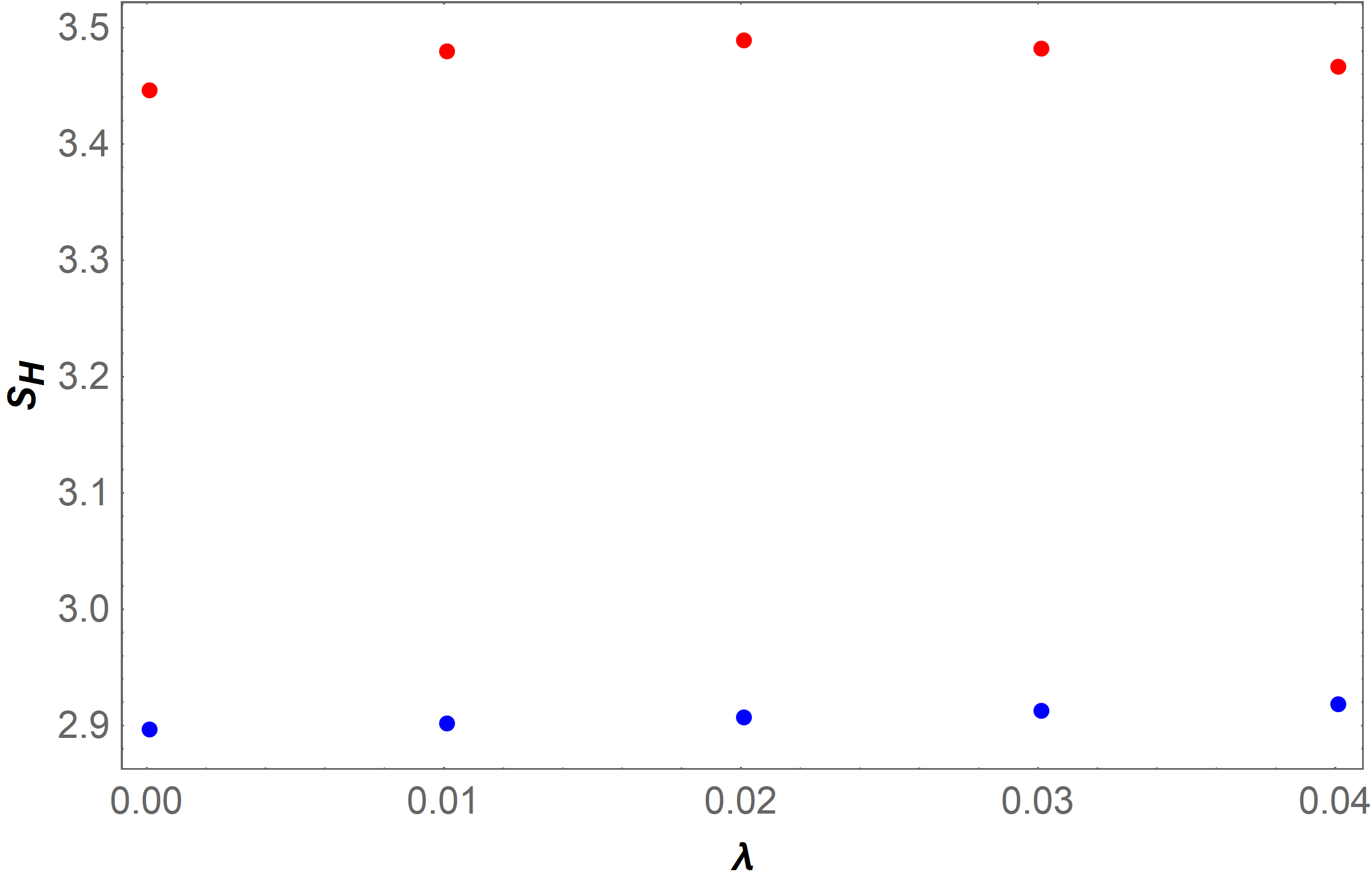}
  \label{fig:image34}
\end{subfigure}
\caption{Shannon entropy of Husimi distribution} (Wehrl entropy). Blue dots denote the ground state, and red dots denote the first excited state.
\label{fig:wehrlentropy} 
\end{figure}

\begin{figure}[H]
\begin{subfigure}{0.245\textwidth}
  \centering
  \includegraphics[width=\linewidth]{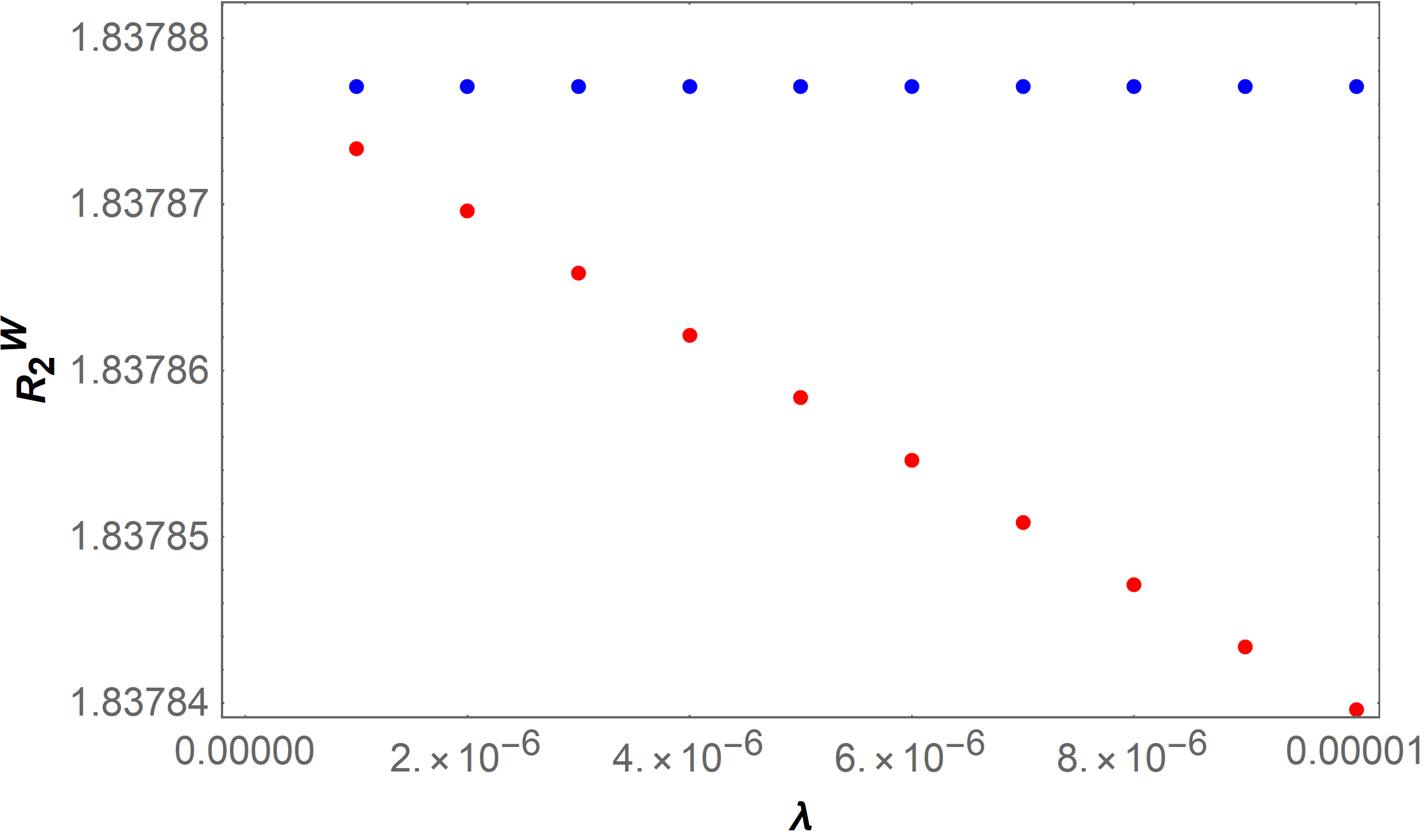}
    \caption{}
  \label{fig:image38}
\end{subfigure}
\begin{subfigure}{0.245\textwidth}
  \centering
  \includegraphics[width=\linewidth]{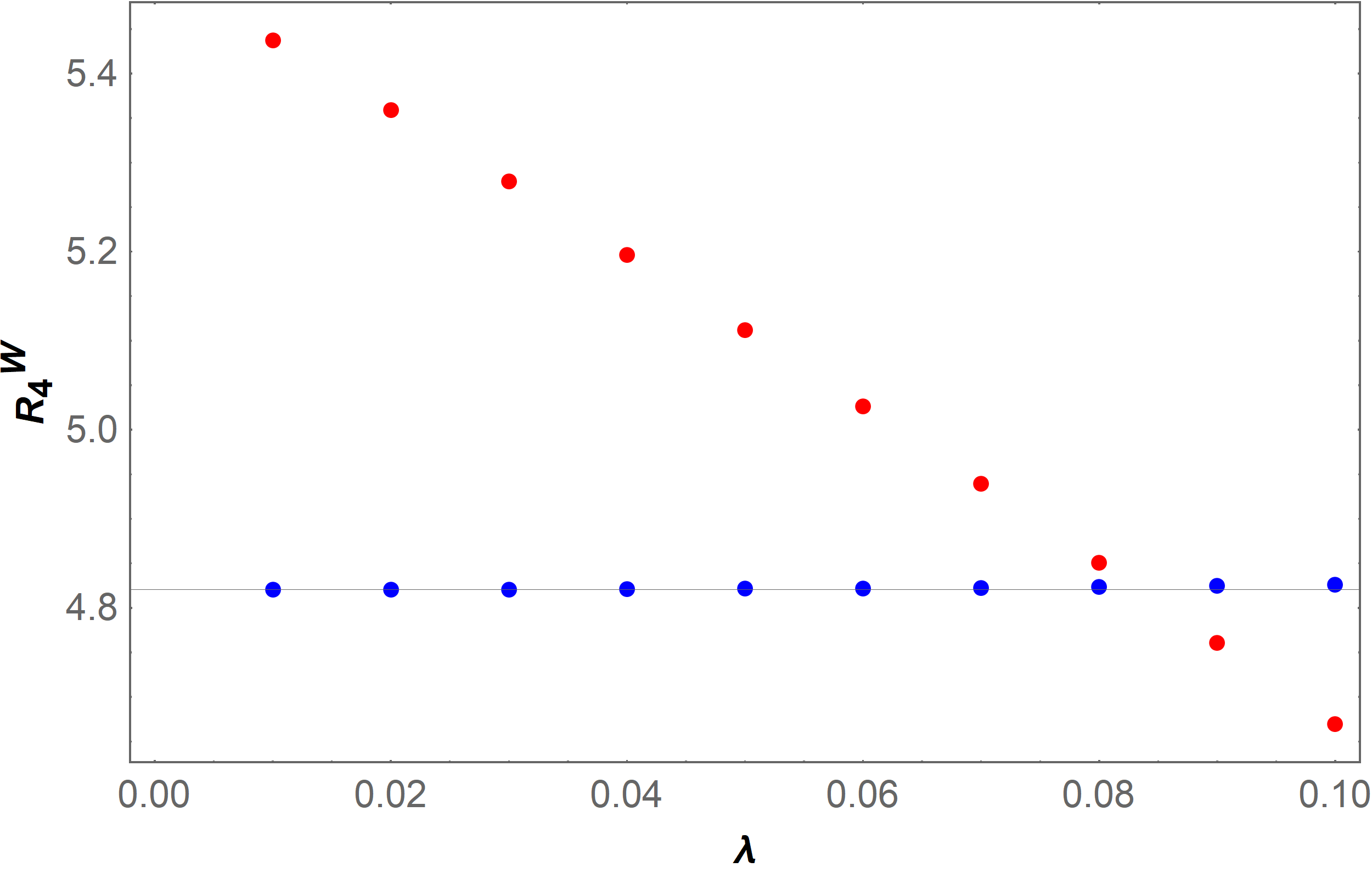}
    \caption{}
  \label{fig:image39}
\end{subfigure}
\begin{subfigure}{0.245\textwidth}
  \centering
  \includegraphics[width=\linewidth]{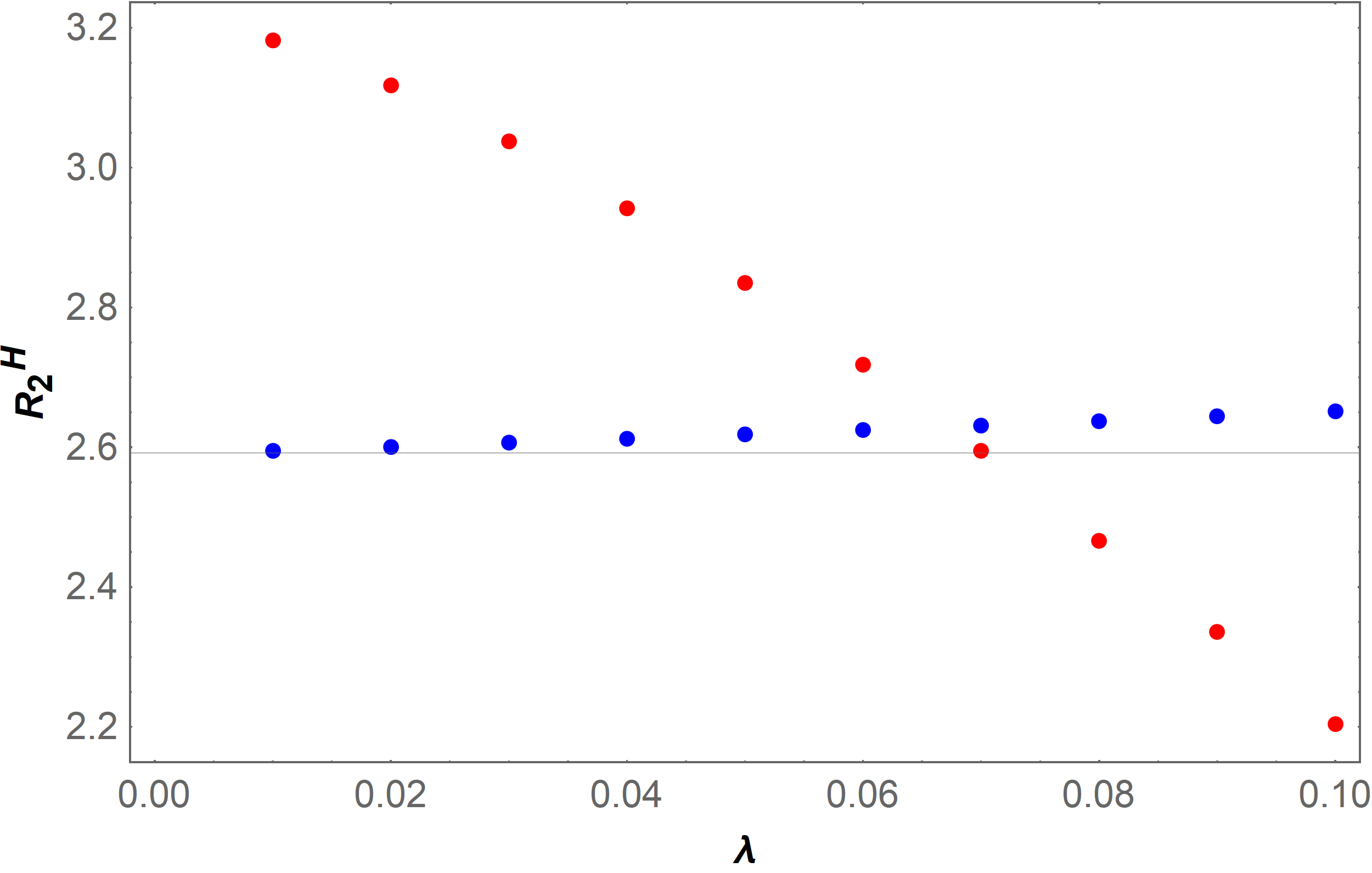}
    \caption{}
  \label{fig:image40}
\end{subfigure}
\begin{subfigure}{0.245\textwidth}
  \centering
  \includegraphics[width=\linewidth]{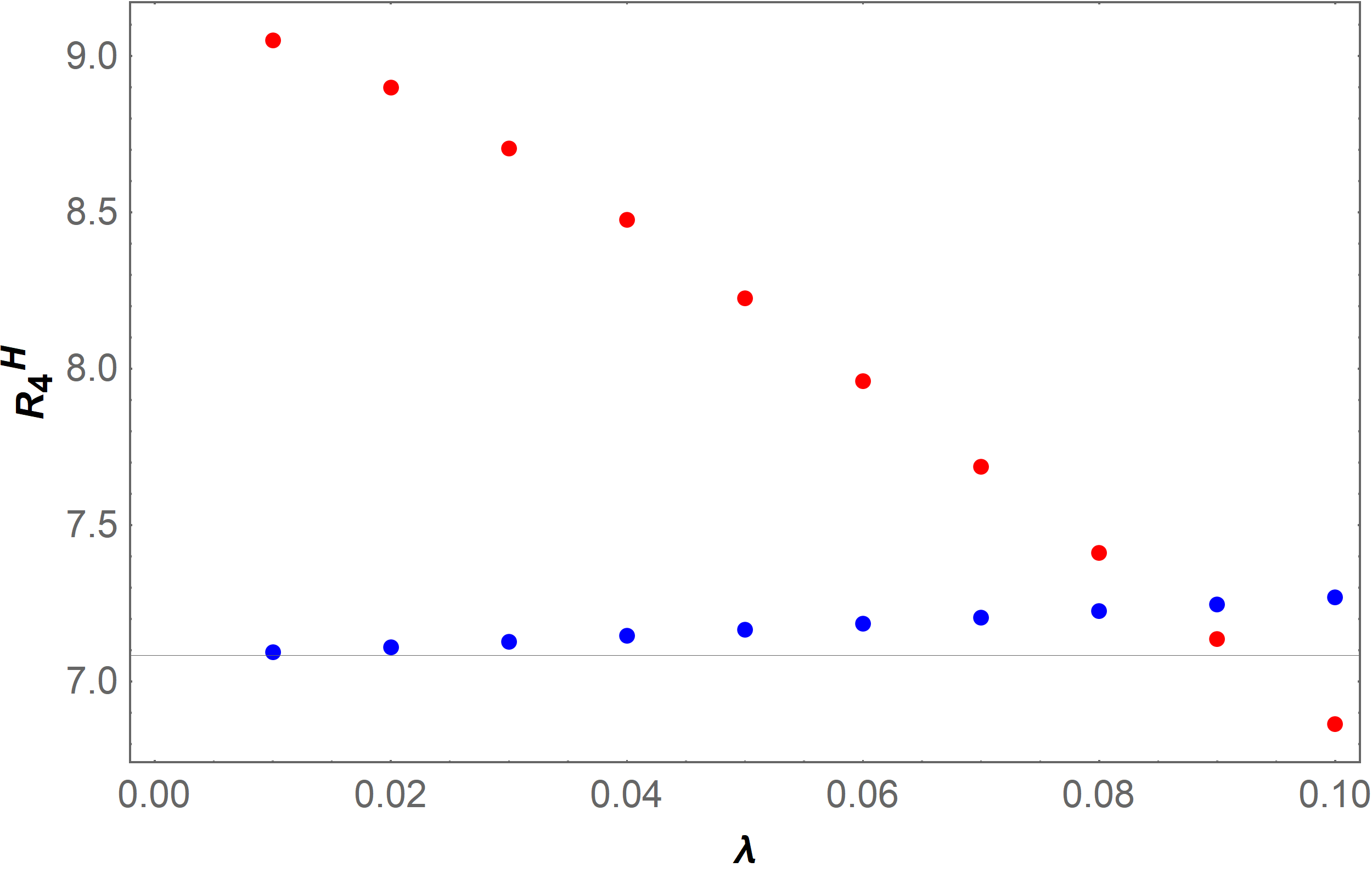}
  \caption{}
  \label{fig:image41}
\end{subfigure}
\caption{Plots (a) and (b) represent the Rényi entropies of the Wigner distribution ($R^{W}_{\alpha}$), while plots (c) and (d) depict the Rényi entropies of the Husimi distribution ($R^{H}_{\alpha}$). Plots (a) and (c) correspond to $\alpha = 2$, whereas plots (b) and (d) correspond to $\alpha = 4$. Blue dots represent the $n=0$ state, while red dots represent the $n=1$ state.
}
\label{fig:renyientropy} 
\end{figure}

After discussing the distributions, we proceed to discuss their respective entropies. We have computed the entropies of both the Wigner distribution and Husimi distribution, along with their marginals. All the marginal distributions exhibit positivity, and there are no complexities associated with their entropies. This facilitates a straightforward comparison between various entropic measures and underscores the distinctions between the entropies of their marginals. In figure \ref{fig:Shannonentropyplots}, we have shown Shannon entropy in coordinate, momentum and combined space. All the entropic measures are plotted using either the wave function, Wigner distribution or Husimi distribution and their marginals with the parameter $\lambda$ for $n=0$ and $n=1$ states. The important observation is that the entropy of the marginals derived from the Wigner distribution consistently remains lower than the corresponding entropy from the Husimi distribution for any value of $n$. Another noteworthy observation is that the entropies obtained using the wave function are almost equal to those obtained from the Wigner distribution. From these plots, we can conclude that the higher entropic values of the Husimi distribution marginals are due to the loss of structure compared to the Wigner distribution marginals.
In plot fig. \ref{fig:wehrlentropy}, we plotted Husimi distribution for $n=0$ and $n=1$ states for different values of $\lambda$. The Shannon entropy of the Wigner distribution becomes imaginary for $n>0$. Nonetheless, in the case of $n=0$ (ground state), the Wigner distribution is real, leading to the equivalence between the Husimi distribution and Wigner distribution for this state. However, for states other than the ground state, the Wigner distribution and Husimi distribution exhibit differences. The Shannon entropy (also known as Wehrl entropy) is real for all values of $n$ because the distribution is positive definite in all states.
Following the discussion on Shannon's entropy, we would like to discuss another significant measure in information theory. It is linked to the uncertainty associated with measurements in the quantum regime, as given in  \cite{uncefisher}. It quantifies the information conveyed by a particular observable regarding a parameter, while taking into account its inherent probability. The Fisher information is given by (based on ref. \cite{falaye2016fisher}),
\begin{align}\label{fisherintro}
     F = \sum_{i=1}^{m} \frac{1}{p_{i}}\bigg[\frac{dp_{i}}{di}\bigg]^{2},
\end{align}
In this context, $p_{i}$ denotes the probability density associated with the system being in micro-state $i$. 
Fisher's information is a measure of the amount of information a particular observable has about a parameter all the while taking into account the probability inherent in it. For a given observable $x$, Fisher information ($F_{x}$) is given by (based on ref. \cite{falaye2016fisher}),
\begin{equation}
    F_{x} = \int_{-\infty}^{\infty} dx \hspace{0.5em}|\psi(x)|^{2}\bigg[\frac{d}{dx}\ln|\psi(x)|^{2}\bigg]^{2}   > 0, 
\end{equation}
In reciprocal space, it is given by
\begin{equation}
    F_{k} = \int_{-\infty}^{\infty} dk\hspace{0.5em}|\phi(k)|^{2}\bigg[\frac{d}{dk}\ln|\phi(k)|^{2}\bigg]^{2}  > 0, 
\end{equation}
Whereas, the Heisenberg's indeterminacy principle in terms of Fisher's information measure is formulated as
\begin{equation}
    F_{x}F_{k} \geq 4. 
\end{equation}
Using these expressions, we generate plots of Fisher's measure in coordinate and momentum space for various values of $\lambda$. Additionally, we derive the uncertainty relation from our dataset. 
\begin{figure}[H]
\begin{subfigure}{0.3\textwidth}
  \centering
  \includegraphics[width=\linewidth]{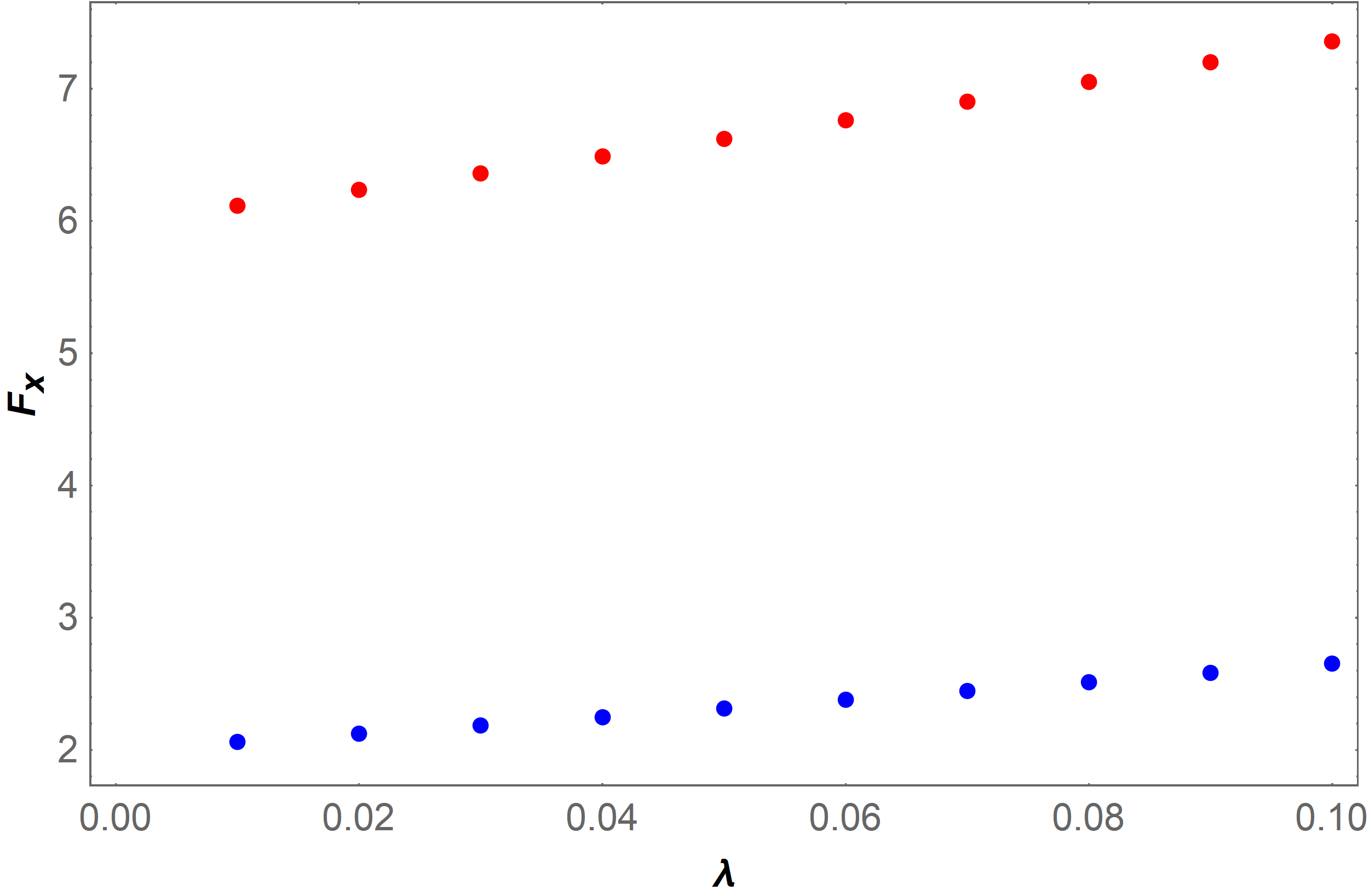}
  \caption{}
  \label{fig:image35}
\end{subfigure}
\begin{subfigure}{0.3\textwidth}
  \centering
  \includegraphics[width=\linewidth]{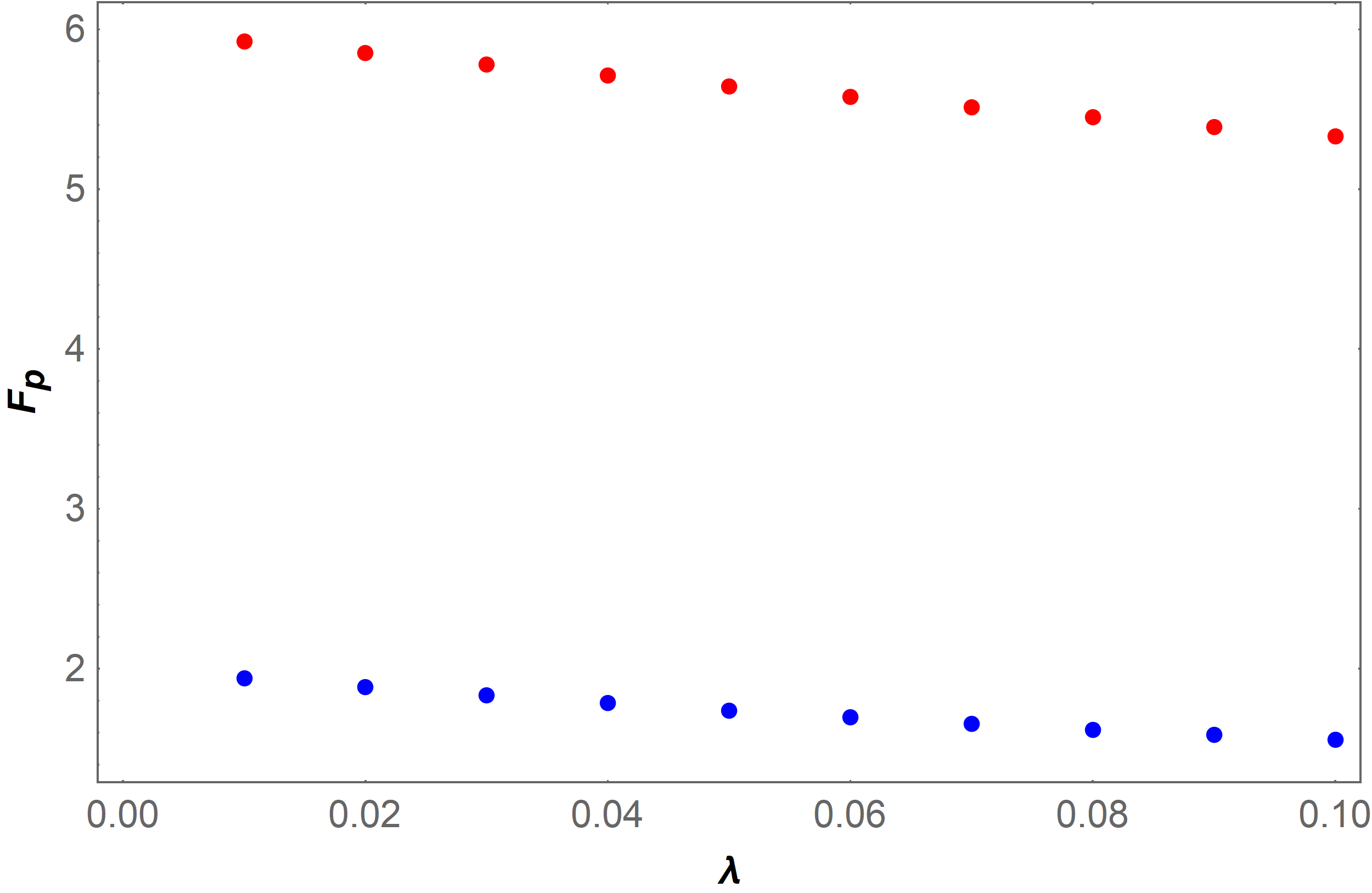}
  \caption{}
  \label{fig:image36}
\end{subfigure}
\begin{subfigure}{0.3\textwidth}
  \centering
  \includegraphics[width=\linewidth]{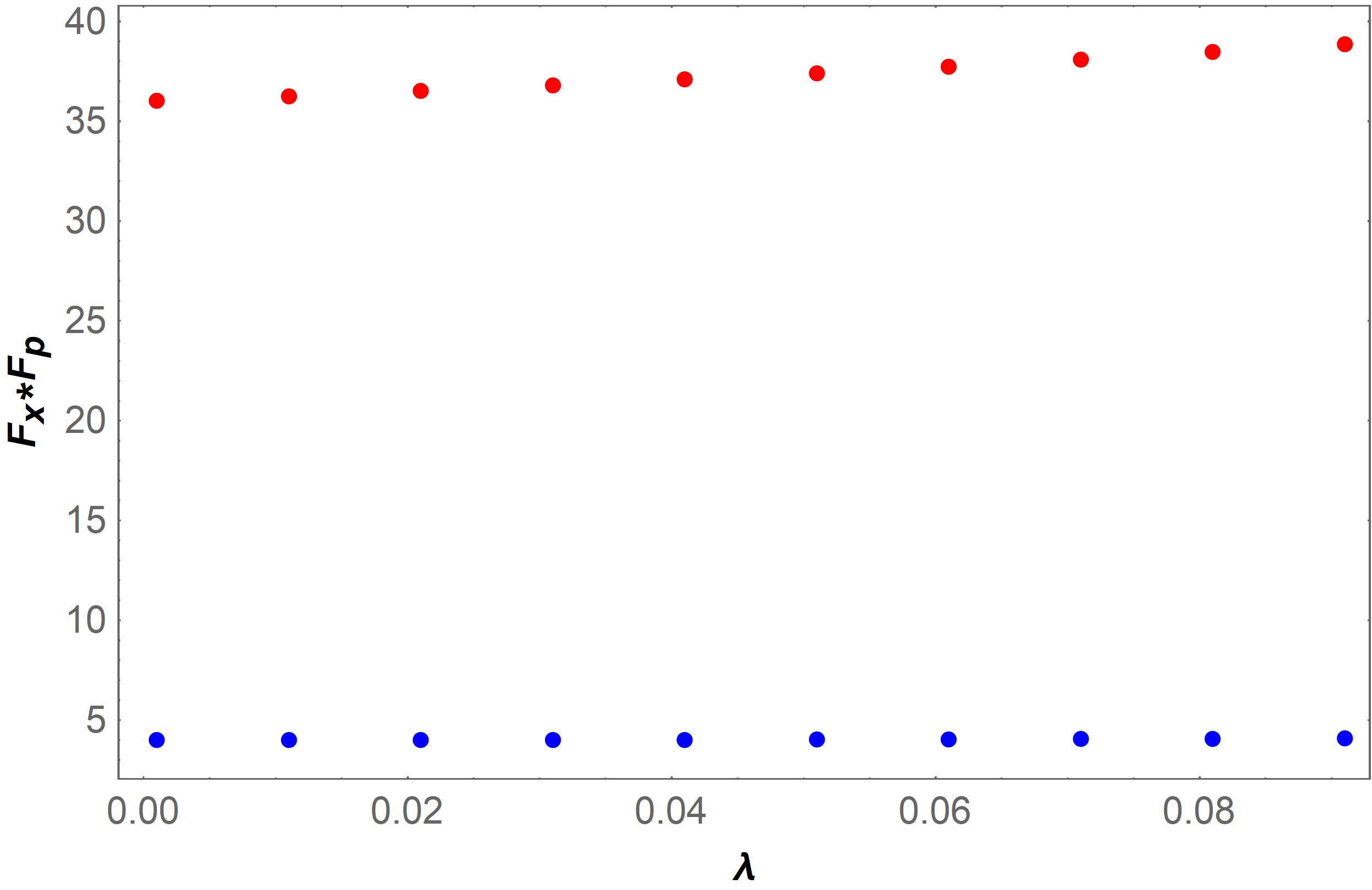}
  \caption{}
  \label{fig:image37}
\end{subfigure}
\caption{(a) Fisher information (FI) measure plot in coordinate space, (b) FI measure plot in momentum space, and (c) Uncertainty relation in the context of FI is $F_{x}\cdot F_{p}\geq 4$. Blue dots denote ground state, and red dots denote first excited state.}
\label{fig:fisherplots}
\end{figure}
We observe that as the value of $n$ increases, the Fisher information also increases. This is calculated using the wave functions in our scenario; however, it can also be extended using the marginals derived from Wigner distribution and Husimi distribution. Next, we proceed to study Rényi entropies. They are shown in fig. \ref{fig:renyientropy}. In contrast to the harmonic oscillator, the value of Rényi entropy remains independent of $n$ for $\alpha = 2$ and $\alpha = 4$. The value of Rényi entropy for $\alpha = 2$ obtained from the Wigner distribution increases for $n=0$, whereas it decreases for $n=1$. The same trend is observed with the Rényi entropies obtained from the Husimi distribution. One of the key differences is that the value of Rényi entropy obtained from the Husimi distribution is higher than that obtained from the Wigner distribution. This clearly shows that there is a loss of information associated with the Husimi distribution, consistent with observations from other entropic measures. Concerning the parameter $\lambda$, we can also infer that the uncertainty relation holds when the perturbing parameter is small, and when it increases at some point the uncertainty relation fails, this indicates that we are going out of the perturbation regime this is present in Fig. \ref{fig:renyientropy} (a). 

\subsection{Understanding Information Measures: KL, Jeffreys, and Rényi Divergence}
\begin{figure}[H]
\begin{subfigure}{0.245\textwidth}
  \centering
  \includegraphics[width=\linewidth]{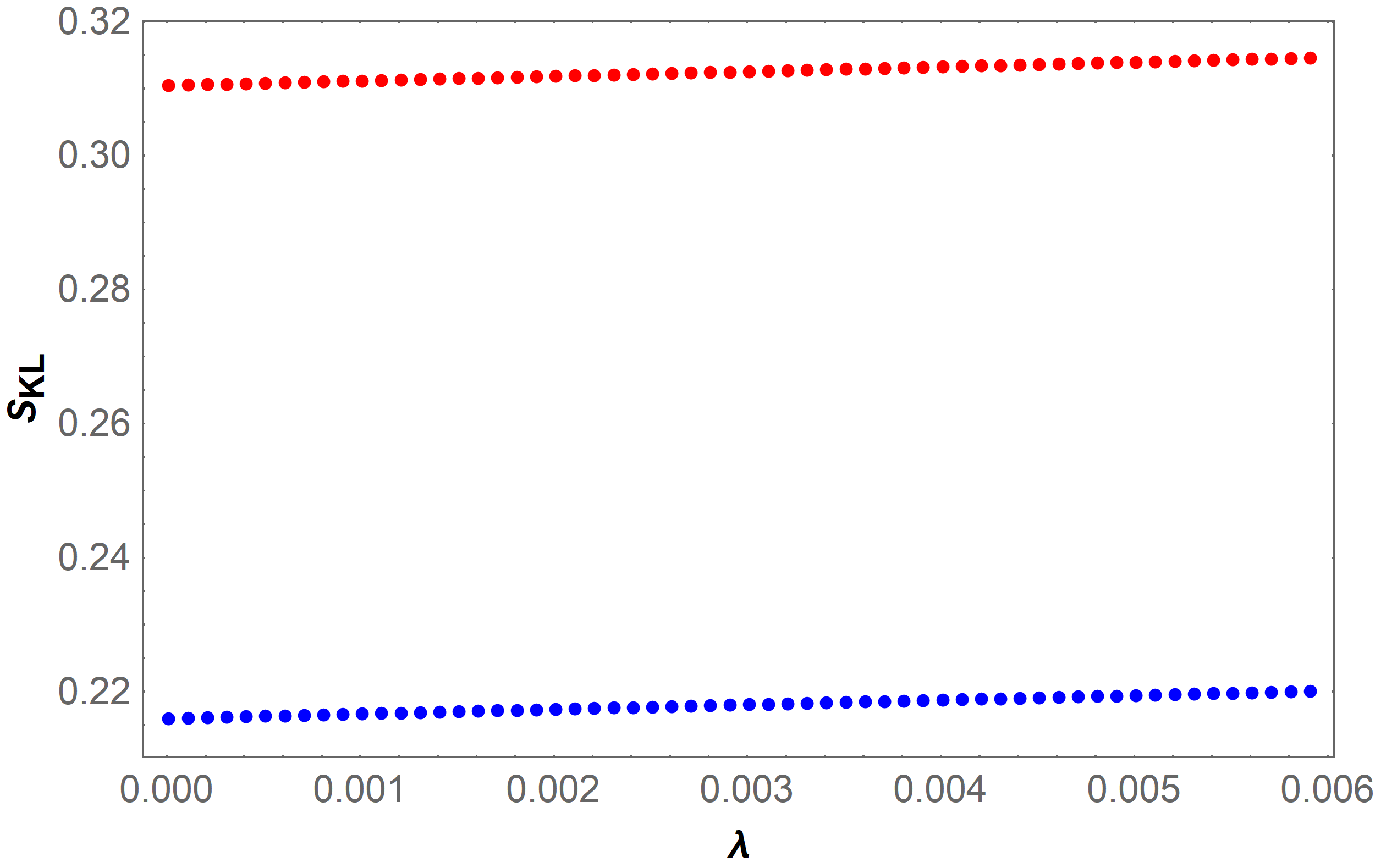}
  \label{fig:image42}
\caption{ }
\label{fig:Shannonplots1} 
\end{subfigure}
\begin{subfigure}{0.245\textwidth}
  \centering
  \includegraphics[width=\linewidth]{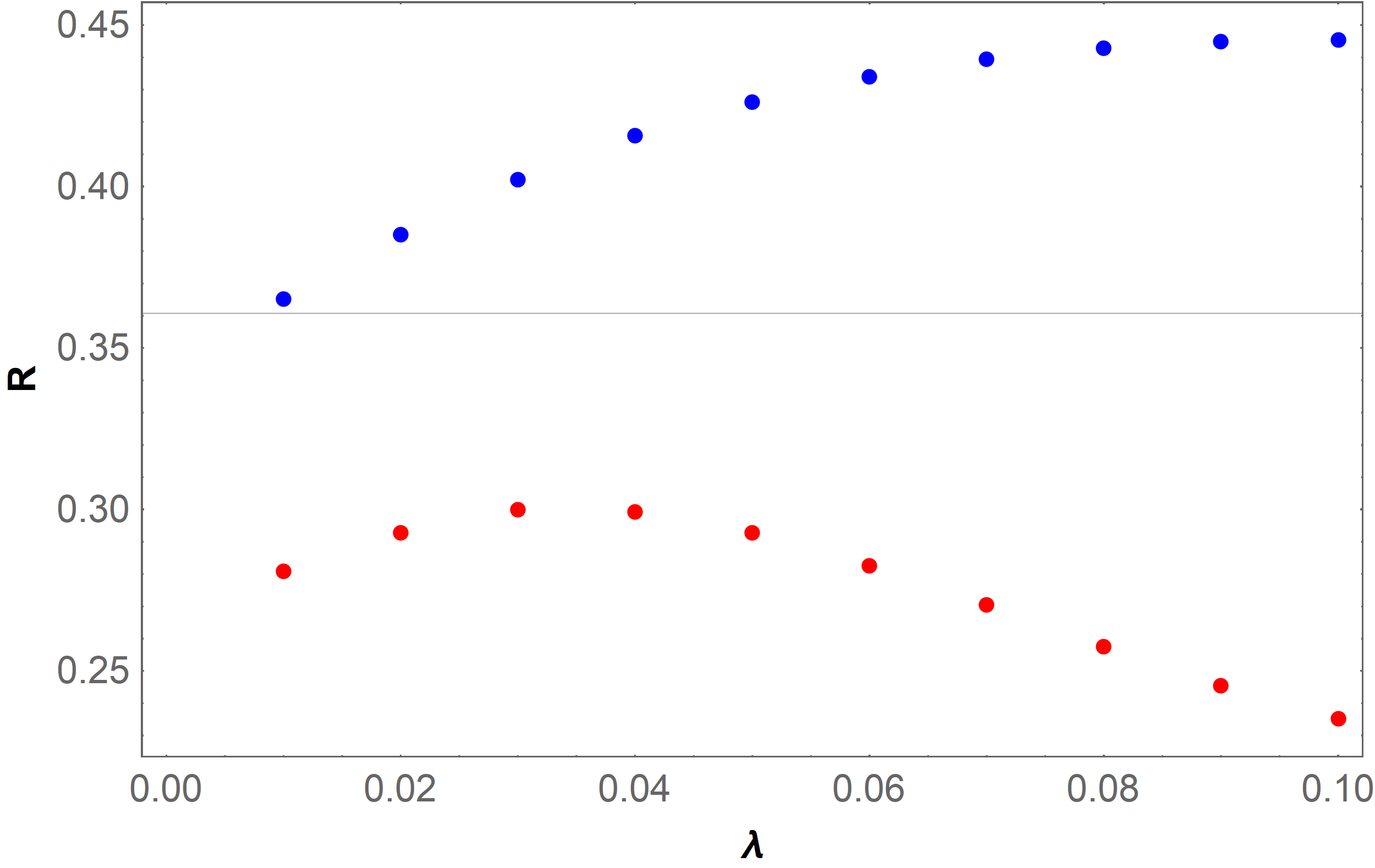}
  \label{fig:image43}
  \caption{}
\end{subfigure} 
\begin{subfigure}{0.245\textwidth}
  \centering
  \includegraphics[width=\linewidth]{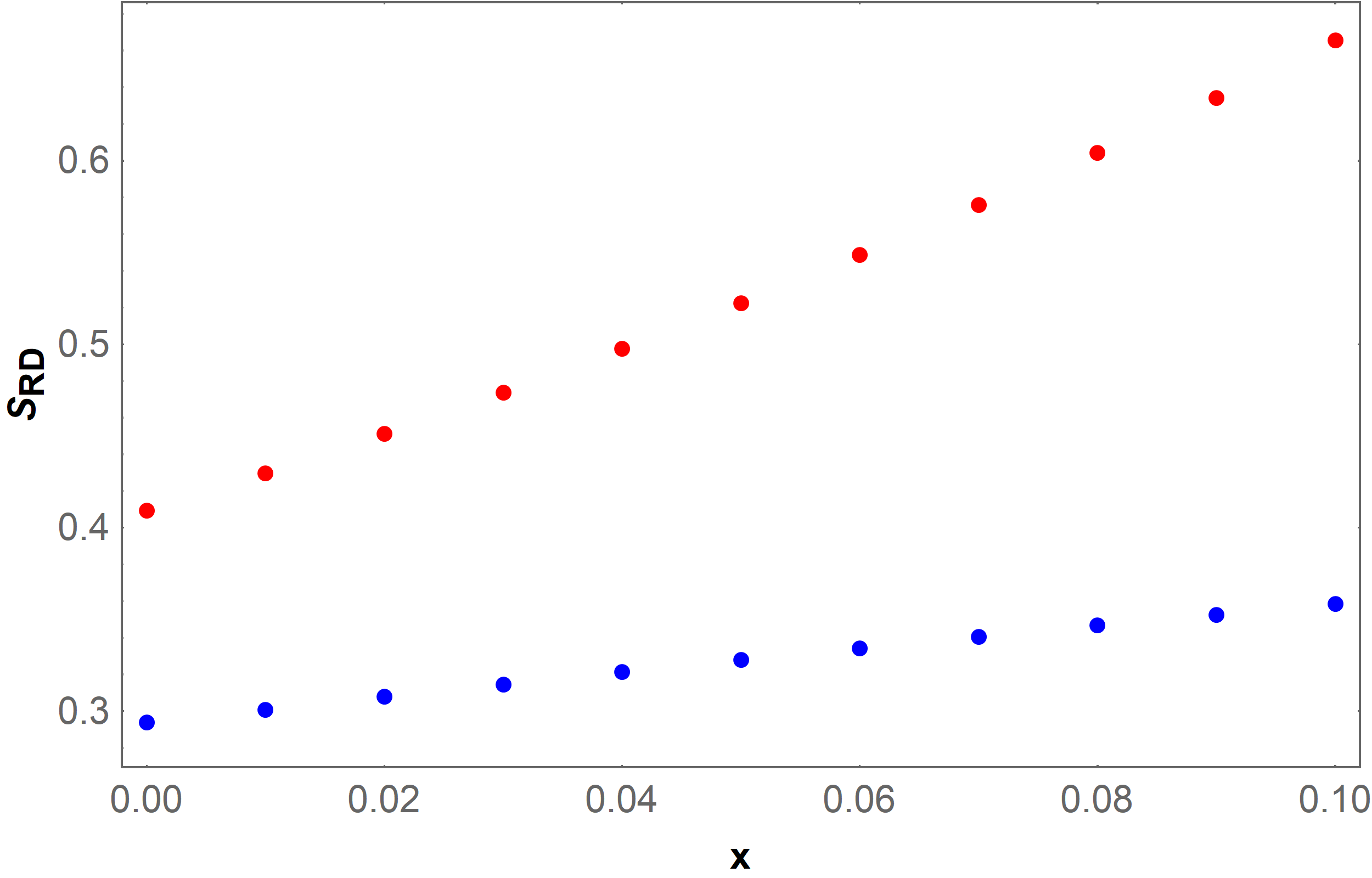}
  \label{fig:image44}
  \caption{}
\end{subfigure}
\begin{subfigure}{0.245\textwidth}
  \centering
  \includegraphics[width=\linewidth]{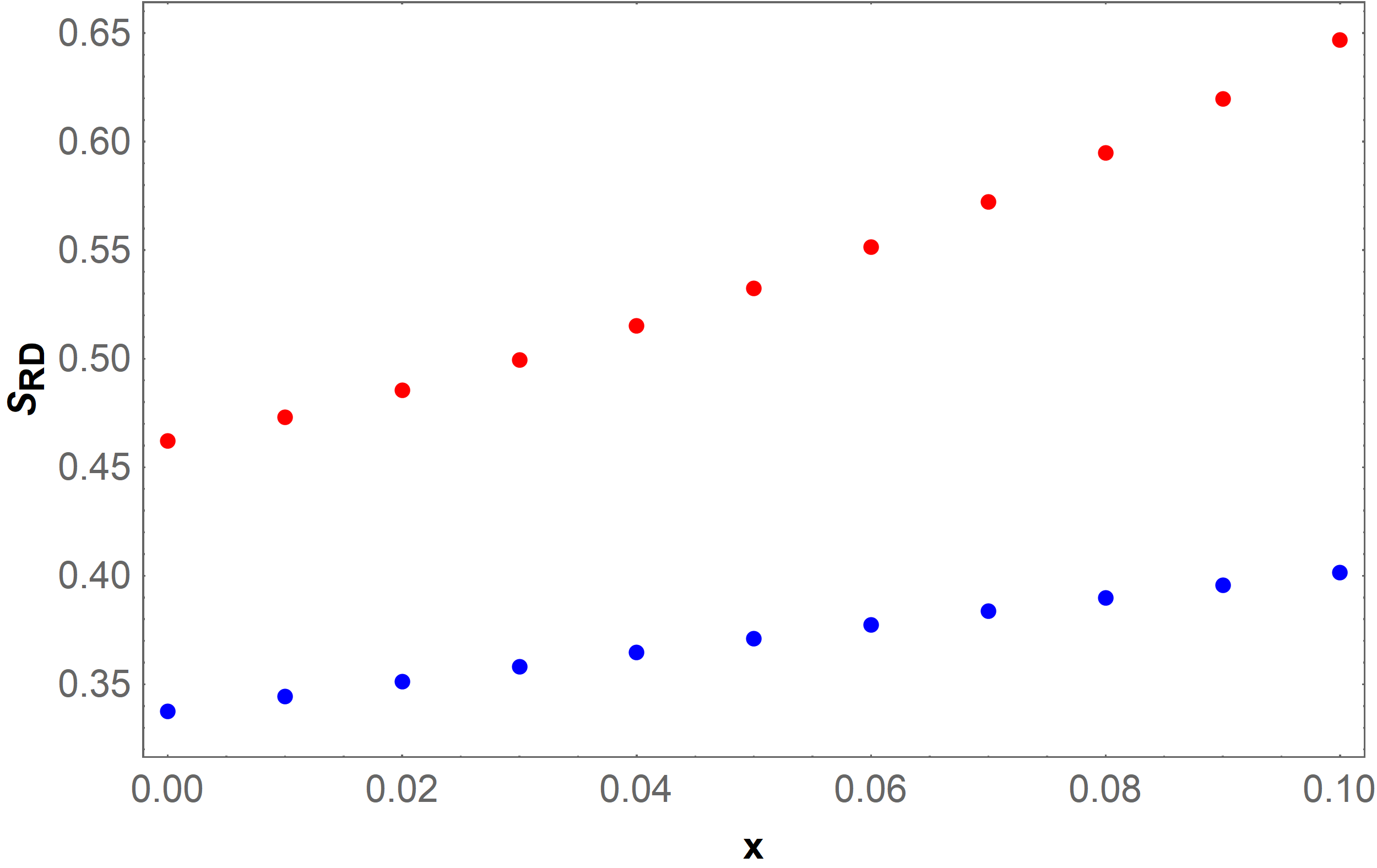}
  \label{fig:image45}
   \caption{}
\end{subfigure}
\caption{Plot (a) is Kullback-Leibler ($S_{KL}$) distance between Wigner and Husimi marginals, plot (b) is cumulative residual Jeffreys
divergence ($R$) distance between Wigner and Husimi marginals, Plots (c) and (d) are Rényi divergence ($D_{R}$ ) distance between Wigner and Husimi marginals. All the distances are plotted against the parameter $\lambda$ for $n=0$ and $n=1$ states. Blue dots denote the $n=0$ state, and red dots denote the $n=1$ state. }
\label{fig:divergenceplots}
\end{figure}

So far, we have obtained various distributions and their marginals. Now, one could eventually ask how these distributions and their respective marginals resemble each other. Specifically, we will examine how much the Wigner distribution resembles the Husimi distribution and how this resemblance depends on varying the parameter $\lambda$ for both ground and first excited states. The feature described in Plot \ref{fig:divergenceplots} highlights this aspect. A clear difference can be observed in the first two plots. The measure $S_{KL}$ indicates that the disparity between the coordinate space density and the Husimi distribution marginal widens as the values of $\lambda$ and $n$ increase. Meanwhile, the $R$ divergence decreases as $\lambda$ and $n$ values increase. This observation is important, as it demonstrates how the marginals and the distribution vary. This can also be observed clearly in figs. \ref{fig:image2main} and \ref{fig:image4main}. In fig. \ref{fig:image4main}, it is evident that as the value of $n$ increases, the value of the Husimi marginals also increases. The lack of structure in the Husimi marginals stands in contrast due to the growing quantity of nodes in the Wigner distribution. This relationship can be understood by examining the plot of $S_{KL}$. The increasing behavior of the Rényi divergences is consistent with the behavior of $S_{KL}$, as demonstrated in our plot.
\subsection{Correlation measures}
The previous sections covered discussions on phase space distributions, their entropies, and the disparities between those entropies and their marginals. These aspects are quantified through the analysis of correlation measures, which gauge the level of association or relationship between two distributions or entropies. We shall discuss some of the results in this subsection. We begin by discussing Cumulative Residual Entropy, which is the entropy corresponding to survival functions. For a constant value of the parameters $b$ and  $\lambda$, the value of cumulative residual entropy decreases as the value of another parameter $a$ increases. The value approaches a constant value for higher values of $b$, which is consistent with its distribution counterpart. The cross cumulative residual entropy is positive definite and it can be another method to resolve the negativity that arises in mutual information through Wigner distribution. The distance measure between two distributions, specifically the Wigner distribution and Husimi distribution, is determined by computing the Cauchy-Schwarz divergence measure (CS). It decreases as the parameter $\lambda$ increases for both $n=0$ and $n=1$. Moreover, it can be deduced, as the value of $n$ rises, divergence also rises.
\begin{figure}[H]
\begin{subfigure}{0.32\textwidth}
  \centering
  \includegraphics[width=\linewidth]{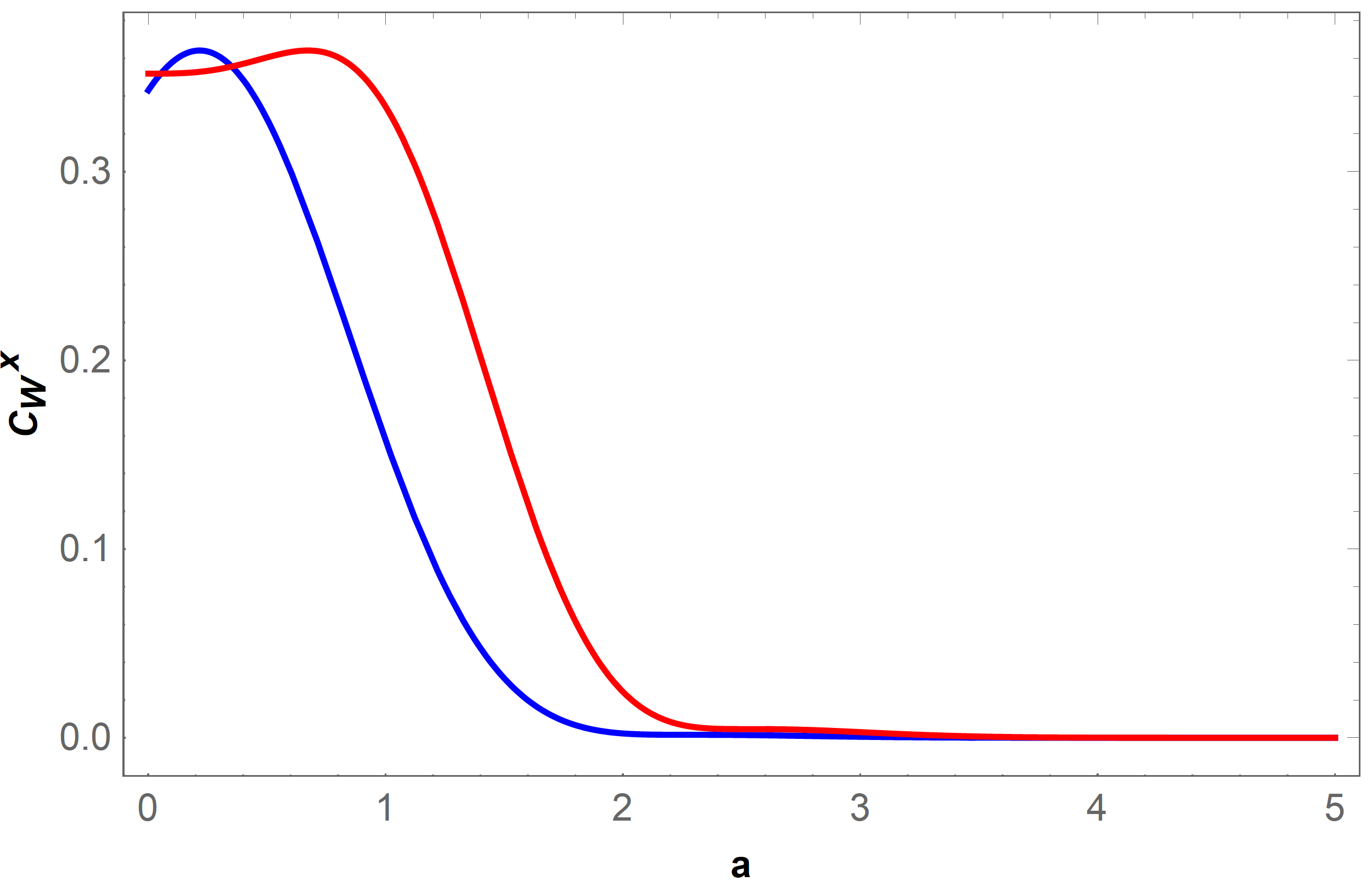}
    \caption{}
  \label{fig:image46}
\end{subfigure}
\begin{subfigure}{0.32\textwidth}
  \centering
  \includegraphics[width=\linewidth]{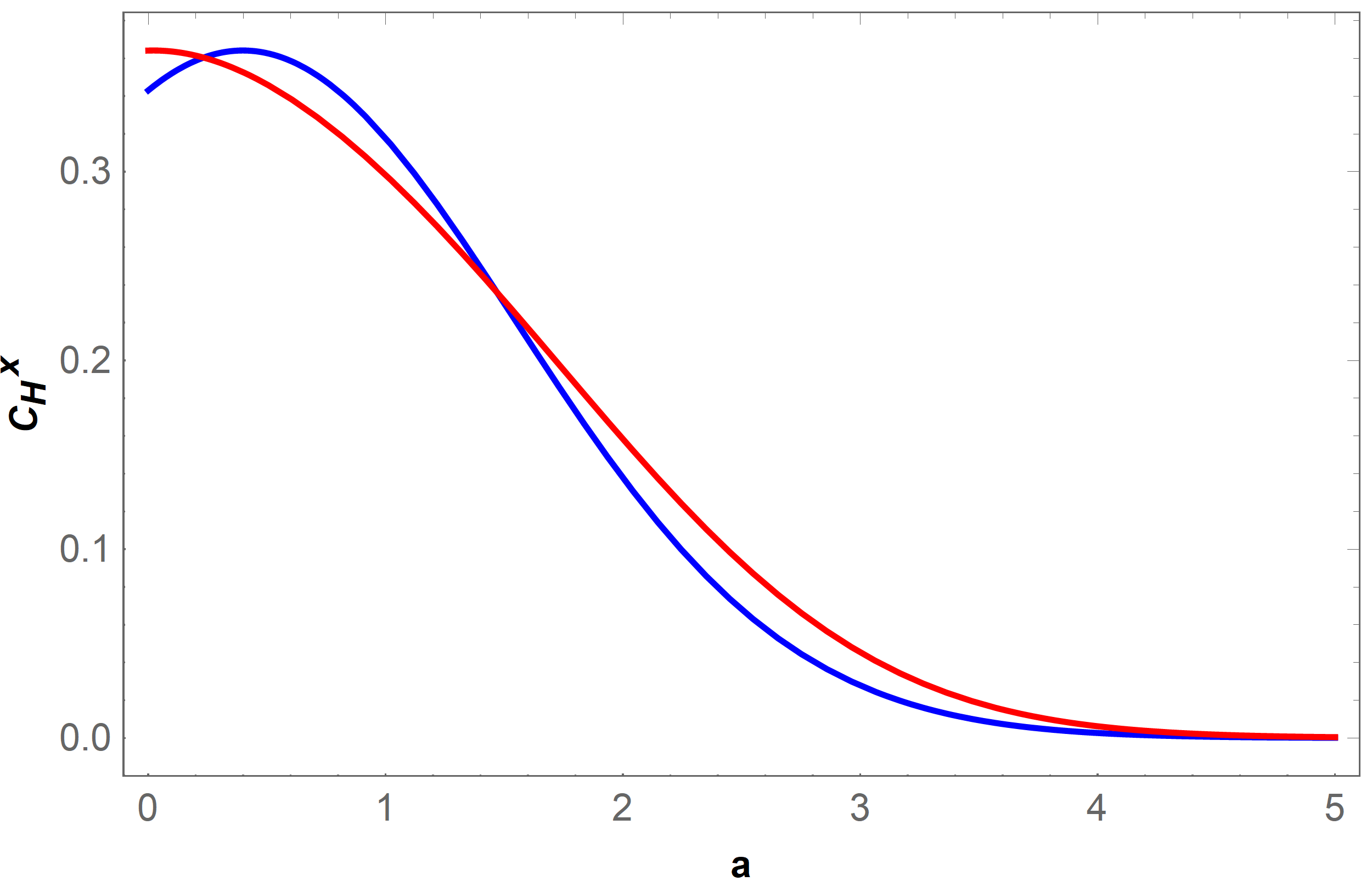}
    \caption{}
  \label{fig:image48}
\end{subfigure}
\begin{subfigure}{0.32\textwidth}
  \centering
  \includegraphics[width=\linewidth]{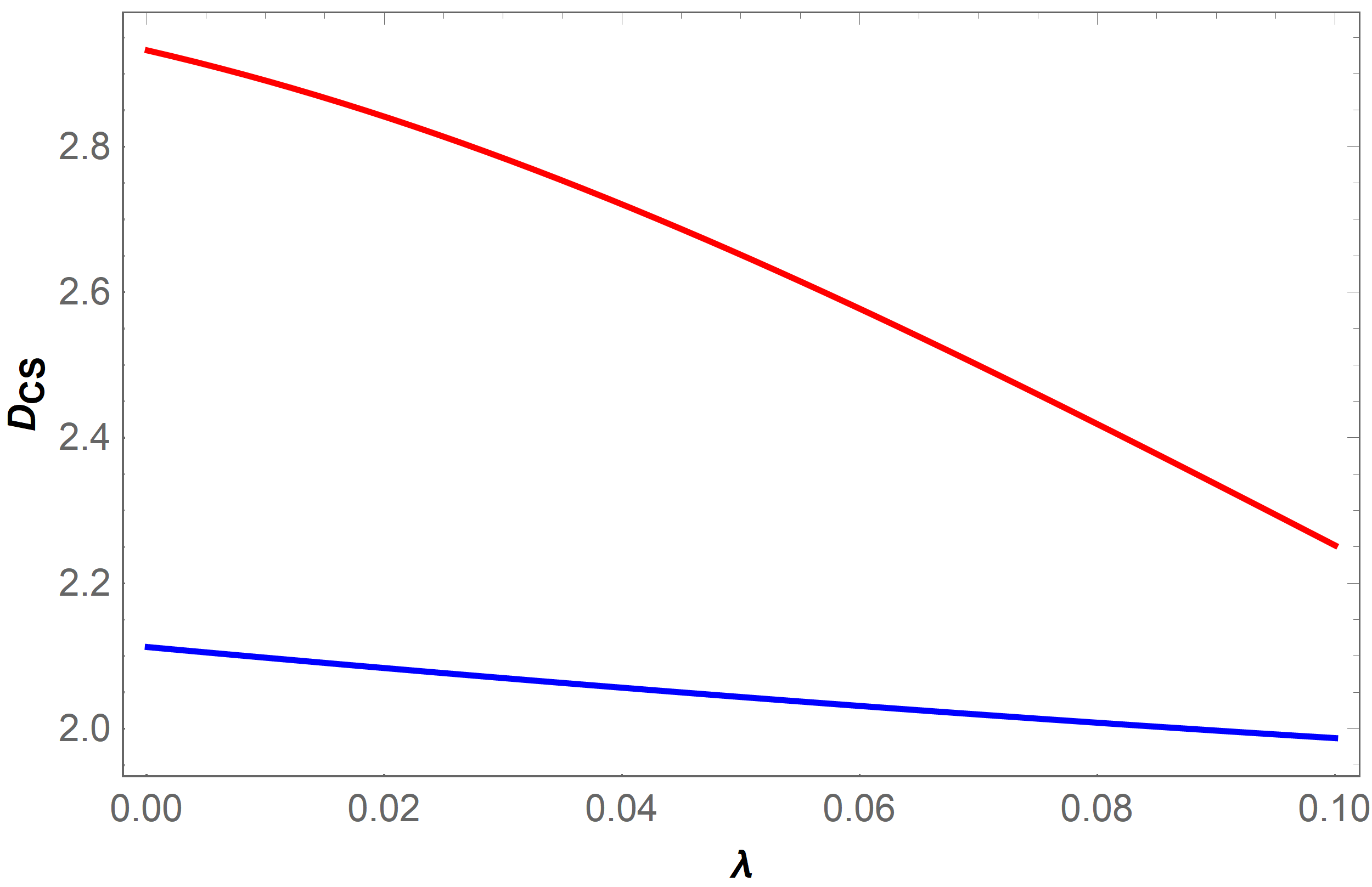}
    \caption{}
  \label{fig:image480}
\end{subfigure}
\caption{Plot (a) illustrates the cumulative residual entropy computed with Wigner distribution, while plot (b) represents the cumulative residual entropy calculated with Husimi distribution. Plot (c) showcases the Cauchy-Schwarz divergence. Blue lines represent the $n=0$ state, and red lines represent the $n=1$ state.
}
\label{fig:CMplots} 
\end{figure}

\begin{figure}[H]
\centering
\begin{subfigure}{0.4\textwidth}
  \centering
  \includegraphics[width=\linewidth]{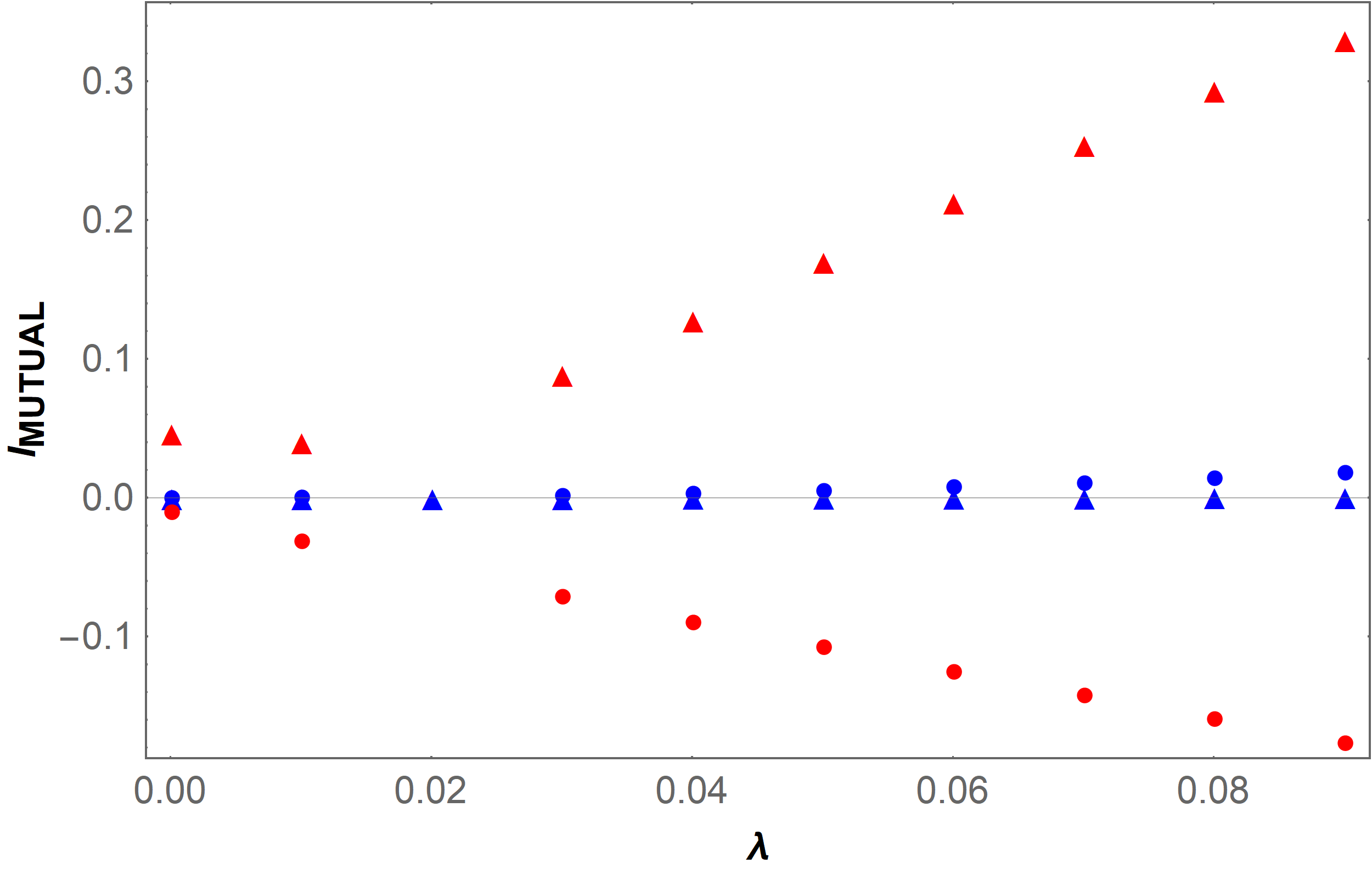}
\caption{}
  \label{fig:image49}
\end{subfigure}
\begin{subfigure}{0.4\textwidth}
  \centering
  \includegraphics[width=\linewidth]{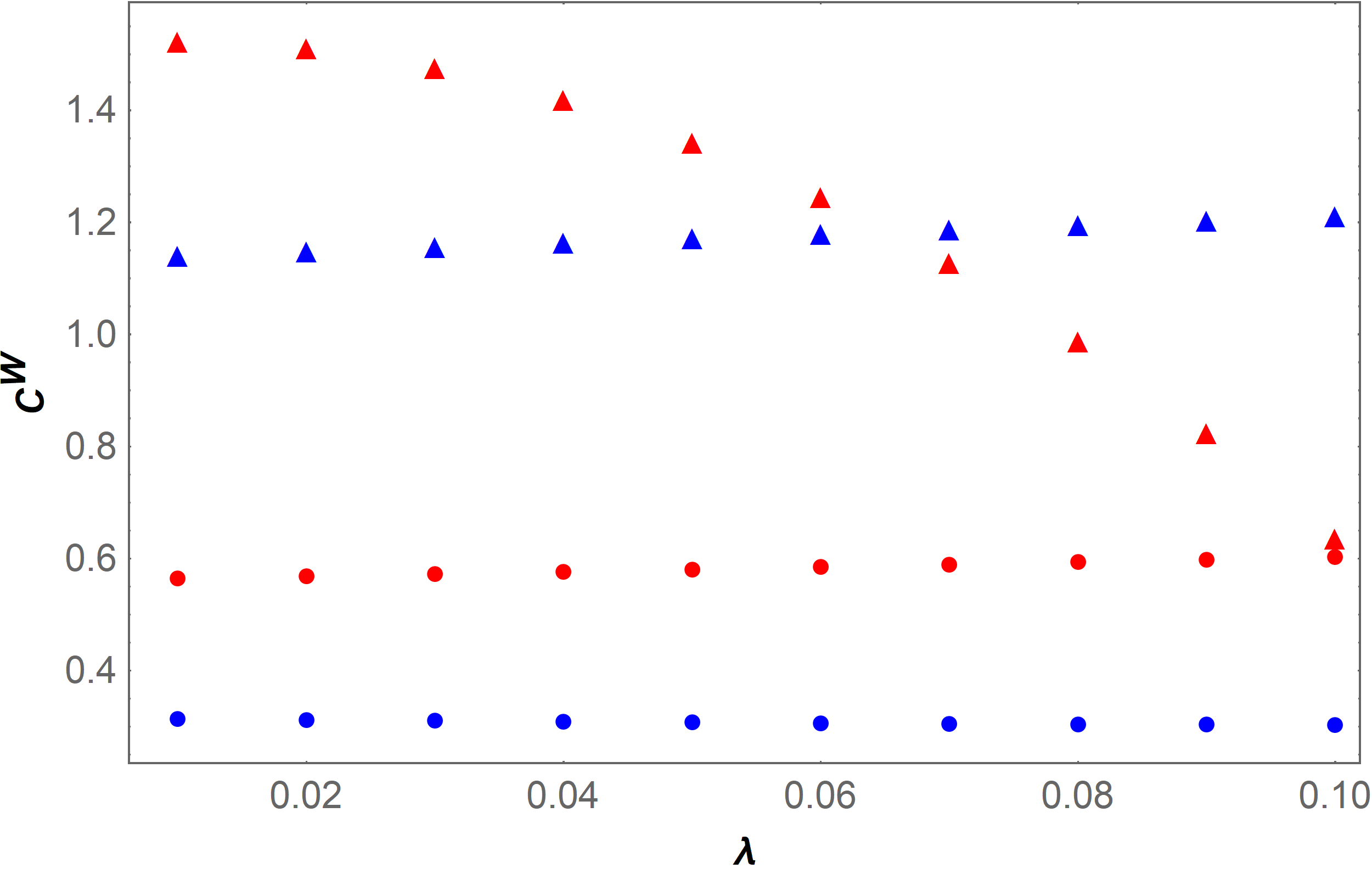}
    \caption{}
  \label{fig:image50}
\end{subfigure}
\caption{Plot (a) is mutual information from both the Wigner distribution and Husimi distribution concerning the parameter $\lambda$. Plot (b) is cross cumulative residual entropy from both the Wigner distribution and Husimi distribution concerning the parameter $\lambda$. Dots denote the values concerning the Wigner distribution and triangles denote the values concerning Husimi distribution. Blue color denotes the $n=0$ state and red color denotes the $n=1$ state. }
\label{fig:corelation} 
\end{figure}

\begin{figure}[H]
\centering
\begin{subfigure}{0.4\textwidth}
  \centering
  \includegraphics[width=\linewidth]{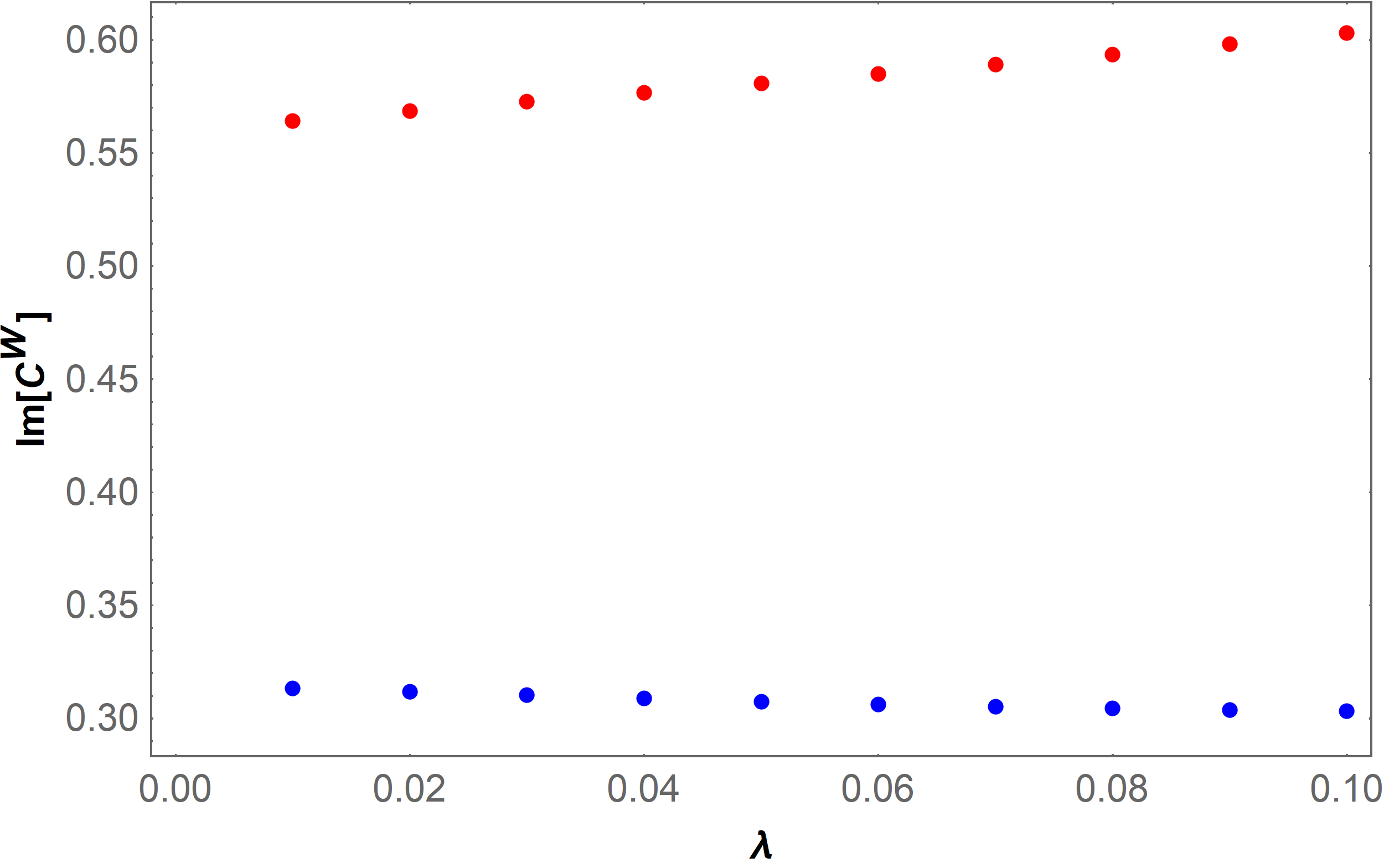}
      \caption{}
  \label{fig:image57}
\end{subfigure}
\begin{subfigure}{0.4\textwidth}
  \centering
  \includegraphics[width=\linewidth]{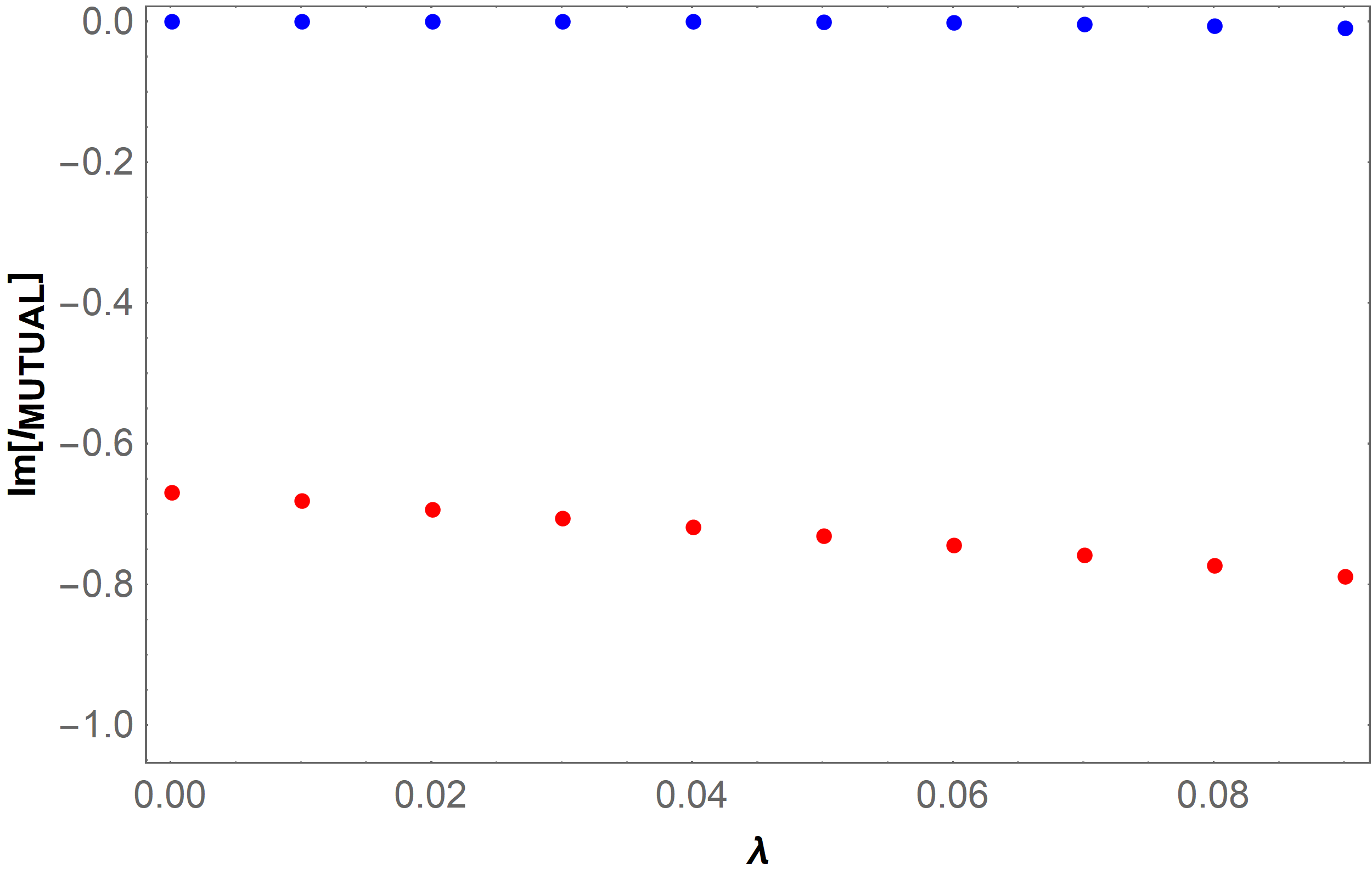}
  \caption{}
  \label{fig:image58}
\end{subfigure}
\caption{Plot (a) shows the Wigner distribution cross-cumulative residual entropy, while plot (b) displays the imaginary component of the Wigner distribution mutual information. The blue dots represent the $n=0$ state, and the red dots represent the $n=1$ state.}
\label{fig:imaginaryplots} 
\end{figure}

The correlation measures obtained with $\lambda$ for both ground and first excited states are given in Fig. \ref{fig:corelation}. Without any perturbation (when $\lambda = 0$), the values of correlation measures are zero, consistent with those of a harmonic oscillator \cite{salazar2023phase}. The correlation measures, namely mutual information and cross-cumulative mutual entropy, are as follows: Mutual information from the Wigner distribution for $n>0$ increases negatively due to the inherent imaginary nature of the Wigner distribution. Unlike the Wigner distribution, the correlation measures of the Husimi distribution increase for $n$. Overall, it remains unaffected by whether real components or absolute values are utilized in the mutual information and cross-cumulative residual entropy measures. 
In fig. \ref{fig:imaginaryplots}, we can observe how the value of mutual information ($I$) and cross cumulative residual entropy ($\mathcal{C}$), as described in equations \eqref{eq:mutualwigner} and \eqref{eq:crosscumulative}, obtained from the imaginary part of the Wigner distribution, increase in the negative direction as the values of $\lambda$ and $n$ increase. This is the particular region that is cut off when we apply Gaussian convolution to the Wigner distribution to obtain the Husimi distribution. By measuring this measure, we gain an insight into the extent of information loss incurred through the Gaussian filtering process applied to the Wigner distribution, as it portrays the actual probability distribution of the system in question.
\section{Conclusion}
This work aims to compare and analyze the actions of different information theoretic metrics while using the Wigner distribution and Husimi distribution. It is indeed a valuable tool for studying certain quantum mechanical systems, particularly those characterized by continuous variables like coordinate and momentum. Many methods have been developed to address its limitations, and we have discussed some of them as well. However, the primary focus is on using Gaussian convolution to obtain the Husimi distribution. Nevertheless, this results in increased entropy, as it is associated with information loss \cite{baez2011characterization}. We have clearly demonstrated how to utilize the Wigner distribution to derive various information-theoretic measures and have explored the available methods for expressing the entropic uncertainty bounds using phase-space distributions. The negative region of it, which is indeed a major limitation, yields Shannon and cumulative residual entropy with negative and imaginary components. We have studied all these aspects concerning the parameter $\lambda$ for the ground state $(n=0)$ and first excited states $(n=1)$. The magnitude (absolute and real values) of Shannon entropies and the other residual entropies increase as the value of $\lambda$ increases and also when the value of $n$ increases. The variations are consistent in this is consistent with marginal densities.\\ \\ 
Conversely, the behavior of Rényi entropies differs when studied with respect to another parameter, $\alpha$. For $\alpha = 2$, the value of Rényi entropy obtained using the Wigner distribution decreases as the value of $n$ increases, which is consistent with the results obtained in the harmonic oscillator \cite{salazar2023phase}. However, a point not discussed in that article is that when using the Husimi distribution, the magnitude of Rényi entropy increases with $n$, and after reaching a certain point, it starts decreasing with the increase in the parameter $\lambda$. It mirrors the situation observed when $\alpha = 4$, similar to the behavior of Rényi entropy obtained from the Husimi distribution for $\alpha = 2$. It was observed that the entropic sums obtained from the Wigner distribution are closer to the actual limit in contrast to the respective ones derived using the Husimi distribution. Even the individual entropies from the Wigner distribution are closer to those obtained from the Husimi distribution. Mutual information, which is a statistical correlation quantifying the coordinate momentum within the regime of Wigner distribution and Husimi distribution was also studied. It was observed to rise with increasing values of $\lambda$ and $n$. The absolute values of the correlations are greater for the correlations obtained from the Wigner distribution than those from the Husimi distribution. We also studied the relative entropy (divergence) of the distributions and their survivals, which rise with increasing values of $\lambda$ and $n$, except for Jeffrey's and Cauchy-Schwarz divergences. We can broaden this work to investigate quantum entanglement. Since the Wigner distribution is capable of analyzing both classical and quantum systems, its use allows us to derive Rényi entropy, which subsequently aids in calculating entanglement entropy \cite{calabrese2004entanglement,cerf1998information,islam2015measuring}. \\
\textit{Declaration:} The authors have equal contributions.
\bibliographystyle{unsrt}
\bibliography{ref}

\end{document}